\documentclass[rmp,reprint,amsmath,amssymb,longbibliography,floatfix,preprintnumbers]{revtex4-1}
\usepackage{hyperref}
\usepackage{tabu}
\usepackage{bm,bbm}
\usepackage{amsthm}
\usepackage{tikz}
\usetikzlibrary{intersections, calc,positioning,decorations.pathreplacing,decorations.pathmorphing,decorations.markings}
\usepackage{pgfplots}
\usepgfplotslibrary{fillbetween}

\theoremstyle{plain}
\newtheorem*{theorem*}{Theorem}

\usepackage{jcolor}
\definecolor{verylightgray}{rgb}{0.92, 0.9, 0.88}

\def\D{{\Delta}}

\def\nslash{\nabla \llap{/\,}}
\def\tr{{\rm tr}}
\def\i{\mathrm{i}}
\def\d{\mathrm{d}}
\def\CA{{\cal A}}
\def\CB{{\cal B}}
\def\CC{{\cal C}}
\def\CD{{\cal D}}

\def\CF{{\cal F}}

\def\CH{{\cal H}}

\def\CK{{\cal K}}
\def\CL{{\cal L}}
\def\CR{{\cal R}}
\def\CM{{\cal M}}
\def\CN{{\cal N}}
\def\CO{{\cal O}}

\def\CS{{\cal S}}
\def\CT{{\cal T}}

\def\BC{\mathbb{C}}

\def\BH{\mathbb{H}}
\def\BR{\mathbb{R}}
\def\BS{\mathbb{S}}
\def\BT{{\mathbb T}}
\def\BZ{\mathbb{Z}}

\def\SU{\mathrm{SU}}
\def\U{\mathrm{U}}
\def\SO{\mathrm{SO}}
\def\O{\mathrm{O}}

\begin{document}

\title{Entanglement entropy: holography and renormalization group}

\author{Tatsuma Nishioka}
\affiliation{Department of Physics, Faculty of Science, The University of Tokyo,
		Bunkyo-ku, Tokyo 113-0033, Japan}
\preprint{UT-18-02}

\begin{abstract}
Entanglement entropy plays a variety of roles in quantum field theory, including the connections between quantum states and gravitation through the holographic principle.
This article provides a review of entanglement entropy from a mixed viewpoint of field theory and holography.
A set of basic methods for the computation is developed and illustrated with simple examples such as free theories and conformal field theories.
The structures of the ultraviolet divergences and the universal parts are determined and compared with the holographic descriptions of entanglement entropy.
The utility of quantum inequalities of entanglement are discussed and shown to derive the $\CC$-theorem that constrains renormalization group flows of quantum field theories in diverse dimensions.
\end{abstract}

\maketitle
\tableofcontents{}

  \section{Introduction}
\label{ss:intro}

Entanglement is one of the most important concepts that distinguish quantum physics from classical physics as the former allows a superposition of states, causing a nonlocal correlation between subsystems far apart from each other.
A measure of quantum entanglement, known as entanglement entropy, has seen an unexpectedly wide range of applications in quantum information theory, condensed matter physics, general relativity, and even in high energy theory in recent years.
The most substantial progress in the subject includes the holographic formula for entanglement entropy proposed by \textcite{Ryu:2006bv,Ryu:2006ef}, which has proved to be a profitable tool to explore various aspects of quantum entanglement in strongly coupled quantum field theories, being a source of inspirational ideas in defining quantum gravity from quantum many-body entangled states \cite{Swingle:2009bg,VanRaamsdonk:2009ar,Maldacena:2013xja}.

The motivation for studying quantum entanglement varies depending on the area of research fields, and we are able only to make a partial list here.
In quantum information theory, quantum entanglement is exploited as an invaluable resource for manipulating computational tasks that are impossible to achieve in classical information theory \cite{nielsen2010quantum,Preskill:1997tq,Vedral:2002zz,ohya2004quantum,Eisert:2006ts,Horodecki:2009zz}.
In condensed matter physics, entanglement is recognized to characterize quantum phases of matter that cannot be distinguished by their symmetries \cite{Kitaev:2005dm,Levin:2006zz} and to diagnose quantum critical phenomena \cite{Vidal:2002rm,jin2004quantum,Calabrese:2004eu} and dynamics of strongly correlated quantum systems \cite{Calabrese:2005in,Calabrese:2007rg,eisler2007evolution}.
In quantum field theory (QFT), entanglement holds an eminent role as nonlocal operators definable for any type of theories that serve as a good probe for a variety of phase transitions such as confinement/deconfinement transition \cite{Nishioka:2006gr,Klebanov:2007ws,Pakman:2008ui}.
Moreover, the monotonic properties of entanglement measures have been successfully applied to derive nontrivial constraints on energy and entropy in recent studies \cite{Balakrishnan:2017bjg,Faulkner:2016mzt,Bousso:2015wca,Bousso:2014uxa,Bousso:2014sda}.

In this review we will review recent developments of entanglement entropy from holographic and field theoretic viewpoints, highlighting its aspect as measures of degrees of freedom under renormalization group (RG) flows in QFTs.
Seeking such a measure is a long-standing activity in theoretical physics. Zamolodchikov's $c$-theorem is one of the most beautiful outcomes proving the existence of the so-called $\CC$-function that monotonically decreases along any RG flow and agrees with the central charge of the conformal field theory (CFT) at fixed points of the flow in (1+1) dimensions \cite{Zamolodchikov:1986gt}.
The $\CC$-function is of theoretical importance for it orders QFTs along RG flows in the theory space and imposes strong constraints on them to rule out their unusual behaviors.

Attempts to extend the $\CC$-theorem to higher dimensions resulted in a conjecture by \textcite{Cardy:1988cwa} who employed a certain type of central charge for conformal anomalies as a measure of degrees of freedom in even spacetime dimensions.
This conjecture named the $a$-theorem was given a proof in four dimensions more recently by \textcite{Komargodski:2011vj}.
In odd dimensions, however, generalizing the $\CC$-theorem faced a significant obstacle as there are no conformal anomalies, hence no central charges.
A major breakthrough was triggered by two novel conjectures.
One is based on the observation that the universal finite part of the entanglement entropy for a spherical region obeys the $\CC$-theorem in a holographic setup in any dimensions \cite{Myers:2010xs,Myers:2010tj}.
The other, now known as the $F$-theorem, states the monotonicity of the free energy on an Euclidean sphere under any RG flow in odd dimensions \cite{Jafferis:2011zi,Klebanov:2011gs}.
These two conjectures were seemingly unrelated at first sight, but they turned out to be the same by showing the equivalence between the universal part of the sphere entanglement entropy and the sphere free energy for CFTs \cite{Casini:2011kv}.

This intriguing connection not only unified the two conjectures, but also was the key to the proof of the $F$-theorem in three dimensions \cite{Casini:2012ei} that shows the monotonicity of the renormalized entanglement entropy interpolating the ultraviolet (UV) and infrared (IR) values of the sphere free energy \cite{Liu:2012eea} based on the strong subadditivity, one of the most stringent inequalities of entanglement entropy, without directly relying on the unitarity of QFT in contrast to the proofs of the $\CC$-theorems in two and four dimensions.

Among the best applications of the $F$-theorem is to constrain the phase diagram of noncompact quantum electrodynamics (QED) coupled with $2N_f$ two-component fermions, which is in the conformal phase with a global symmetry $\SU (2N_f)$ for $N_f$ above a critical value $N_\text{crit}$, but is believed to flow for $N_f\le N_\text{crit}$ to the chiral symmetry broken phase that is described by $2N_f^2$ Nambu-Goldstone bosons and a free Maxwell field due to the spontaneous symmetry breaking to the subgroup $\SU (N_f)\times \SU (N_f) \times \U (1)$ at the IR fixed point.
The analyses using the $F$-theorem by \textcite{Grover:2012sp,Giombi:2015haa} exclude the possibility of any RG flow from the conformal to the broken symmetry phase for $N_f \gtrapprox 4.4$, hence suggesting the upper bound $N_\text{crit} \le 4$ that can be used as a benchmark for the estimates by the other methods \cite{DiPietro:2015taa,Karthik:2015sgq,Gusynin:2016som,Karthik:2016ppr,Herbut:2016ide,DiPietro:2017kcd}.\footnote{A recent study of QED$_3$ with $N_f =1$ shows the global symmetry is enhanced to $\O(4)$ that indicates $N_\text{crit} \le 1$ \cite{Benini:2017dus}.
We thank Igor Klebanov for pointing this out to us.
}

The same sort of argument with the quantum inequality of entanglement was extended more recently \cite{Casini:2017vbe,Lashkari:2017rcl} and yielded monotonic functions along RG flows in higher dimensions, which provides an alternative proof of the $a$-theorem in four dimensions \cite{Casini:2017vbe}.
These monotonic functions are of particular interest in themselves as a $\CC$-function due to their UV finiteness in any dimensions.
They certainly deserve further investigation regardless of their applications to proving the conjectures for the higher-dimensional $\CC$-theorem \cite{Myers:2010xs,Myers:2010tj,Klebanov:2011gs,Giombi:2014xxa}.

\subsection{Outline}
This review is intended to give a relatively self-contained exposition of the recent applications of the quantum entanglement inequalities to the dynamics of the RG flows in QFT.
We try to streamline various approaches to the QFT entanglement so that the reader can be quickly acquainted with the modern techniques used in literatures.

In Sec. \ref{ss:BasicEE}, we start with reviewing the fundamentals of bipartite entanglement in quantum mechanical (QM) systems. 
After defining the notion of separable and entangled states, we introduce several measures of quantum entanglement, such as entanglement and R{\'e}nyi entropies to quantify how much entanglement a quantum state possesses for a given bipartition of the system.
We discuss the relations between a few entanglement measures that we adopt in due course and summarize some of the most important inequalities they satisfy for later use in the QFT applications.

From Secs. \ref{ss:RTapproach}-\ref{ss:CFT}, we consider the entanglement entropy associated with a subregion of a constant time slice in QFT, whose evaluation needs more sophisticated techniques than in quantum mechanics due to the continuity of spacetime.
The real time formalism given in Sec. \ref{ss:RTapproach} is the most straightforward generalization of the quantum mechanical one where the spacetime is discretized on lattice and the entanglement entropy is calculated by taking the partial trace of the Hilbert space of the lattice system.
This approach can be implemented easily for free field theories and is best suited for the numerical calculations.

Section \ref{ss:Euclidean} takes an alternative approach that employs the so-called replica trick to reduce the calculation of the entanglement entropy to the partition function on a certain type of singular manifold in Euclidean QFT.
In the Euclidean formalism, the Lorentz invariance of the theory is manifest in contrast to the real time approach, and one can resort to the conventional QFT methods for the entropy calculation to study the UV divergent structures of the entanglement entropy by using the effective action on a curved background.
In Sec. \ref{ss:HeatKernel} we describe a way to fix a few coefficients of the UV divergent terms in the entanglement entropy by adapting the heat kernel method to a manifold with a singular locus of codimension-two.
Several useful identities obtained there will be applied to the derivation of important formulas for entanglement entropy in later sections.

Section \ref{ss:CFT} is concerned with the entanglement entropy in CFT, a class of QFTs invariant under the conformal symmetry that emerges at the fixed points of RG flows in the theory space of QFTs.
The conformal symmetry will be exploited to extract the universal parts of the entanglement entropy free from the ambiguity caused by the renormalization scheme of the UV divergences.
When the conformal anomalies exist, the case in even spacetime dimensions, we show the universal parts are characterized by the central charges and the shape of the codimension-two hypersurface surrounding the subregion to define the entanglement entropy in a time slice.

Section \ref{ss:Holography} begins with the quick overview of the AdS/CFT correspondence, the equivalence between the classical gravitational theory on the ($d+1$)-dimensional anti-de Sitter (AdS) space and CFT with a large number of degrees of freedom living on the $d$-dimensional boundary of the AdS space.
Given the AdS/CFT dictionary, we derive the holographic formulas of entanglement and R{\'e}nyi entropies and show they fulfill the characteristic properties of entanglement such as the strong subadditivity inequalities.
An intriguing relation of the entanglement entropy across a sphere to thermal entropy is established by making a coordinate transformation to the AdS black hole geometry where the holographic formula turns out to evaluate the black hole entropy proportional to the area of horizon.

In Sec. \ref{ss:RGflow} we explore the dynamics of RG flows in QFTs with the aid of quantum entanglement.
We use the mixture of the techniques in field theory and holography developed in the previous sections.
We first outline the motivation and current situation for the $\CC$-theorem that orders theories along RG flows in the space of QFTs.
Then we show the quantum inequalities of entanglement, in conjugation with the Lorentz invariance, and provide strong constraints on the RG flows that are enough to prove the entropic $c$- and $F$-theorems in (1+1) and (2+1) dimensions, respectively.
After briefly examining the implications for the dynamics of RG flows, the validity of the $F$-theorem is exemplified by explicit calculations of the entanglement entropy of a free massive scalar field in the large mass limit.
We compare the large mass expansion with the numerical results and see the agreement, but will find the nonstationary behavior at the UV fixed point that questions the stationarity of entanglement entropy under the relevant perturbation.
We comment on the apparent puzzle between the free scalar result and the conformal perturbation theory of entanglement entropy, and suggest a possible resolution by pointing out that the conformal symmetry is broken for a certain class of theories even at the UV fixed point.

To gain further insight from different viewpoints, we consider a few examples of holographic RG flows and evaluate the holographic entanglement entropies.
In an asymptotically AdS space describing a holographic gapped system, 
we find a topology change of the Ryu-Takayanagi hypersurface for the holographic entanglement entropy, which is interpreted as a confinement/deconfinement phase transition and indicates the prominent role of entanglement entropy as an order parameter of quantum phase transitions.
A small test of the $F$-theorem in the holographic RG models is carried out under the assumptions of the null energy condition as the bulk counterpart of the unitarity in QFT.
We conclude this review with a comment on the exact results on entanglement entropy and its generalizations in supersymmetric field theories.

\paragraph*{Conventions $:$}
Throughout this review we use natural units $\hbar= c =  k_B  =1$ for the Planck constant, the speed of light and the Boltzmann constant,
the mostly plus sign convention $(-,+,\cdots, +)$ for the Lorentzian metric in $((d-1) +1)$ dimensions, and the all plus sign convention $(+,+,\cdots, +)$ for the Euclidean metric in $d$ dimensions.
The imaginary unit is denoted by the roman letter ``$\i$."
The calligraphic letters $\CM$ and $\CB$ stand for a $d$-dimensional manifold for QFT and a $(d+1)$-dimensional manifold for gravitational theory, respectively.

\subsection{References to related subjects}
We assume the reader has background knowledge about the basics of QFT and general relativity.
Some familiarity with conformal field theory in higher dimensions and QFT on a curved spacetime, covered by the standard textbooks \cite{DiFrancesco:1997nk} and \cite{Birrell:1982ix}, would also be helpful.

The interested reader may refer to the other sources of literature listed for more comprehensive treatments of the subjects that we will or will not touch on in this review.

An introductory account of quantum entanglement in finite-dimensional systems is given by \textcite{nielsen2010quantum} and \textcite{Preskill:1997tq} with the emphasis on the application to quantum information theory.
The implementation of the real time approach is described for free lattice models by \textcite{peschel2009reduced} and for free field theories by \textcite{Casini:2009sr} that also  compares the approach with the Euclidean formalism.
The UV divergent structures of entanglement entropy are discussed for lattice models by \textcite{eisert2010colloquium} and for QFT on a curved space by employing the heat kernel method by \textcite{Solodukhin:2011gn}.
The condensed matter applications of entanglement being a probe of quantum critical phenomena and quantum quench dynamics are elaborated in the reviews by \textcite{Calabrese:2009qy,Calabrese:2016xau} by means of the CFT methods [see also \textcite{Laflorencie:2015eck}].
An accessible approach to more rigorous mathematical aspects of the entanglement properties in QFT can be found in the recent review by \textcite{Witten:2018zxz}.
The developments of the holographic method in the early days are summarized by \textcite{Nishioka:2009un} and more recent ones by \textcite{Takayanagi:2012kg}.
The role of quantum entanglement in the black hole information problem was emphasized by \textcite{Harlow:2014yka}.
Recent attempts to build up spacetime geometry from the entanglement structure in QFT can be found in \textcite{VanRaamsdonk:2016exw}.
The exact results for the $F$-theorem in supersymmetric field theories are presented by \textcite{Pufu:2016zxm}.
Finally, we recommend \textcite{Rangamani:2016dms} for a comprehensive overview of the entire subjects.

  \section{Entanglement in quantum mechanical system}\label{ss:BasicEE}

We introduce the notion of bipartite entanglement for pure states in finite-dimensional quantum mechanical systems and classify the states into two types depending on whether they contain nontrivial entanglement entropy that quantifies the amount of quantum entanglement.
A set of inequalities of entanglement entropy, which will play crucial roles in the latter sections, will be given without proofs.
Other entanglement measures frequently used in the literature will also be introduced and compared with entanglement entropy.

\subsection{Bipartite entanglement}
\label{ss:def EE}
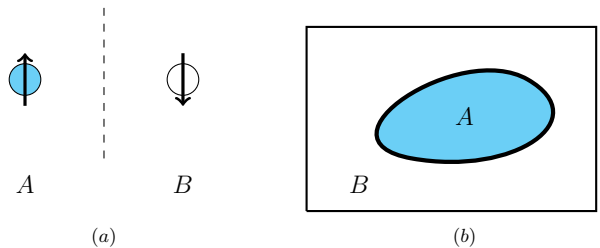
\begin{figure}[htbp]
\centering
	\begin{tikzpicture}[scale=.7, every node/.style={scale=0.8}]
		\draw[dashed] (0, -1.5) -- (0, 1.5);
		\filldraw[fill=cyan!50] (-1.5, 0) circle (0.3);
		\draw (1.5, 0) circle (0.3);
		\draw[very thick, ->] (-1.5, -0.5) -- (-1.5, 0.5);
		\draw[very thick, ->] (1.5, 0.5) -- (1.5, -0.5);
		\node at (-1.5, -2) {\large $A$};
		\node at (1.5, -2) {\large $B$};
		\node at (0, -3) {$(a)$};
	\end{tikzpicture}
	\qquad\qquad
	\begin{tikzpicture}[scale=.7, every node/.style={scale=.8}]
		\draw[thick] (-3,-2) -- (2.5, -2) -- (2.5, 1.5) -- (-3, 1.5) -- (-3, -2);
		\filldraw[ultra thick, fill=cyan!50] (-1, -1) .. controls +(170:2) and +(150:1.5) .. (1.2, 0.5) .. controls +(-30:1.5) and +(-10:2) .. (-1, -1);
		\node at (0,-0.2) {\large $A$};
		\node at (-2,-1.5) {\large $B$};
		\node at (0, -2.5) {$(b)$};
	\end{tikzpicture}
\caption{
\label{fig:spinchain}
The decomposition of the system into a subsystem $A$ shown in blue and its complement $B$.
$(a)$ Two spin systems. The subsystems $A$ and $B$ are left and right spins, respectively.
$(b)$ In $d$-dimensional quantum field theory, a spatial region at a given time slice is split into the subsystems $A$ and $B$ whose common boundary $\partial A = \partial B$ is always a codimension-two hypersurface in $d$ dimensions.
}
\end{figure}

Given a lattice model or QFT, suppose the system is in a
pure ground state $|\Psi\rangle$, i.e., 
the density matrix for the Hilbert space $\CH_\text{tot}$ is given in the form\footnote{We normalize the ground state as $\langle \Psi |\Psi\rangle = 1$ so that $\tr_\text{tot} (\rho_\text{tot}) = 1$.}
\begin{align}\label{pure}
\rho_\text{tot}=|\Psi\rangle \langle
\Psi| \ . 
\end{align}
We then divide the total system into two subsystems $A$ and $B = \bar A$ complementary to each other as in Fig.\ \ref{fig:spinchain}.

In the spin chain example,
we cut off the chain in between the sites and divide the lattice points into two groups. 
Note that this cutting procedure is an imaginary process without changing the system at all.
In what follows, the total Hilbert space will be assumed to take a direct product form of two Hilbert spaces of the subsystems,\footnote{This assumption is not necessarily valid for QFTs in general, especially with gauge symmetries.
We, however, will ignore this issue for simplicity in the reminder of this review.
For discussions and attempts to define entanglement entropy in gauge theories, see, e.g., \textcite{Casini:2013rba,Radicevic:2014kqa,Donnelly:2014gva,Donnelly:2014fua,Huang:2014pfa,Ghosh:2015iwa,Hung:2015fla,Aoki:2015bsa,Chen:2015kfa,Donnelly:2015hxa,Radicevic:2015sza,Pretko:2015zva,Soni:2015yga,Ma:2015xes,VanAcoleyen:2015ccp} and a review by \textcite{Pretko:2018yxl}.
}
\begin{align}
	\CH_\text{tot} = \CH_{A}\otimes \CH_{B} \ .
\end{align}

Let $\{ |i\rangle_B, i=1,2,\cdots\}$ be an orthonormal basis in $\CH_B$, and define
the {\it reduced density matrix} $\rho_A$ of the system $A$ by taking the partial trace over the system $B$,
\begin{align}\label{ReducedDM}
\rho_A \equiv \tr_{B}( \rho_\text{tot} ) \equiv \sum_{i} {}_B\langle i |\, \rho_\text{tot} \,| i\rangle_B \ .
\end{align}
Note that this definition depends on the choice of the subsystem, but not on the choice of the orthonormal basis $\{ |i\rangle_B\}$.
For example, if $\rho_\text{tot}$ is the tensor product of the density matrices $\rho_A$ and $\rho_B$ describing the subsystems, $\rho_\text{tot} = \rho_A \otimes \rho_B$, the partial trace simply recovers them, $\tr_B(\rho_\text{tot}) = \rho_A$ and $\tr_A(\rho_\text{tot}) = \rho_B$.

An alternative characterization of the reduced density matrix is the condition
\begin{align}
	\tr\left(\rho_\text{tot}\, \CO \right) = \tr_A\left( \rho_A \,\CO_A\right)\ ,
\end{align}
for any operator $\CO$ of the form $\CO = \CO_A \otimes {\bf 1}_B$ with the identity operator ${\bf 1}_B$ in $\CH_B$.
In this sense, the reduced density matrix $\rho_A$ has enough information for $\CH_A$ to reconstruct every correlation function in the subregion $A$.
The total density matrix needs not be pure in this characterization, but one may wonder, once the complete information $\rho_A$ about a system $A$ is given, if it is possible to find a pure density matrix in an enlarged Hilbert space of $\CH_A$ whose partial trace recovers $\rho_A$.
Indeed the answer is affirmative, and one can always construct such an enlarged Hilbert space and the density matrix in the following way.
The most general density matrix is of the form
\begin{align}
	\rho_A = \sum_i\, p_i\, |i\rangle_A {}_A\langle i | \ ,
\end{align}
where $\{ |i\rangle_{\bar A} \}$ is an orthonormal basis of $\CH_A$ and the coefficients $p_i \ge 0$ sum up to $1$, $\sum_i \,p_i = 1$.
We then copy $\CH_A$ into another Hilbert space $\CH_{\bar A}$ with the basis given by $\{ |i\rangle_{\bar A} \}$ and define a pure density matrix $\rho$ by
\begin{align}\label{Def_Purification}
	\rho = |\chi\rangle \langle \chi| \ , \qquad |\chi\rangle \equiv \sum_{i}\, \sqrt{p_i} \, |i\rangle_A \otimes | i \rangle_{\bar A} \ ,
\end{align}
in the enlarged Hilbert space $\CH = \CH_A\otimes \CH_{\bar A}$.
It is straightforward to check that this construct correctly reproduces $\rho_A$ under the partial trace over $\bar A$.
The operation is called \emph{entanglement purification} as it constructs a pure unentangle state from a mixed entangled state by enlarging the Hilbert space.
In Sec. \ref{ss:RindlerTFD}, we see that the entanglement purification turns out to be a fundamental concept in understanding what an observer restricted to a subregion feels like in a thermal system in QFT.

Finally let us introduce a measure of entanglement, the {\it entanglement entropy} of the subsystem $A$, by the von Neumann entropy of the reduced density matrix $\rho_A$,
\begin{align}\label{EE_def}
S_A =
- \mathrm{tr}_{A}\left[
\rho_{A} \log \rho_{A}\right] \ .
\end{align}
Note that the entanglement entropy of the total system always vanishes, $S_\text{tot} = 0$ for a pure ground state \eqref{pure}. 
Entanglement entropy remains finite in a finite-dimensional quantum system, but it suffers from UV divergences in QFT due to the short range interaction near the boundary $\partial A$ of the subsystem $A$ as discussed in detail in Sec. \ref{ss:Euclidean}.

\subsection{Separable and entangled states}
Having introduced a notion of entanglement entropy as the von Neumann entropy of the density matrix of a subsystem $A$, we want to understand what it really measures for a given system.
To proceed with the discussion, consider a pure ground state $|\Psi\rangle$ in a general form:
\begin{align}\label{PureProductState}
|\Psi\rangle = \sum_{i,\mu} c_{i\mu}\, |i\rangle_A  \otimes |\mu \rangle_B  \ ,
\end{align}
where $|i\rangle_A$ and $|\mu\rangle_B$ are orthonormal bases for $\CH_A = \{ |i\rangle_A, i = 1, \cdots, d_A\}$ and $\CH_B = \{ |\mu\rangle_B, \mu = 1, \cdots, d_B\} $, respectively, and the coefficient $c_{i\mu}$ is a $d_A \times d_B$ matrix with complex entries.
There are two different cases depending on the type of the coefficient matrix $c_{i\mu}$.

\subsubsection{Separable state}
When $c_{ij}$ factorizes, $c_{ij} = c_i^A c_\mu^B$, the ground state $|\Psi\rangle$ is called a \emph{separable} state (a pure product state) and can be recast into the product form,
\begin{align}
|\Psi\rangle = |\Psi_A \rangle \otimes |\Psi_B\rangle \ , 
\end{align}
where $|\Psi_{A} \rangle \equiv \sum_i c_i^{A} |i\rangle_{A}$ and $|\Psi_{B} \rangle \equiv \sum_\mu c_\mu^{B} |\mu\rangle_{B}$.
This is the case where the reduced density matrix, Eq.\,\eqref{ReducedDM}, also becomes pure $\rho_A = |\Psi_A\rangle\langle\Psi_A|$.
Thus a separable state has vanishing entanglement entropy
\begin{align}
	S_A = 0 \ .
\end{align}
Moreover, one can show entanglement entropy vanishes if and only if the pure ground state is separable as we will see.

\subsubsection{Entangled state} 
The ground state is called an {\it entangled} (or inseparable) state if it is not separable (with the coefficient matrix $c_{i\mu} \neq c_i^A c_\mu^B$).
This is the case where the entanglement entropy takes a positive value.

Indeed, we can simplify Eq.\,\eqref{PureProductState} by changing the bases into the \emph{Schmidt decomposition} form,
\begin{align}\label{PureDiag}
|\Psi\rangle = \sum_{k=1}^{\text{min}(d_A, d_B)} \sqrt{p_k}\, |\psi_k\rangle_A\otimes  |\psi_k \rangle_B \ ,
\end{align}
where $p_k$ are non-negative real numbers satisfying $\sum_k p_k = 1$ and
$|\psi_k\rangle_{A,B} $ are new orthonormal bases for the subsystems $A$ and $B$.
Note that this decomposition works for any $d_A \times d_B$ rectangular matrix $c_{i\mu}$.
To see how it actually works, we ``diagonalize" the coefficient matrix by the singular-value decomposition,
\begin{align}
c = U \, \Sigma\, V^\dagger \ ,
\end{align}
where $U$ and $V$ are $d_A \times d_A$ and $d_B\times d_B$ unitary matrices.
$\Sigma$ is a diagonal $d_A\times d_B$ real matrix given by
\begin{align}
	\Sigma = 
		\begin{cases}
			\left(\begin{array}{ccc|c}
			\sqrt{p_1} & & &  \\
			& \ddots & & \bf 0 \\
			& & \sqrt{p_{d_B}} & 
			\end{array}\right)  &\qquad  (d_A < d_B) \ , \\
			\\
			\left(\begin{array}{ccc}
			\sqrt{p_1} & &  \\
			& \ddots &  \\
			& & \sqrt{p_{d_A}}\\
			\hline
			 & \bf 0 &
			\end{array}\right) &\qquad  (d_A \ge d_B) \ , 
		\end{cases}
\end{align}
with non-negative real entries $\sqrt{p_k}\ge 0$, $k=1,\cdots, \text{min}(d_A, d_B)$.
The square root for the eigenvalues is purely conventional for later use.
The new orthonormal bases  $|\psi_k\rangle_{A,B}$ are the unitary transformation of the original ones $|i\rangle_{A,B}$ as $|\psi_k\rangle_A = \sum_{i} U_{ik} |i\rangle_A$ and $|\psi_k\rangle_B = \sum_{\mu} V_{k\mu} |\mu\rangle_B$.

The Schmidt decomposition Eq.\,\eqref{PureDiag} is particularly nice as it yields a mixed state density matrix with the probability distribution $\{ p_k\}$ for the reduced density matrix,
\begin{align}\label{Red_DM_A}
\begin{aligned}
	\rho_A &= \sum_{i=1}^{d_B} {}_B\langle \psi_i | \Psi \rangle \langle \Psi | \psi_i \rangle_B \ ,\\
		&= \sum_{k=1}^{\text{min}(d_A, d_B)}\, p_k\, |\psi_k\rangle_A\, {}_A\langle \psi_k | \ .
\end{aligned}
\end{align}
In our case, the condition $\sum_k\, p_k = 1$ follows from the normalization $\langle \Psi |\Psi\rangle = 1$.
We see that the reduced density matrix of a subsystem can be mixed even if the total system is in a pure ground state.
This should be compared with the discussion for entanglement purification around Eq.\,\eqref{Def_Purification} where we saw the opposite.
In the Schmidt decomposition the entanglement entropy 
\begin{align}\label{SchmidtEE}
S_A = - \sum_{k=1}^{\text{min}(d_A, d_B)} p_k \,\log p_k
\end{align}
is nothing but the Shannon entropy of the probability distribution $\{ p_k\}$.
Under the constraint $\sum_k p_k = 1$, it takes the maximum value
\begin{align}\label{MES}
S_A|_\text{max} = \log \text{min}(d_A, d_B)
\end{align}
at $p_k = 1/\text{min}(d_A, d_B)$ for any $k$.\footnote{To prove Eq.\,\eqref{MES}, one can introduce the Lagrange multiplier $x\,(\sum_{k}p_k -1)$ to the entropy Eq.\,\eqref{SchmidtEE} and extremize it with respect to $p_k$.}

\bigskip
In summary, an entangled state is a superposition of several quantum states.
An observer who can  access only a subsystem $A$ will find him or herself in a mixed state when the pure ground state $|\Psi\rangle$ in the total system is entangled:
\begin{align}
\begin{aligned}
	 |\Psi\rangle:& ~\text{separable} \qquad &\longleftrightarrow \qquad &\rho_A: \text{pure state} \ , \\
	 |\Psi\rangle:& ~\text{entangled} \qquad &\longleftrightarrow \qquad &\rho_A: \text{mixed state} \ .
\end{aligned}
\end{align}

Entanglement entropy measures how much a given state differs from a separable state.
It reaches the maximum value when a given state is a superposition of all possible quantum states with an equal weight.

\paragraph{Two spin system}\label{TwoSpin}
Let us illustrate a simple example of an entangled state.
Consider a system of two particles $A$ and $B$ with spin $1/2$.
The Hilbert spaces $\CH_A$ and $\CH_B$ are spanned by two states: $\CH_{A,B} = \{ |0\rangle_{A,B}, |1\rangle_{A,B}\}$.
We let them be orthonormal bases satisfying ${}_{A,B}\langle i | j \rangle_{A,B} = \delta_{ij}$ for $i,j = 0,1$.
Since the total Hilbert space is the tensor product of the two subsystems $\CH_\text{tot} = \CH_A \otimes \CH_B$, it has the four-dimensional orthonormal basis: $\CH_\text{tot} = \{ |00\rangle, |01\rangle, |10\rangle, |11\rangle \}$ where $|ij\rangle \equiv |i\rangle_A \otimes |j\rangle_B$ are tensor product states.

Suppose the ground state is given by
\begin{align}\label{TwoSpinPure}
|\Psi\rangle = \frac{1}{\sqrt{2}} \Big( |01\rangle - |10\rangle \Big) \ .
\end{align}
The reduced density matrix for the particle $A$ is obtained by taking the partial trace over $\CH_B$ of the total density matrix Eq.\,\eqref{pure}, 
\begin{align}
\begin{aligned}
	\rho_A &=  \frac{1}{2}\Big( |0\rangle_A \,{}_A\langle 0 |   +  |1 \rangle_A \,{}_A\langle 1 |\Big) \ .
\end{aligned}
\end{align}
It is convenient to write it in a matrix form acting on the two-dimensional vector space $\CH_A$,
\begin{align}\label{TwoSpinRD}
	\rho_A = \left( 
		\begin{array}{cc}
			1/2 & 0 \\
			0 & 1/2
		\end{array}
		\right) \ .
\end{align}
It shows that $\rho_A$ is not pure and 
the entanglement entropy does not vanish,
\begin{align}\label{EE_TwoSpin}
\begin{aligned}
S_A &= -\tr_A \left[ \left( 
		\begin{array}{cc}
			1/2 & 0 \\
			0 & 1/2
		\end{array}
		\right)  \left( 
		\begin{array}{cc}
			\log(1/2) & 0 \\
			0 & \log(1/2)
		\end{array}
		\right)    \right] \ , \\
	&=\log 2 \ .
\end{aligned}
\end{align}
This is a maximally entangled state for it saturates the upper bound Eq.\,\eqref{MES} with $d_A = d_B = 2$.

A more general state to consider is
\begin{align}
|\Psi\rangle =  \cos\theta\, |0 1 \rangle - \sin\theta\, |10\rangle \ ,
\end{align}
parametrized by a parameter $\theta$ ranging from $0$ to $\pi/2$, that reduces the state Eq.\,\eqref{TwoSpinPure} at $\theta=\pi /4$.
By a similar calculation we have
\begin{align}
	S_A = - \cos^2 \theta \log ( \cos^2 \theta) - \sin^2\theta \log (\sin^2\theta) \ ,
\end{align}
which shows the states at $\theta = 0, \pi /2$ are pure product states with vanishing entanglement entropy while the state Eq.\,\eqref{TwoSpinPure} at $\theta= \pi/4$ is maximally entangled.

\paragraph{Thermofield double state}
A more nontrivial example of an entangled state is the thermofield double state defined by
\begin{align}\label{TFD}
	| \Psi\rangle = \frac{1}{\sqrt{Z} }\sum_n\,e^{-\beta E_n/2} |n\rangle_A \otimes |n\rangle_B \ ,
\end{align}
where we normalize the state with the partition function $Z = \sum_n e^{-\beta E_n}$.
The peculiarity of the thermofield double state becomes manifest in taking the partial trace over the subsystem $B$.
Namely the reduced density matrix for the subsystem $A$ becomes a Gibbs state of inverse temperature $\beta$:
\begin{align}\label{Gibbs_State}
\begin{aligned}
	\rho_A  &=  \frac{1}{Z} \sum_{n} e^{-\beta E_n} |n\rangle_A {}_A\langle n | \ ,\\
		&= \frac{1}{Z}\, e^{-\beta H_A} \ .
\end{aligned}
\end{align}
In the second line, we introduced the (modular) Hamiltonian $H_A$ such that $H_A |n\rangle_A = E_n |n\rangle_A$.
Actually the thermofield double state is the entanglement purification of a thermal state with the Boltzmann weight $p_i = e^{-\beta\,E_i}/Z$ in Eq.\,\eqref{Def_Purification}.
Namely we can purify the thermal system $A$ in the extended Hilbert space $\CH_A\otimes \CH_B$ by copying the state vectors $\{ |n\rangle_B\}$ from $\CH_A$ to $\CH_B$.
Then every expectation value of local operators in the thermal system $A$ is representable in the thermofield double state Eq.\,\eqref{TFD} of the total system $A\cup B$.
In this example, the entanglement entropy measures the thermal entropy of the subsystem $A$:
\begin{align}
\begin{aligned}
	S_A &= -\tr_A\left[ \rho_A\, (-\beta H_A - \log Z)\right] \ , \\
		&= \beta \,(\langle H_A \rangle -F)   \ ,
\end{aligned}
\end{align}
where $F$ is the thermal free energy $\beta\, F = - \log Z$.

The thermofield double state is also important to understand the thermal nature of black holes when we consider QFT on a background geometry with a horizon in Sec. \ref{ss:RindlerTFD}.

\subsection{Bell states}
In the two spin system, we saw the state Eq.\,\eqref{TwoSpinPure} is maximally entangled.
Actually there are totally four independent maximally entangled states in the two qubit system:
\begin{align}\label{BellStates}
\begin{aligned}
	|B_1\rangle &= \frac{1}{\sqrt{2}} \Big( |00\rangle + |11\rangle\Big) \ ,\\
	|B_2\rangle &= \frac{1}{\sqrt{2}} \Big(  |00\rangle - |11\rangle\Big) \ ,\\
	|B_3\rangle &= \frac{1}{\sqrt{2}} \Big(  |01\rangle + |10\rangle\Big) \ ,\\
	|B_4\rangle &= \frac{1}{\sqrt{2}} \Big(  |01\rangle - |10\rangle\Big) \ .
\end{aligned}
\end{align}
These are known as the Bell states or Einstein-Podolsky-Rosen pairs in quantum information theory.
These states manifest their quantum mechanical aspects in the sense that they violate the Bell's inequalities holding in a local hidden variable theory accounting for the probabilistic features of quantum mechanics with a hidden variable and a probability density.

For a system of $n$ qubits, there are entangle states called the Greenberger-Horne-Zeilinger (GHZ) states \cite{greenberger1989going,greenberger1990bell}:
\begin{align}
	|\text{GHZ}\rangle = \frac{1}{\sqrt{2}} \left( |0\rangle^{\otimes n} + |1\rangle^{\otimes n} \right) \ .
\end{align}
Another type of entangled states is called the $W$ state \cite{Dur:2000zz},
\begin{align}
	|W \rangle = \frac{1}{\sqrt{n}} \left( |10\cdots00\rangle +  |010\cdots 0\rangle +\cdots + |00\cdots01\rangle  \right) \ .
\end{align}
These two types of states are inequivalent as the GHZ state is fully separable while the $W$ state is not as seen next.

\paragraph*{Tripartite system $:$}
In a tripartite system ($n=3$), the GHZ and $W$ states become
\begin{align}\label{GHZstate}
	|\text{GHZ}\rangle = \frac{1}{\sqrt{2}} \left( |000\rangle + |111\rangle \right) \ ,
\end{align}
and 
\begin{align}
	|W\rangle = \frac{1}{\sqrt{3}} \left( |001\rangle + |010\rangle + |100\rangle\right) \ .
\end{align}
We denote the three subsystems by $A,B$ and $C$. 
Tracing out the Hilbert space of the subsystem $C$, the reduced density matrices for the system $A\cup B$ are 
\begin{align}
	\begin{aligned}
		\rho_{A\cup B}^{(\text{GHZ})} &=\frac{1}{2}\left( |00\rangle \langle 00| + |11\rangle \langle 11| \right)\ , \\
		\rho_{A\cup B}^{(W)} &= \frac{2}{3}|B_3\rangle \langle B_3 | + \frac{1}{3} |00\rangle \langle 00|  \ .
	\end{aligned}
\end{align}
The $\rho_{A\cup B}^{(\text{GHZ})}$ is fully separable in a sense that it can be written in the form
\begin{align}
	\rho_{A\cup B}^{(\text{GHZ})} = \sum_{i=1}^k p_i \,\rho_A^{(i)} \otimes \rho_B^{(i)} \ ,
\end{align}
where $k=2$, $p_i = 1/2$ and $\rho_{A,B}^{(1)} = |0\rangle \langle  0|$, $\rho_{A,B}^{(2)} = |1 \rangle \langle1|$.
On the other hand, the $\rho_{A\cup B}^{(W)}$ cannot be written in such a form because of the appearance of the Bell state $|B_3\rangle$.
This implies that the $W$ state is still entangled even after the partial trace.

\subsection{Properties of entanglement entropy}
\label{ss:prop EE}
Entanglement entropy has several useful properties that we summarize without proofs.
The interested reader can refer to \textcite{nielsen2010quantum} for the derivations and the other properties.
\begin{itemize}
\item If a ground state wave function is pure, the entanglement entropy of the subsystem $A$ and its complement $B= \bar A$ are the same:
\begin{align} 
S_A=S_B\ . \label{ext}
\end{align}
This follows from the symmetry of the Schmidt decomposition under the exchange of $A$ and $B$ and also from the result Eq.\,\eqref{SchmidtEE}.
However, $S_A$ is no longer equal to $S_B$ when the total system is in a mixed state e.g., at finite temperature.

\item Given two disjoint subsystems $A$ and $B$, the entanglement entropies satisfy the subadditivity,
\begin{align}\label{Sub}
 S_{A\cup B}\le S_{A}+S_{B} \ .
\end{align}
Also it satisfies the triangle inequality or the Araki-Lieb inequality \cite{Araki:1970ba},
\begin{align}\label{Araki_Lieb}
|S_A - S_B| \le S_{A\cup B} \ .
\end{align}
which is symmetric between $A$ and $B$.

\item
For any three disjoint subsystems $A$, $B$ and $C$,
the following inequalities hold:
\begin{align}\label{SSA}
\begin{aligned}
 S_{A\cup B\cup C} + S_{B} &\le S_{A\cup B}+S_{B\cup C} \ ,  \\
 S_{A} + S_{C} &\le S_{A\cup B}+S_{B\cup C} \ .
\end{aligned}
\end{align}
These are known as the {\it strong subadditivity}, the most fundamental inequalities for entanglement entropy.
The two inequalities are shown to be equivalent to each other \cite{Araki:1970ba}.
The proof is based on a convexity of a function built from the density matrix that is Hermitian when a system is unitary \cite{Lieb:1973cp,Narnhofer:1985aq}.
The subadditivity Eq.\,\eqref{Sub} and the Araki-Lieb inequality Eq.\,\eqref{Araki_Lieb} are derivable from the strong subadditivity.
\end{itemize}

The strong subadditivity of entanglement will play vital roles in the entropic proofs of the $c$- and $F$-theorems for renormalization group flows in QFT as we see in Sec. \ref{ss:RGflow}.

\subsection{Relations between entanglement measures}
Among several measures of quantum entanglement we list a few, relative entropy, mutual information and R{\'e}nyi entropy, that are often used in the applications to QFT, and discuss their features and connections to each other.

\subsubsection{Relative entropy}
Given two states described by the density matrices $\rho$ and $\sigma$, one can define the relative entropy by \cite{umegaki1962conditional}
\begin{align}\label{RelEnt}
S(\rho || \sigma) = \tr \left[ \rho\, (\log \rho -\log \sigma) \right]\ .
\end{align}
It measures the ``distance'' between the two states with several (defining) properties \cite{ohya2004quantum,Vedral:2002zz},
\begin{align}
S(\rho|| \rho) &= 0 \ , \\
S(\rho_1 \otimes \rho_2|| \sigma_1\otimes \sigma_2) &= S( \rho_1|| \sigma_1) + S( \rho_2 || \sigma_2)\ , \\
S(\rho|| \sigma) &\ge \frac{1}{2} || \rho - \sigma ||^2 \ , \label{RelEn}\\
S(\rho|| \sigma) &\ge S(\tr_p\, \rho | \tr_p\, \sigma) \ , \label{RelEnMon}
\end{align}
where $\tr_p$ is a partial trace with respect to the subsystem $p$, and the norm means
\begin{align}
|| \rho || = \tr\, (\sqrt{\rho^\dagger \rho}) \ .
\end{align}
With particular choices of the density matrices, the relative entropy reduces to entanglement entropy,
\begin{align}\label{RelEnMutual}
\begin{aligned}
S(\rho_A || {\bf 1}_A/d_A) &= \log d_A - S_A \ , 
\end{aligned}
\end{align}
where ${\bf 1}_A$ is the $d_A \times d_A$ unit matrix for the $d_A$-dimensional Hilbert space of the region $A$.

The relative entropy is always non-negative, $S(\rho ||\sigma) \ge 0$, bounded from below by the inequality Eq.\,\eqref{RelEn}.
Various inequalities for the other entanglement measures follow from the monotonicity of the relative entropy given by Eq.\eqref{RelEnMon} (see Table \ref{tab:EM_relations}).
For example, the strong subadditivity Eq.\,\eqref{SSA} is derivable as follows.

Let $\rho_{A\cup B\cup C}$ be the density matrix for the total system $A\cup B\cup C$, and we denote its restrictions to the subsystems $A\cup B, B\cup C$ and $B$ by $\rho_{A\cup B}, \rho_{B\cup C}$ and $\rho_B$, respectively.
Since the reduced density matrices have the property such that $\tr_{A
\cup B\cup C}[\rho_{A\cup B\cup C}\, (\CO_{A\cup B} \otimes {\bf 1}_{C}/d_C)] = \tr_{A\cup B} (\rho_{A\cup B}\, \CO_{A\cup B})$,
we can show the identities
\begin{align}\label{SSAproof}
\begin{aligned}
&S(\rho_{A\cup B\cup C} || {\bf 1}_{A\cup B\cup C}/d_{A\cup B\cup C}) \\
	&\quad = S(\rho_{A\cup B} || {\bf 1}_{A\cup B}/d_{A\cup B}) + S(\rho_{A\cup B\cup C} || \rho_{A\cup B} \otimes {\bf 1}_C/d_C) \ , \\
&S(\rho_{B\cup C} || {\bf 1}_{B\cup C}/d_{B\cup C}) \\
	&\quad = S(\rho_B|| {\bf 1}_B/d_B) + S(\rho_{B\cup C}|| \rho_B \otimes {\bf 1}_C/d_C) \ .
\end{aligned}
\end{align}
On the other hand, the monotonicity Eq.\,\eqref{RelEnMon} implies
\begin{align}
S(\rho_{A\cup B\cup C}|| \rho_{A\cup B} \otimes {\bf 1}_C/d_C) \ge S(\rho_{B\cup C} || \rho_B \otimes {\bf 1}_C/d_C) \ , 
\end{align}
and combined with the equalities Eq.\,\eqref{SSAproof}, we arrive at the inequality,
\begin{align}\label{Rel_SSA}
\begin{aligned}
&S(\rho_{A\cup B\cup C} || {\bf 1}_{A\cup B\cup C}/d_{A\cup B\cup C}) + S(\rho_B|| {\bf 1}_B/d_B) \\
	&\quad \ge S(\rho_{A\cup B} || {\bf 1}_{A\cup B}/d_{A\cup B}) + S(\rho_{B\cup C} || {\bf 1}_{B\cup C}/d_{B\cup C}) \ .
\end{aligned}
\end{align}
Translating the relative entropy to entanglement entropy through Eq.\,\eqref{RelEnMutual} and noting the dimensional relations such as $d_{A\cup B\cup C} = d_A d_B d_C$, we find Eq.\,\eqref{Rel_SSA} is nothing but the strong subadditivity of entanglement entropy given by the first line of Eq.\,\eqref{SSA}.
We do not bother to derive the second inequality in Eq.\,\eqref{SSA} from the monotonicity of the relative entropy as it is equivalent to the first \cite{Araki:1970ba}.

\begin{table}[htbp]
\caption{\label{tab:EM_relations} Equivalence between the entanglement inequalities of various measures}
\begin{ruledtabular}
\begin{tabular}{ccc}
	Entanglement entropy & Mutual information & Relative entropy \\ \hline
	Subadditivity & Positivity & Positivity\\
	Eq.\,\eqref{Sub} & Eq.\,\eqref{Mutual_Pos} & Eq.\,\eqref{RelEn}\\
	\rule{0pt}{3ex}    
	Strong subadditivity & Monotonicity & Monotonicity \\
		Eq.\,\eqref{SSA} & Eq.\,\eqref{MI_Mon} & Eq.\,\eqref{RelEnMon}
\end{tabular}
\end{ruledtabular}
\end{table}

One can estimate a lower bound for the relative entropy by combining Eq.\,\eqref{RelEn} with the Schwarz inequality $|| X ||\ge \tr (XY)/|| Y ||$, 
\begin{align}\label{RE_bounded}
S(\rho|| \sigma) \ge \frac{1}{2} \frac{(\langle \CO\rangle_\rho - \langle \CO \rangle_\sigma)^2}{||\CO||^2} \ ,
\end{align}
where $\langle \CO\rangle_\rho$ is the expectation value of an operator $\CO$ with the density matrix $\rho$.

\subsubsection{Mutual information}
\begin{figure}[htbp]
\centering
	\begin{tikzpicture}[thick]
			\draw[fill=mizuiro] (1,1) node {$A$} circle (0.7);
			\draw[fill=kakiiro] (5,2) node {$B$} circle (1);
	\end{tikzpicture}
	\caption{\label{fig:MI} The mutual information between two disjoint regions $A$ (blue) and $B$ (red).}
\end{figure}
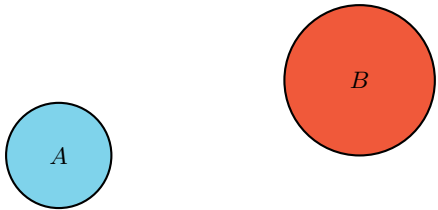

The mutual information $I(A,B)$ of two systems $A$ and $B$ (see Fig.\,\ref{fig:MI}) is defined by 
\begin{align}\label{MI}
I(A,B)\equiv S_A+S_B-S_{A\cup B} \ .
\end{align}
It measures how much the two subsystems are correlated.
It is symmetric under the exchange of $A$ and $B$ by definition and free from ultraviolet divergences in QFT while entanglement entropy generically diverges as discussed in Sec. \ref{ss:UVstructure}.
The subadditivity Eq.\,\eqref{Sub} guarantees the mutual information to be non-negative: 
\begin{align}\label{Mutual_Pos}
	I(A,B) \ge 0\ .
\end{align}
In addition, the strong subadditivity Eq.\,\eqref{SSA} leads to the monotonicity
\begin{align}\label{MI_Mon}
	I(A, B \cup C) \le I(A, B) \ ,
\end{align}
for any region $C$.

The mutual information can be written by the relative entropy as
\begin{align}\label{RelMutual}
	 I(A,B) &= S(\rho_{A\cup B} || \rho_A \otimes \rho_B)\ .
\end{align}
This relation shows that the mutual information between two subsystems quantifies how much 
the state $\rho_{A\cup B}$ for the union $A\cup B$ differs from the separable state $\rho_A \otimes \rho_B$ and can be a good measure of global correlations over spatially disconnected regions in QFT.

Combining the third inequality of Eq.\,\eqref{RelEn} and the expression of the mutual information Eq.\,\eqref{RelMutual} with the Schwarz inequality,
one obtains the lower bound for the mutual information \cite{wolf2008area}
\begin{align}
I(A, B) \ge \frac{1}{2} \frac{( \langle \CO_A \CO_B \rangle - \langle \CO_A \rangle \langle \CO_B \rangle)^2}{|| \CO_A ||^2 || \CO_B||^2} \ ,
\end{align}
where $\CO_A$ and $\CO_B$ are any bounded operators in the regions $A$ and $B$, respectively.

\subsubsection{R{\'e}nyi entropy}
The R{\'e}nyi entropy is a one-parameter generalization of entanglement entropy labeled by an integer $n$, called the replica parameter \cite{rrnyi1961measures}
\begin{align}\label{RenyiDef}
S_n (A) = \frac{1}{1-n} \log\, \tr_A ( \rho_A^n) \ .
\end{align}
In the $n\to 1$ limit with the normalization $\tr_A( \rho_A) = 1$, the R{\'e}nyi entropy reduced to entanglement entropy,
\begin{align}\label{RenyiEE}
S_A = \lim_{n\to 1} S_n(A) \ .
\end{align}
The R{\'e}nyi entropy provides us more information about the eigenvalues of the reduced density matrix than entanglement entropy.
It is defined for an integer $n$, but one assume an analytic continuation of $n$ to a real number in taking the limit Eq.\,\eqref{RenyiEE}.
This analytic continuation will be useful to compute entanglement entropy of QFT by the replica method in Sec. \ref{ss:Euclidean}.

\paragraph{Thermal interpretation of R{\'e}nyi entropic inequalities}

The R{\'e}nyi entropy Eq.\,\eqref{RenyiDef} is shown to satisfy several inequalities between different values of $n$ \cite{zyczkowski2003renyi},
\begin{align}
\partial_n S_n &\le 0 \ , \label{RenyiIneq1}\\
\partial_n \left( \frac{n-1}{n} S_n\right) &\ge 0 \ ,\label{RenyiIneq2}\\
\partial_n \left[ (n-1)S_n\right] &\ge 0 \ , \label{RenyiIneq3}\\
\partial_n^2 \left[ (n-1)S_n\right] &\le 0 \ . \label{RenyiIneq4}
\end{align}
To grasp the physical meaning of these inequalities we introduce the \emph{modular Hamiltonian} $H_A$ by
\begin{align}
	\rho_A \equiv e^{- 2\pi H_A}\ ,
\end{align}
and consider a ``thermal" partition function
\begin{align}\label{Thermal_PF}
	Z(\beta) \equiv \tr_A (\rho_A^n) = \tr_A (e^{- \beta H_A}) \ ,
\end{align}
at inverse temperature $\beta \equiv 2\pi n$ where $H_A$ is regarded as a Hamiltonian.
In analogy to statistical mechanics we can define the \emph{modular energy} $E(\beta)$, the \emph{modular entropy}  $\tilde S(\beta)$, and the \emph{modular capacity} $C(\beta)$\footnote{The modular capacity $C(\beta)$ is the same as the capacity of entanglement defined by \textcite{Yao:2010woi}.} by the canonical relations,
\begin{align}\label{ModularEntropy}
	\begin{aligned}
		E(\beta) &\equiv - \partial_\beta \log Z(\beta) \ ,\\
		\tilde S(\beta) &\equiv (1-\beta \partial_\beta) \log Z(\beta) \ ,\\
		C(\beta) &\equiv \beta^2 \partial_\beta^2 \log Z(\beta) \ .
	\end{aligned}
\end{align}
With these ``thermodynamic" quantities, the inequalities Eq.\,\eqref{RenyiIneq2}, \eqref{RenyiIneq3} and Eq.\,\eqref{RenyiIneq4} are rewritten in more concise forms \cite{Hung:2011nu,Nakaguchi:2016zqi}:
\begin{align}\label{Modular_Ineq}
	\begin{aligned}
		\tilde S(\beta) &\ge 0 \ , \\
		E(\beta) & \ge 0 \ , \\
		C(\beta) &\ge 0 \ .
	\end{aligned}
\end{align}
The inequality Eq.\,\eqref{RenyiIneq1} is not independent from the others and can be derived from Eq.\,\eqref{RenyiIneq4}.
We thus find the inequalities of the R{\'e}nyi entropy have a good interpretation as the \emph{stability of the thermal system} when the replica parameter $n$ is regarded as the inverse temperature for the modular Hamiltonian.

There is a one-to-one correspondence between the modular entropy and the R{\'e}nyi entropy,
\begin{align}\label{Modular_entropy}
	\tilde S_n \equiv n^2 \partial_n \left( \frac{n-1}{n} S_n\right) \ ,
\end{align}
where we changed the notation for $\tilde S(\beta)$ to emphasize the $n$ dependence.
Inverting this relation by integration, we can reconstruct the R{\'e}nyi entropy from the modular entropy,
\begin{align}\label{Renyi_Modular}
	S_n = \frac{n}{n-1} \int_1^n \d\hat n\, \frac{1}{\hat n^2}\, \tilde S_{\hat n} \ .
\end{align}

We discuss the implications of the inequalities Eq.\,\eqref{Modular_Ineq} for a gravitational system when a holographic description of the R{\'e}nyi entropy is available in Sec. \ref{ss:HRE}.

\paragraph{Relative R{\'e}nyi entropy}
The relative R{\'e}nyi entropy is an analog of the R{\'e}nyi entropy for the relative entropy, defined by \cite{muller2013quantum,wilde2013strong}
\begin{align}
S_n( \rho ||\sigma) =
\frac{1}{n-1} \log\left[ \tr (\sigma^{(1-n)/(2n)} \rho \,\sigma^{(1-n)/(2n)})^n \right] \ , 
\end{align}
for $n \in (0,1) \cup (1,\infty)$ and 
\begin{align}
\begin{aligned}
	S_1( \rho ||\sigma) &= S( \rho ||\sigma)\ , \\
	S_\infty( \rho ||\sigma) &= \log || \sigma^{-1/2} \rho\, \sigma^{-1/2} ||_\infty  \ .
\end{aligned}
\end{align}
The relative R{\'e}nyi entropy is shown to be monotonic under the partial trace operation $\tr_p$,
\begin{align}
S_n( \rho ||\sigma) \ge S_n( \tr_p \,\rho || \tr_p\, \sigma) \ ,
\end{align}
for $n\ge 1/2$ as is the relative entropy \cite{frank2013monotonicity,beigi2013sandwiched}.

The relative R{\'e}nyi entropy reduces to the R{\'e}nyi entropy in taking a special state as $\sigma$ in the same way as in Eq.\,\eqref{RelEnMutual}, 
\begin{align}
\begin{aligned}
	S_n(\rho_A || {\bf 1}_A/d_A) &= \log d_A - S_n(A) \ .
\end{aligned}
\end{align}
It is also shown to be monotonic with respect the parameter $n$ \cite{muller2013quantum,beigi2013sandwiched},
\begin{align}
\begin{aligned}
	\partial_n S_n(\rho ||\sigma) &\ge 0 \ ,
\end{aligned}
\end{align}
which is considered as a generalization of the inequality Eq.\,\eqref{RenyiIneq1} for the R{\'e}nyi entropy.

One may be tempted to define the R{\'e}nyi mutual information in a similar manner to the mutual information Eq.\,\eqref{MI}.
This naive construction, however, can give rise to a negative value for a certain class of states when $n\neq 1$ in general and does not allow the interpretation as an entanglement measure of quantum information \cite{adesso2012measuring}.\footnote{An example of such states for $n=2$ is
\begin{align}
	\rho_{A\cup B} = x\, |00\rangle \langle 00| + y\, |01\rangle \langle 01| + (1-x-y)\, |10\rangle \langle 10| \ ,
\end{align}
with the parameter ranges $0< x < 1/2$ and $0< y < (1/2 -x)/(1-x)$.
}

Instead, one can introduce the $n$-R{\'e}nyi mutual information by \cite{beigi2013sandwiched}
\begin{align}
	I_n(A, B) &=  \text{min}_{\sigma_B} S_n(\rho_{A\cup B} || \rho_A \otimes \sigma_B) \ ,
\end{align}
where the minimum is taken over all density matrices $\sigma_B$.
It is always non-negative by definition and reduces to the mutual information when $n=1$.

\subsection{Entanglement entropy at finite temperature}

The entanglement entropy $S_A(T)$ at finite temperature $T=\beta^{-1}$ can be defined just by replacing the total density matrix Eq.\,\eqref{pure} with the thermal density matrix
\begin{align}
	\rho_\text{thermal}=\frac{e^{-\beta H}}{\tr(e^{-\beta H})}\ ,
\end{align}
where $H$ is the Hamiltonian of the total system.
By definition, $S_A(\beta)$ equals the thermal entropy when $A$ is the total system
\begin{align}
	S_\text{tot} = S_\text{thermal}  \ .
\end{align}
Let $|\Psi\rangle$ be the ground state with no energy $H|\Psi\rangle = 0$, and $|\phi\rangle$ be the first excited (normalized) state with the energy $H|\phi\rangle = E_\phi |\phi\rangle$.
Then the density matrix has an expansion around the ground state at zero temperature,
\begin{align}
	\rho_\text{thermal}=\frac{|\Psi\rangle \langle\Psi |  + |\phi\rangle \langle \phi |\, e^{-\beta E_\phi} + \cdots}{1 + e^{-\beta E_\phi} + \cdots}\ .
\end{align}
The reduced density matrix allows a similar expansion,
\begin{align}
	\rho_A = \rho_{0A} + e^{-\beta E_\phi}\, (\rho_{\phi A}  - \rho_{0A}) + \cdots \ ,
\end{align}
where we defined $\rho_{0A} \equiv \tr_B (|\Psi\rangle \langle \Psi |)$ and $\rho_{\phi A} \equiv  \tr_B (|\phi\rangle \langle \phi |)$.
It follows from the $n\to 1$ limit of the R{\'e}nyi entropy Eq.\,\eqref{RenyiDef} that 
the entanglement entropy receives the universal thermal contribution from the excited state \cite{Cardy:2014jwa,Herzog:2014fra},
\begin{align}
	S_A (T) = S_A (T=0) + e^{-\beta E_\phi} \Delta \langle H_{0A}\rangle + \cdots \ ,
\end{align}
where $H_{0A} \equiv - \log \rho_{0A}$ is the modular Hamiltonian for the ground state and $\Delta \langle H_{0A}\rangle$ stands for the difference of the modular Hamiltonian,
\begin{align}
	\Delta \langle H_{0A}\rangle  \equiv \tr_A\left[ H_{0A} (\rho_{\phi A} - \rho_{0A})\right] \ .
\end{align}

  \section{Real time formalism}\label{ss:RTapproach}
We illustrate the real time approach for calculating the reduced density matrix of a subsystem in the Hamiltonian description.
This method is suitable for the numerical computation once we discretize the spacetime to lattice.
We describe only the bosonic case here and defer the fermionic case to the Appendix \ref{ss:RTFermion}.

\subsection{Two coupled harmonic oscillators}
To illustrate the real time approach, we start with a simple example of two coupled harmonic oscillators \cite{Srednicki:1993im,Bombelli:1986rw}; see Fig.\,\ref{fig:HarmChain}.
Suppose this system is described by the Hamiltonian,
\begin{align}\label{TwoHarmH}
H = \frac{1}{2}\left[ p_A^2 + p_B^2 + k(x_A^2 + x_B^2) + l (x_A - x_B)^2 \right] \ ,
\end{align}
where $x_{A,B}$ and $p_{A,B}$ are the positions and the conjugate momenta of the two oscillators, and $k$ and $l$ are related to their mass and the coupling.
First, we want to find the ground state wave function of the system.
To this end, we introduce new variables $x_\pm = (x_A \pm x_B)/\sqrt{2},~ \omega_+ = k^{1/2}$, and $\omega_- = (k + 2l)^{1/2}$.
In these new variables, the Hamiltonian is for two \emph{uncoupled} harmonic oscillators and the Schr{\"o}dinger equation becomes
\begin{align}
	 \frac{1}{2}\left[ - \partial_+^2 - \partial_-^2 + \omega_+^2 x_+^2 + \omega_-^2 x_-^2\right] \Psi (x_+, x_-) = E\, \Psi (x_+, x_-)  \ .
\end{align}
Then the ground state wave function is the product of two copies of the ground state wave function of one harmonic oscillator,
\begin{align}\label{TwoHarmGS}
\Psi  (x_+, x_-) = \frac{(\omega_+ \omega_-)^{1/4}}{\pi^{1/2}} \exp\left[ - \frac{\omega_+ x_+^2 + \omega_- x_-^2}{2}\right] \ ,
\end{align}
with the energy $E = (\omega_+ + \omega_-)/2$.
The overall factor is chosen so that the wave function is normalized to be $\int_{-\infty}^\infty \d x_A \int_{-\infty}^\infty \d x_B |\Psi (x_A, x_B)|^2 = 1$ where we denote the wave function in the original variables $x_A$ and $x_B$ by the same symbol $\Psi (x_A, x_B)$.

Next we trace out the oscillator at $x_B$ and construct the reduced density matrix $\rho_A$ for the oscillator at $x_A$.
If we represent the ground state wave function as $\Psi (x_A, x_B) = (\langle x_A| \otimes \langle x_B|)| \Psi\rangle$,
then the reduced density matrix follows as
\begin{align}\label{rhoATwoHarm}
\begin{aligned}
\rho_A (x_A, x_A') &=  \langle x_A| \tr_B \left(|\Psi\rangle \langle \Psi|\right) | x_A'\rangle \ ,\\
&= \int_{-\infty}^\infty \d x_B \, \Psi(x_A, x_B)\,\Psi^\ast (x_A', x_B) \ , \\
&= \sqrt{\frac{\gamma - \beta}{\pi}}\, \exp\left[ -\frac{\gamma}{2}(x_A^2 + x_A'^2) + \beta\, x_A x_A'\right] \ ,
\end{aligned}
\end{align}
where we introduced the parameters 
\begin{align}
\beta = \frac{(\omega_+ - \omega_-)^2}{4(\omega_+ + \omega_-)} \ , \qquad \gamma = \frac{2\omega_+ \omega_-}{\omega_+ + \omega_-} + \beta \ .
\end{align}

Finally, we need the eigenvalues of the reduced density matrix to compute the entanglement entropy.
The eigenfunction $f_n$ with the eigenvalue $p_n$ must satisfy
\begin{align}
\int_{-\infty}^\infty \d x'\, \rho_A (x, x')\,f_n(x') = p_n\, f_n (x) \ ,
\end{align}
and the solution is given by 
\begin{align}\label{EigenTwoHarm}
\begin{aligned}
f_n (x) &= H_n (\alpha^{1/2} x)\, e^{-\frac{\alpha x^2}{2}} \ ,\qquad (n=0,1,2,\cdots) \ ,
\end{aligned}
\end{align}
where $\alpha = (\omega_+ \omega_-)^{1/2}$ and $H_n$ is the Hermite polynomial.
The eigenvalue $p_n$ is 
\begin{align}\label{Two_harmonic_eigen}
	p_n = (1-\xi)\,\xi^n \ , \qquad \xi = \frac{\beta}{\alpha + \gamma} \ .
\end{align}
One way to derive the eigenfunction is to expand the reduced density matrix by the Hermite polynomial,
\begin{align}
\begin{aligned}
	\rho_A (x, x') &= \sqrt{\frac{\alpha}{\pi}}\,(1-\xi)\,e^{-\frac{\alpha}{2}(x^2 + x'^2)} \\
		&\qquad \cdot \sum_{n=0}^\infty \frac{\xi^n}{2^n n!}  \,H_n(\alpha^{1/2} x) \,H_n(\alpha^{1/2} x') \ ,
\end{aligned}
\end{align}
and use the orthogonality,
\begin{align}
	\int_{-\infty}^\infty \d x\, e^{-x^2} H_n(x)\, H_m(x) = \sqrt{\pi}\,2^n  n!\,\delta_{nm} \ .
\end{align}
Having diagonalized the reduced density matrix, we are ready to obtain the entanglement entropy from the eigenvalue Eq.\,\eqref{Two_harmonic_eigen},
\begin{align}
\begin{aligned}
S_A &=- \sum_{n=0}^\infty\, p_n \log p_n \ , \\
	&= - \log(1-\xi) - \frac{\xi}{1-\xi}\log\xi \ .
\end{aligned}
\end{align}

\subsection{$N$-coupled harmonic oscillators}
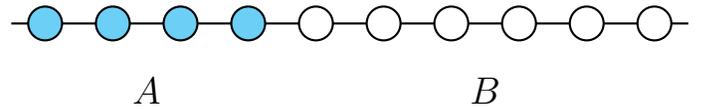
\begin{figure}[htbp]
\centering
\begin{tikzpicture}[thick, scale=0.45]
	\draw (-1,0) --++(20,0);
	\draw[fill=cyan!50] (0,0) circle (0.5);
	\draw[fill=cyan!50] (2,0) circle (0.5);
	\draw[fill=cyan!50] (4,0) circle (0.5);
	\draw[fill=cyan!50] (6,0) circle (0.5);
	\draw[fill=white] (8,0) circle (0.5);
	\draw[fill=white] (10,0) circle (0.5);
	\draw[fill=white] (12,0) circle (0.5);
	\draw[fill=white] (14,0) circle (0.5);
	\draw[fill=white] (16,0) circle (0.5);
	\draw[fill=white] (18,0) circle (0.5);
	\node at (3,-2) {\Large $A$};
	\node at (13, -2) {\Large $B$};
\end{tikzpicture}
\caption{\label{fig:HarmChain} A system of $N$-coupled harmonic oscillators. Here the total size of the system is $N=10$ and the subsystem $A$ (blue) consists of four oscillators from the left.}
\end{figure}

It is straightforward to generalize the model Eq.\,\eqref{TwoHarmH} of two harmonic oscillators to a system of $N$-coupled harmonic oscillators (see Fig.\,\ref{fig:HarmChain})
\begin{align}\label{NHarm}
H = \frac{1}{2}\sum_{i=1}^N \pi_i^2 + \frac{1}{2}\sum_{i,j=1}^N \phi_i \,K_{ij}\, \phi_j \ ,
\end{align}
where $K$ is a real symmetric matrix and $\pi_i$ is a canonical conjugate momentum of the oscillator $\phi_i$ satisfying the commutation relation,\footnote{The oscillator and conjugate momentum are considered as Hermitian operators.}
\begin{align}\label{CCR_phi}
	[\phi_i, \pi_j]=\i\,\delta_{ij} \ . 
\end{align}
It reduces to the previous model Eq.\,\eqref{TwoHarmH} for $N=2$, $K_{11}=K_{22} = k+l$ and $K_{12} = -l$.
Instead of repeating the arguments in the literature \cite{Srednicki:1993im,Bombelli:1986rw},
we adopt a different method that suites for numerical computation.

First, we introduce creation and annihilation operators $a_I$ and $a_I^\dagger$ satisfying the commutation relation $[a_I, a_J^\dagger] = \delta_{IJ}$.
Note that the new indices $I$ and $J$ are different from the indices $i$ and $j$ labeling the lattice site.
They parametrize the Fock space generated by $a_I$ and $a_I^\dagger$.
We assume the operators $a_I$ and $ a_I^\dagger$ are linearly related to the original oscillators $\phi_i$ and $ \pi_j$ with indices $i$ and $j$ in the region $A$ by \cite{Casini:2009sr},
\begin{align}\label{Phi_to_a}
\begin{aligned}
	\phi_i &= \alpha_{iI} (a_{I}^\dagger + a_{I}) \ , \\
	\pi_j &= -\i\, \beta_{jJ} (a_J^\dagger - a_J) \  ,
\end{aligned}
\end{align}
where $\alpha_{iI}$ and $\beta_{jJ}$ are real matrices with the mixed indices of the coordinate space and the Fock space.
These matrices are not independent of each other due to the commutation relation Eq.\,\eqref{CCR_phi} and have to be real matrices satisfying
\begin{align}\label{alphabet}
\alpha\, \beta^T = -\frac{1}{2} \ .
\end{align}
We now {\it assume} that there exists a (modular) Hamiltonian bilinear in the creation and annihilation operators,
\begin{align}
	H_A= \sum_{I} \epsilon_I\, a^\dagger_I a_I \ ,
\end{align}
which generates the reduced density matrix \cite{peschel1999density,chung2000density},
\begin{align}\label{rhoA_ansatz}
\rho_A = \prod_{I\in A} (1 - e^{-\epsilon_I}) \, e^{-H_A} \ .
\end{align}
In order to determine the spectrum $\epsilon_I$, we compare the two-point functions $\langle \phi_i\, \pi_j\rangle = \i\, \delta_{ij}/2$, $X_{ij} \equiv \langle \phi_i \,\phi_j\rangle$ and $P_{ij} = \langle \pi_i \,\pi_j\rangle$ in the subsystem $A$ with those evaluated with the ansatz Eq.\,\eqref{rhoA_ansatz},
\begin{align}\label{XPD}
\begin{aligned}
X_{ij} &= \tr_A (\rho_A\, \phi_i \,\phi_j) \ , \\
P_{ij} &= \tr_A (\rho_A\, \pi_i\, \pi_j) \ , \\
\frac{\i}{2}\delta_{ij}  &= \tr_A (\rho_A\, \phi_i\, \pi_j) \ .
\end{aligned}
\end{align}
The right hand sides can be diagonalized by the coefficient matrices $\alpha$ and $\beta$ in Eq.\,\eqref{Phi_to_a}:
\begin{align}\label{XPDeqs}
\begin{aligned}
X &= \alpha\, (2n+1)\, \alpha^T \ , \\
P &= \beta\, (2n+1)\, \beta^T \ ,
\end{aligned}
\end{align}
where $n$ is the diagonal matrix with entries,
\begin{align}
n_{IJ} = \tr_A (\rho_A\, a_I^\dagger a_J ) = \frac{\delta_{IJ}}{e^{\epsilon_I} - 1} \ .
\end{align}
We can read off the spectrum $\epsilon_I$ of the reduced density matrix from the matrix product $XP$ that is also diagonalized with Eq.\,\eqref{alphabet} as
\begin{align}
XP = \frac{1}{4} \alpha\, (2n+1)^2 \alpha^{-1} \ .
\end{align}
Comparing the eigenvalues of the matrices in both sides, we find
\begin{align}\label{EigenRel1}
\nu_I = \frac{1}{2}\coth \frac{\epsilon_I}{2} \ ,
\end{align}
where $\nu_I$ are the eigenvalues of $C=\sqrt{XP}$.

The entanglement entropy follows from the limit of the R{\'e}nyi entropy Eq.\,\eqref{RenyiEE}.
Since the trace of the $n^{th}$ power of $\rho_A$ is given by
\begin{align}
\begin{aligned}
\tr_A( \rho_A^n) &= \prod_{I}\frac{(1- e^{-\epsilon_I})^n}{1- e^{-n \epsilon_I}} \ ,\\
	&= \prod_{I}\left[ \left(\nu_I + \frac{1}{2}\right)^n - \left(\nu_I - \frac{1}{2}\right)^n \right]^{-1} \ ,
\end{aligned}
\end{align}
we can represent the entanglement entropy in terms of the matrix $C$,
\begin{align}\label{EE_Lattice}
\begin{aligned}
S_A &= \tr_A \left[ \left( C + \frac{1}{2}\right) \log\left( C + \frac{1}{2}\right)\right. \\
		&\qquad \qquad \qquad \left.- \left( C - \frac{1}{2}\right) \log\left( C - \frac{1}{2}\right)\right] \ ,
\end{aligned}
\end{align}
where $\tr_A$ means the indices of the matrix $C_{ij}$ are restricted to the subsystem $A$, $i,j\in A$.

While Eq.\,\eqref{EE_Lattice} is enough to determine the entanglement entropy of the subsystem on lattice, it is cumbersome to calculate the matrix $C$ from the two-point functions $X_{ij}$ and $P_{ij}$.
Actually there is a shortcut for the system of the coupled harmonic oscillators Eq.\,\eqref{NHarm}, where $X_{ij}$ and $P_{ij}$ are representable by the interaction matrix $K$ as follows.
First consider the case where the $K$ matrix is a diagonal matrix $D = \text{diag}(\omega_1^2, \omega_2^2, \cdots,\omega_N^2)$.
One can then define creation and annihilation operators 
	\begin{align}
		\begin{aligned}
			b_i &= \frac{1}{\sqrt{2\omega_i}} \left(\omega_i\, \phi_i + \i\, \pi_i\right) \ , \\
			b_i^\dagger &= \frac{1}{\sqrt{2\omega_i}} \left(\omega_i\, \phi_i  - \i\,\pi_i\right) \ ,
		\end{aligned}
	\end{align}
with the commutation relations $[b_i, b_j^\dagger] = \delta_{ij}$.
With them, the Hamiltonian Eq.\,\eqref{NHarm} with the interaction matrix $K=D$ is recast into the following standard form:
\begin{align}
	H = \sum_{i=1}^N \omega_i \left( b_i^\dagger b_i + \frac{1}{2}\right) \ .
\end{align}
Let $|0\rangle$ be a ground state annihilated by all $b_i$, then the two-point correlation functions are
	\begin{align}\label{XandP}
	\begin{aligned}
		X_{ij} &= \langle 0| \phi_i \,\phi_j |0\rangle = \frac{1}{2\omega_i}\delta_{ij} \ , \\
		P_{ij} &= \langle 0| \pi_i \,\pi_j |0\rangle = \frac{\omega_i}{2}\delta_{ij} \ .
	\end{aligned}
	\end{align}
Since any real symmetric matrix $K$ can be diagonalized by an orthogonal matrix $O$, $K = O^T D\,O$, and $K^n = O^T D^n\,O$ for any real $n$, applying the derivation to the new basis $\tilde\phi_i = O_{ij} \phi_j$ leads to
\begin{align}
X_{ij} = \frac{1}{2}(K^{-1/2})_{ij} \ , \qquad P_{ij} = \frac{1}{2}(K^{1/2})_{ij} \ .
\end{align}

The matrix $C=\sqrt{XP}$ is constant, $C = 1/2$, when $A$ is the total system, and the entanglement entropy Eq.\,\eqref{EE_Lattice} clearly vanishes in this case as expected.
On the other hand, it no longer vanishes for a subsystem $A$ specified by $i=1,\cdots, N_A$ where $N_A < N$ because $C_{ij} = \left[\sum_{k=1,\cdots, N_A} (K^{-1/2})_{ik} (K^{1/2})_{kj}\right]^{1/2}/2$ is not necessarily $1/2$ in general.

\subsection{Free massive scalar fields}
The argument used in the previous section can be extended to higher-dimensional theories without difficulty if the system has rotational symmetry.
Namely, for a spherical entangling surface, the system can be effectively reduced to $(1+1)$-dimensional theories on the radial coordinate \cite{Srednicki:1993im,Lohmayer:2009sq,Huerta:2011qi}.

Consider a free massive real scalar field with the action\footnote{We may use a nonminimally coupled scalar field with a term $\CR \phi^2$ in the action, but the result is the same as the minimally coupled case \cite{Casini:2014yca,Herzog:2016bhv}.}
\begin{align}\label{Scalar_action_L}
	I = -\frac{1}{2} \int \d^d x\, \sqrt{-g}\,\left[ (\partial_\mu\phi)^2 + m^2 \phi^2 \right] \ .
\end{align}
We put the theory on the radial coordinates, 
\begin{align}\label{RadialLattice}
	\d s^2 = -\d t^2 + \d r^2 + r^2 \d\Omega_{d-2}^2 \ ,
\end{align}
where $\d\Omega_{d-2}^2$ is the metric for a unit $(d-2)$-sphere.
Rescaling the scalar field $\phi$ by $r^{d/2-1}$ and Fourier decomposing along the angular directions simplifies the Hamiltonian,
\begin{align}
\begin{aligned}
	H &= \frac{1}{2} \sum_{l=0}^\infty \,g_s (d-2,l) \\
		&\qquad\cdot \int_0^\infty \d r \, \left\{  \pi^2_l (r) 
	 + r^{d-2}\left[ \partial_r \left( \frac{\phi_l(r)}{r^{d/2 -1}} \right)\right]^2 \right.\\
	&\left. \qquad \qquad \qquad + \left( m^2 + \frac{l(l+d-3)}{r^2}\right) \phi_l^2 (r) \right\} \ ,
\end{aligned}
\end{align}
where $\pi_l$ are the conjugate momenta of the field $\phi_l$ with the orbital angular momentum $l$, satisfying the commutation relation
\begin{align}
	[\phi_l(r), \pi_{l'}(r')] = \i\, \delta_{l l'} \delta(r - r') \ .
\end{align}
The factor $g_s (d,l)$ is the degeneracy of the $l^{th}$ angular mode of a scalar field on unit $d$-sphere $\BS^d$,
\begin{align}\label{ScalarDeg}
	g(d,l) = \frac{(2l + d -1) \Gamma(l + d - 1)}{l!\, \Gamma(d)} \ .
\end{align}

We discretize the radial coordinate $r$ to $N$ sites parametrized by $i$ with lattice spacing $a$ that leads to the following replacement rule:
\begin{align}
\begin{aligned}
	r& ~\to~ i\, a \ , \qquad \delta(r - r') ~ \to ~ \frac{\delta_{ii'}}{a} \ ,\\
	 \phi_l (r)& ~\to~ \phi_{l,i} \ , \qquad  \pi_l(r) ~\to~ \frac{\pi_{l,i}}{a} \ .
\end{aligned}
\end{align}
for the terms without derivatives and 
\begin{align}
\begin{aligned}
	r& ~\to~ \left(i + \frac{1}{2}\right)\,a \ , \\
	 \partial_r \left( \frac{\phi_l(r)}{r^{d/2 -1}} \right) &~\to~ \frac{1}{a^{d/2}}\left[\frac{\phi_{l,i+1}}{(i+1)^{d/2-1}} - \frac{\phi_{l,i}}{i^{d/2-1}} \right]\ , 
\end{aligned}
\end{align}
for the terms with derivatives.\footnote{Among several discretization schemes we exclusively use Srednicki lattice \cite{Srednicki:1993im} in this article.}
The resulting Hamiltonian on the lattice takes a similar form as Eq.\,\eqref{NHarm} for the $N$-coupled harmonic oscillator:
\begin{align}
	H = \frac{1}{2a} \sum_{l=0}^\infty g_s(d-2,l)  \left[ \sum_{i=1}^N \pi_{l,j}^2 + \sum_{i,j=1}^N \phi_{l,i} \,K_l^{ij} \,\phi_{l,j}\right] \ .
\end{align}
The symmetric matrix $K_l$ is chosen to be
\begin{align}
\begin{aligned}
		K_l^{11} &= \left( \frac{3}{2}\right)^{d-2} + l (l+ d - 3) + (m\,a)^2 \ , \\
		K_l^{ii} &= \frac{1}{i^{d-2}}\left[ \left( i - \frac{1}{2}\right)^{d-2} + \left( i + \frac{1}{2}\right)^{d-2} \right] \\
			&\qquad \qquad  + \frac{l (l+ d - 3)}{i^2} + (m\,a)^2\ , \\
		K_l^{i,i+1} &= K_l^{i+1,i} = - \frac{(i+1/2)^{d-2}}{i^{d/2-1} (i+1)^{d/2-1}}  \ .
\end{aligned}
\end{align}
Hence the entanglement entropy of a free massive scalar field for a spherical system amounts to the summation of those over each angular mode,
\begin{align}\label{EE_Lattice_Scalar}
S_A =  \sum_{l=0}^\infty\, g_s(d-2,l)\,S_l \ ,
\end{align}
where $S_l$ is the entropy for the $l^{th}$ mode of the form Eq.\,\eqref{EE_Lattice} ,
\begin{align}
\begin{aligned}
	S_l &= \tr_A \left[ \left( C_l + \frac{1}{2}\right) \log\left( C_l + \frac{1}{2}\right)\right. \\
		&\qquad \qquad \qquad\left.- \left( C_l - \frac{1}{2}\right) \log\left( C_l - \frac{1}{2}\right)\right] \ ,
\end{aligned}
\end{align}
with $C_l = (C_l)_{ij}$ for $i,j \in A$ being
\begin{align}
(C_l)_{ij} = \frac{1}{2}\left[\sum_{k\in A} (K_l^{-1/2})_{ik} (K_l^{1/2})_{kj}\right]^{1/2} \ .
\end{align}
For a spherical entangling surface of radius $R$, the indices $i,j,k \in A$ range from $1$ to $N_A$ where $N_A$ is fixed by $R = (N_A + 1/2)a$.

  \section{Euclidean formalism}\label{ss:Euclidean}
In this section, we rewrite the definition of entanglement entropy by introducing an auxiliary (replica) parameter.
After giving a representation of the reduced density matrix in terms of the path integral, we derive an expression of the entanglement entropy given by the Euclidean partition function on a singular manifold. 
This approach is sometimes easier than the real time approach in performing analytic computations in QFT.

\subsection{Replica trick}
We adopt the definition Eq.\,\eqref{EE_def} of entanglement entropy with Eqs.\,\eqref{RenyiDef} and \eqref{RenyiEE},
\begin{align}\label{EE_Replica}
\begin{aligned}
S_A &= -\lim_{n\to 1}\frac{\log \tr_A( \rho_A^n)}{n-1} \ ,\\
	&= -\lim_{n\to 1}\partial_n \log \tr_A (\rho_A^n) \ ,
\end{aligned}
\end{align}
where we used the fact that the reduced density matrix is normalized $\tr_A (\rho_A) = 1$.
Although the R{\'e}nyi entropy is defined only for an integer $n$, the analytic continuation is assumed in taking the $n\to 1$ limit.
This method is called the replica trick that is often employed for the entanglement entropy calculation in QFT.

\paragraph*{Two spin system with the replica trick $:$}
We revisit the two spin system Eq.\,\eqref{TwoSpinPure} in Sec. \ref{TwoSpin} with the replica trick.
The reduced density for the spin $A$ was given by Eq.\,\eqref{TwoSpinRD}, hence we immediately have the trace of the $n^{th}$ power,
\begin{align}
	\tr_A (\rho_A^n)  = 2^{1-n} \ .
\end{align}
Plugging this into Eq.\,\eqref{EE_Replica} yields the entanglement entropy
\begin{align}
	S_A = - \lim_{n\to 1} \partial_n \log 2^{1-n} = \log 2 \ ,
\end{align}
which agrees with the previous calculation Eq.\,\eqref{EE_TwoSpin}.

\subsection{Path integral representation in quantum mechanics}
We have seen how the reduced density matrix for entanglement entropy is calculated in the Hamiltonian formulation so far.
There is an equivalent, but complementary approach using the path integral formulation of quantum mechanics which we are going to deal with and will be further extended to quantum field theory in the next section.

We first recap the basic facts about quantum mechanics and the path integral formulation.
In the Schr\"odinger picture a state evolves under the Schr\"odinger equation,
\begin{align}
	\i\, \frac{\d}{\d t}\, |\psi (t)\rangle = \hat H\,  |\psi (t)\rangle \ ,
\end{align}
while operators such as a position operator $\hat x$ do not.
Hence the eigenvector $\hat x |x\rangle = x| x\rangle$ for the operator is also time independent.

In the Heisenberg picture, operators evolve under the Heisenberg equation
\begin{align}
	\i\, \frac{\d}{\d t}\, \hat A(t) = [\hat H, \hat A(t)] \ ,
\end{align}
but the state vectors are time independent.
The operator $\hat x(t)$ and the eigenvector $|x; t \rangle$ are related to those in the Schr{\"o}dinger picture as
\begin{align}
	\hat x(t) &= e^{\i \hat H t}\, \hat x\,e^{-\i \hat H t} \ , \qquad
	|x; t \rangle = e^{\i \hat H t} |x\rangle \ .
\end{align}
The transition amplitude from an initial state $|x_i; t_i\rangle$ at time $t=t_i$ to a final state $|x_f; t_f\rangle$ at $t=t_f (>t_i)$ is given by their overlap $\langle x_f; t_f |x_i; t_i\rangle$ in the Heisenberg picture.
Thus the transition amplitude in the Schr\"odinger picture is
\begin{align}
	\langle x_f; t_f |x_i; t_i\rangle = \langle x_f |\, e^{-\i \hat H(t_f - t_i)}\,|x_i\rangle \ .
\end{align}
A standard argument for splitting the time interval to a number of small intervals and inserting the complete bases at each time slice yields the path integral representation,
\begin{align}\label{QMkernel}
	\langle x_f | e^{-\i \hat H(t_f - t_i)}|x_i\rangle = \int_{x(t_i) = x_i}^{x(t_f) = x_f} [\CD x(t)]\, e^{\i \int_{t_i}^{t_f}\, \d t \,L(x, \dot x)} \ ,
\end{align}
where $L(x, \dot x)$ is the Lagrangian for the system of the given Hamiltonian and $[\CD x(t)]$ is the path integral measure.

We represent the ground state wave function in the path integral language.
It is most easily achieved by analytically continuing the Lorentzian time $t$ to the Euclidean time $\tau$ by the Wick rotation $t = - \i\,\tau$.
The transition amplitude becomes
\begin{align}\label{QMkernelE}
\begin{aligned}
	\langle x_f; \tau_f |x_i; \tau_i\rangle &= \langle x_f | \,e^{-\hat H(\tau_f - \tau_i)}\,|x_i\rangle \ ,\\
		&= \int_{x(\tau_i) = x_i}^{x(\tau_f) = x_f} [\CD x(\tau)]\, e^{- I_E(x)} \ ,
\end{aligned}
\end{align}
where $I_E(x)$ is the Euclidean action.
We specialize the propagator Eq.\,\eqref{QMkernelE} to the case with $(x_f, \tau_f) = (x,0)$ and $(x_i, \tau_i) = (y,T)$, and insert a complete set of the energy eigenstates $\hat H |n\rangle = E_n |n\rangle$ of the Hamiltonian in the left-hand side to obtain
\begin{align}\label{EuclideanAmp}
\begin{aligned}
	\langle x; 0 | y; T\rangle &= \langle x | \,e^{\hat H T}\,|y\rangle \ , \\
			&= \sum_{n} \psi_n (x) \,\psi_n^\ast (y)\, e^{E_n T} \ ,
\end{aligned}
\end{align}
where $\psi_n(x) \equiv \langle x | n\rangle$ is the wave function of the $n^\text{th}$ eigenstate with energy $E_n$.
The transition amplitude Eq.\,\eqref{EuclideanAmp} is dominated by the ground state wave function in $T \to -\infty$ limit,
\begin{align}
	\langle x; 0 | y; -\infty \rangle = \lim_{T\to -\infty} \psi_0(x)\, \psi_0^\ast (y)\, e^{E_0 T} \ .
\end{align}
Multiplying $\psi_0(y)$ and integrating over $y$ using $\int \d y\, |\psi_0(y)|^2 = 1$, 
we obtain the path integral representation of the ground state wave function $\psi_0 (x)$ at $\tau =0$
\begin{align}
	\psi_0 (x) =  \int dy \int_{x(-\infty) = y}^{x(0) = x} [\CD x(\tau)]\, e^{- I_E(x)}\,\psi_0(y) \ , 
\end{align}
where we set $E_0 = 0$ for simplicity.
We make a change of the notation and use $\Psi(x) \equiv \psi_0(x)$ for the ground state wave function from now on.
Also we write the path integral without specifying the boundary condition at $\tau = -\infty$,
\begin{align}\label{GSWFQM}
	\Psi (x) = \int_{\tau=-\infty}^{\tau = 0,\, x(0) = x} [\CD x(\tau)]\, e^{- I_E(x)} \ .
\end{align}
Similarly, the complex conjugate of the ground state wave function is given by the path integral from $\tau = 0$ with a given boundary condition to $\tau = \infty$,
\begin{align}\label{GSWFQMC}
	\Psi^\ast (x) = \int^{\tau=\infty}_{\tau = 0,\, x(0) = x} [\CD x(\tau)]\, e^{- I_E(x)} \ .
\end{align}

We are now in position to apply the path integral representation to calculating the entanglement entropy.
Suppose the system is at zero temperature and in a pure (not necessarily normalized) ground state $|\Psi\rangle$.
The Hilbert space is described by the density matrix $\rho_\text{tot} = |\Psi\rangle\langle \Psi | / Z$ with the partition function $Z\equiv \langle \Psi | \Psi\rangle$ and the reduced density matrix $\rho_A$ becomes
\begin{align}\label{RedDM}
\rho_A = \frac{1}{Z}\,\tr_B (|\Psi\rangle \langle \Psi |) \ .
\end{align}
We consider a bipartite system of two particles whose coordinates are denoted by $x_A$ and $x_B$.
Then a state in the total system is spanned by a tensor product state $|x_A, x_B\rangle \equiv |x_A\rangle\otimes |x_B\rangle$, with which
the matrix element of the reduced density matrix $\rho_A$ is written as
\begin{align}\label{DM_pathinte}
\begin{aligned}
	\langle x_A| \,\rho_A\, |x_A' \rangle &= \frac{1}{Z} \int \d x_B\, \langle x_A, x_B |\Psi\rangle \langle \Psi |x_A', x_B\rangle \ ,
\end{aligned}
\end{align}
where the partial trace over the Hilbert space $B$ is taken by the integration over $x_B$.
We substitute into Eq.\,\eqref{DM_pathinte} the path integral representations Eqs.\,\eqref{GSWFQM} and \eqref{GSWFQMC} of the ground state, but slightly shift the Euclidean time from $\tau = 0$ to $\tau = -\epsilon$ for $\Psi(x)$ and $\tau = \epsilon$ for $\Psi^\ast(x)$ with a small parameter $\epsilon>0$ as a regularization,
\begin{align}
\begin{aligned}
	\langle x_A| &\,\rho_A\, |x_A' \rangle 
		= \frac{1}{Z} \int \d x_B\\ 
		& \cdot \int_{\tau=\epsilon, \text{bc}^+}^{\tau=\infty}\int^{\tau=-\epsilon, \text{bc}^-}_{\tau=-\infty} [\CD y_A(\tau)\,\CD y_B(\tau)]\, e^{- I_E (y_A, y_B)} \ ,
\end{aligned}
\end{align}
where $\text{bc}^\pm$ are the boundary conditions at $\tau = \pm \epsilon$ specified by
\begin{align}
\begin{aligned}
	\text{bc}^+ &=\{ y_A(\epsilon) = x_A', y_B(\epsilon) = x_B\} \ , \\
	\text{bc}^- &= \{y_A(-\epsilon) = x_A, y_B(-\epsilon) = x_B \}\ .
\end{aligned}
\end{align}
Performing the $x_B$ integral and letting $\epsilon$ be zero amounts to the path integral for $y_B$ over the entire Euclidean time while the boundary conditions for $y_A$ at $\tau = \pm \epsilon$ can be implemented by inserting delta functions
\begin{align}\label{QM_rhoA}
\begin{aligned}
	\langle x_A| \,\rho_A\, |x_A' \rangle = \frac{1}{Z} \int^{\tau=\infty}_{\tau=-\infty} &[\CD y_A(\tau)\,\CD y_B(\tau)]\,e^{- I_E (y_A, y_B)}\\
	&\cdot \delta \left[ y_A(0^+) - x_A'\right]\,\delta \left[ y_A(0^-) - x_A\right]\ , 
\end{aligned}
\end{align}
where we introduced the notation $0^\pm =\lim_{\epsilon\to +0}\pm\epsilon$.
Eq.\,\eqref{QM_rhoA} leads to the path integral representation of the entanglement entropy, but we postpone the derivation until the next section where we focus on the QFT case.

\subsection{Path integral representation of entanglement entropy in quantum field theory}
The real time formalism is not effective for QFT unless we discretize the space to lattices, and the reduced density matrix is not so straightforward to obtain in QFT.
We want to make use of the path integral representation and represent the entanglement entropy in terms of the partition function on a manifold with a singularity on the entangling surface.

\begin{table}[htbp]
\caption{\label{tab:QM_QFT}Correspondence between quantum mechanics (QM) and quantum field theory (QFT).}
\begin{ruledtabular}
\begin{tabular}{ccc}
  & QM & QFT\\ \hline
 Variable & $\hat x$ & $\hat \phi(\vec x)$\\
 State & $|x\rangle$ & $|\phi (\vec x)\rangle$ \\
 Wave function & $\Psi(x)$ & $\Psi[ \phi(\vec x) ]$
\end{tabular}
\end{ruledtabular}
\end{table}

To this end, we need a representation of the ground state wave function in the path integral form as in the quantum mechanical case.
In QFT$_d$, the variable corresponding to the coordinate $\hat x$ in QM is a quantum field $\hat\phi(\vec x)$ living on a $(d-1)$-dimensional space parametrized by the coordinate vector $\vec x$ and the state vector obeying the Schr{\"o}dinger equation is the eigenvector of the field operator $\hat \phi (t,\vec{x})$ at $t=0$: $\hat \phi (0,\vec{x}) |\phi_0(\vec{x}) \rangle = \phi_0(\vec{x})|\phi_0(\vec{x}) \rangle$ (see Table \ref{tab:QM_QFT}).

\begin{figure}[h]
\centering
	\begin{tikzpicture}[scale = 0.8]
			\draw[fill=verylightgray, thick] (-3, 0) -- node[below] {$t=0$} node[above]{$\phi_0$} (-1, 0) -- (-1, -2) -- node[below]{$t=-\infty$} (-3, -2) -- (-3, 0);
			\draw[dashed, thick] (-3, 0) -- (-3, 2) -- node[above] {$t=\infty$} (-1, 2) -- (-1, 0);
			\node at (-4, 0) {$\langle \phi_0 |\Psi\rangle =$};
			\draw[->, thick] (-3.1, -3) -- (-0.9, -3) node[right] {$\vec{x}$};
			\draw[->, thick] (-0.5, -2) -- (-0.5, 2.1) node[above] {$t$};			
	\end{tikzpicture}
	\hspace*{0.5cm}
	\begin{tikzpicture}[scale = 0.8, xshift=-0.5cm]
			\draw[fill=verylightgray, thick] (3, 0) -- node[above] {$t=0$} node[below]{$\phi_0'$} (5, 0) -- (5, 2) -- node[above]{$t=\infty$} (3, 2) -- (3, 0);
			\draw[dashed, thick] (3, 0) -- (3, -2) -- node[below] {$t=-\infty$} (5, -2) -- (5, 0);
			\node at (2, 0) {$\langle \Psi | \phi_0' \rangle =$};
			\draw[->, thick] (2.9, -3) -- (5.1, -3) node[right] {$\vec{x}$};
			\draw[->, thick] (5.5, -2) -- (5.5, 2.1) node[above] {$t$};	
	\end{tikzpicture}
\caption{\label{fig:wavefunc} Path integral representations of a wave function.
The integral is performed over the shaded regions.}
\end{figure}
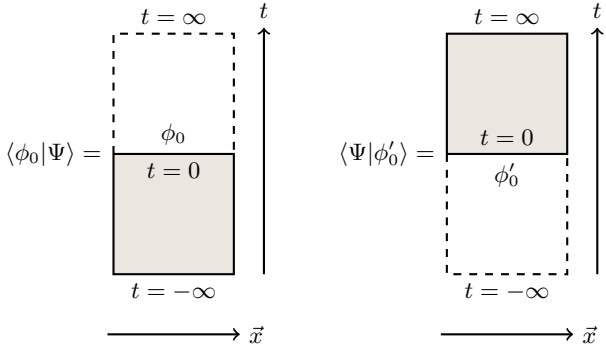

In QFT the wave function of the ground state $|\Psi\rangle$ is given by the wave functional $\Psi \left[\phi_0(\vec{x})\right] = \langle \phi_0(\vec{x}) |\Psi\rangle$ whose Euclidean path integral representation is
\begin{align}
\begin{aligned}
\Psi \left[\phi_0(\vec{x})\right] &= \langle \phi_0(\vec{x}) |\Psi\rangle \ ,\\
	&= \int_{t=-\infty}^{t=0, \, \phi(t=0,\vec{x}) = \phi_0 (\vec{x})} [\CD\phi (t,\vec{x})] \,e^{-I_E[\phi]} \ .
\end{aligned}
\end{align}
Similarly, we can represent the conjugate of the wave functional by
\begin{align}
\begin{aligned}
\Psi^\ast \left[\phi_0'(\vec{x})\right] &=\langle\Psi |\phi_0'(\vec{x})\rangle \ ,\\
	&= \int_{t=0,\,\phi(t=0,\vec{x}) = \phi_0' (\vec{x})}^{t=\infty} [\CD\phi (t,\vec{x})] \,e^{-I_E[\phi]} \ .
\end{aligned}
\end{align}
Figure \ref{fig:wavefunc} shows the pictorial representations of the wave functionals.
The partition function $Z$ is the path integral over the entire Euclidean space as it is written by
\begin{align}
	Z = \int [\CD \phi_0(t=0, \vec x)]\, \langle \Psi | \phi_0 \rangle \,\langle\phi_0| \Psi\rangle \ .
\end{align}
In this pictorial expression, taking the partial trace over the subsystem $B=\bar A$ is equivalent to gluing the edges of the two sheets $\langle\Psi|\phi_0\rangle$ and $\langle \phi_0 |\Psi\rangle$ along $B$.
In other words, this is implemented by integrating the total density matrix over every state $\phi^B(t=0,\vec{x})$ with support only on $\vec{x}\in B$: 
\begin{align}
\rho_A = \frac{1}{Z} \int [\CD \phi^B(t=0, \vec{x}\in B)] \langle \phi^B| \Psi \rangle \langle \Psi | \phi^B\rangle \ .
\end{align}
The reduced density matrix has two indices $[\rho_A]_{ab} = \langle \phi_a^A |\rho_A |\phi_b^A\rangle$ where $\phi_a^A(\vec{x}\in A)$ and $\phi_b^A(\vec{x}\in A)$ specify the boundary conditions on $A$ at $t=0^+$ and $0^-$, respectively.\footnote{Here the order of the indices of $\rho_A$ is chosen so that the path integral representation looks natural. One may use the opposite order and still obtain the same result.}
Therefore, $\rho_A$ is represented by the path integral on the Euclidean space with a cut along the subsystem $A$,
\begin{align}\label{QFT_RD_PIR}
\begin{aligned}
 &[\rho_A]_{ab} \\
&=
 \frac{1}{Z} \int [\CD \phi^B(t=0, \vec{x}\in B)]\,  \left(\langle \phi_a^A| \langle \phi^B|\right) |\Psi\rangle \langle \Psi |\left( | \phi_b^A \rangle  | \phi^B \rangle \right) \ , \\
&=\frac{1}{Z} \int [\CD \phi^B(t=0, \vec{x}\in B)]\quad  
\begin{tikzpicture}[baseline={([yshift=-.5ex]current bounding box.center)}]
			\draw[fill=verylightgray, thick] (-1, -1.5) -- ++ (3, 0) --++  (0, 1.4) -- node [below] {\textcolor{kakiiro}{$\phi_a^A$}} ++ (-3, 0) --++ (0, -1.4);
			\draw[fill=verylightgray, thick] (-1, 0.1) -- node[above] {\textcolor{hanadairo}{$\phi_b^A$}} ++ (3, 0) --++  (0, 1.4) --++ (-3, 0) --++ (0, -1.4);
			\filldraw [fill=black] (0,0.1) node [above left] {$\phi^B$} circle (0.05);
			\filldraw [fill=black] (1,0.1) node [above right] {$\phi^B$} circle (0.05);			
			\filldraw [fill=black] (0,-0.1) node [below left] {$\phi^B$} circle (0.05);
			\filldraw [fill=black] (1,-0.1) node [below right] {$\phi^B$} circle (0.05);
			\filldraw [fill=black] (0,-1.8) circle (0.05);
			\filldraw [fill=black] (1,-1.8) circle (0.05);
			\node[right] at (2, 0.2) {$t = 0^+$};
			\node[right] at (2, -0.2) {$t = 0^-$};
			\draw[dashed] (0, -0.1) -- (0, -1.8);
			\draw[dashed] (1, -0.1) -- (1, -1.8);
			\draw[thick, ->] (-1, -1.8) -- node[below] {$B\quad~~ A \quad~~B$} (2.1, -1.8) node[right] {$\vec{x}$};
\end{tikzpicture} \\
&=\frac{1}{Z}\quad \begin{tikzpicture}[baseline={([yshift=-.5ex]current bounding box.center)}]
			\draw[fill=verylightgray, thick] (-1, -1.5) -- (2, -1.5) --  (2, 1.5) -- (-1, 1.5) -- (-1, -1.5);
			\draw[thick, ->] (-1, -1.8) -- node[below] {$B\quad~~ A \quad~~B$} (2.1, -1.8) node[right] {$\vec{x}$};
			\filldraw [fill=black] (0,-1.8) circle (0.05);
			\filldraw [fill=black] (1,-1.8) circle (0.05);
			\draw[fill=white, thick] (0, -0.1) -- node[below] {\textcolor{kakiiro}{$\phi_a^A$}} (1, -0.1) -- (1, 0.1) -- node[above] {\textcolor{hanadairo}{$\phi_b^A$}} (0, 0.1) -- (0, -0.1);
			\draw[dashed] (1, -0.1) -- (2, -0.1);
			\draw[dashed] (1, 0.1) -- (2, 0.1);
			\node[right] at (2, 0.2) {$t = 0^+$};
			\node[right] at (2, -0.2) {$t = 0^-$};
			\draw[dashed] (0, -0.1) -- (0, -1.8);
			\draw[dashed] (1, -0.1) -- (1, -1.8);
\end{tikzpicture} \\
&= \frac{1}{Z} \int
[\CD \phi(t,\vec{x})]\, e^{-I_E[\phi]} \\
	&\qquad \cdot  \prod_{\vec{x} \in A} \delta\left( \phi(0^+, \vec{x}) - \phi_b^A(\vec{x})\right)\,\delta\left( \phi(0^-, \vec{x}) - \phi_a^A(\vec{x})\right) \ .
\end{aligned}
\end{align}
While we derived it pictorially here it takes exactly the same form as the quantum mechanical case, Eq.\,\eqref{QM_rhoA}.
In other words, we are able to arrive at the path integral representation Eq.\,\eqref{QFT_RD_PIR} of the reduced density matrix in QFT from the previous result Eq.\,\eqref{QM_rhoA} in QM using the dictionary in Table \ref{tab:QM_QFT}.

It follows from this representation that the trace of $n^{th}$ power of the reduced density matrix 
is given by the partition function $Z_n$ on the $n$-fold cover $\CM_n$ of the original spacetime that is constructed by gluing $n$ copies of the sheet with a cut along $A$ \cite{Holzhey:1994we,Calabrese:2004eu},
\begin{align}\label{Replica_DM_PF}
\begin{aligned}
\tr_A (\rho_A^n) &= \frac{1}{(Z)^n}\qquad 
	\begin{tikzpicture}[baseline={([yshift=-.5ex]current bounding box.center)},scale=0.75, every node/.style={scale=0.85}]
				\draw[fill=verylightgray, thick] (-2, 4) -- (1, 4) -- (2, 5.5) --(-1, 5.5) -- (-2, 4);
				\draw[fill=white] (-0.7, 4.7) -- node[below] {\textcolor{kakiiro}{$\phi_1^A$}}++(1.2, 0) --++ (0.1, 0.15) -- node[above] {\textcolor{hanadairo}{$\phi_2^A$}} ++(-1.2, 0) --++ (-0.1, -0.15);
				\draw[fill=verylightgray, thick] (-2, 2) -- (1, 2) -- (2, 3.5) --(-1, 3.5) -- (-2, 2);
				\draw[fill=white] (-0.7, 2.7) -- node[below] {\textcolor{hanadairo}{$\phi_2^A$}} ++(1.2, 0) --++ (0.1, 0.15) -- node[above] {\textcolor{enjiiro}{$\phi_3^A$}} ++(-1.2, 0) --++ (-0.1, -0.15);
				\draw[dotted, thick] (-0.2, 1.8) -- (-0.2, 0.2);
				\draw[fill=verylightgray, thick] (-2, -1.5) -- (1, -1.5) -- (2, 0) --(-1, 0) -- (-2, -1.5);
				\draw[fill=white] (-0.7, -0.8) -- node[below] {\textcolor{ebicha}{$\phi_n^A$}} ++(1.2, 0) --++ (0.1, 0.15) -- node[above] {\textcolor{kakiiro}{$\phi_1^A$}} ++(-1.2, 0) --++ (-0.1, -0.15);
				\draw [decorate,decoration={brace,amplitude=10pt,raise=4pt},yshift=0pt]
(2, 5.5) -- (2, -1.5) node [black,midway,right=0.5] {$n$ copies $\equiv Z_n$};
			\end{tikzpicture} \\
	&= \frac{Z_n}{(Z)^n} \ ,
\end{aligned}
\end{align}
where we denote the partition function on the $n$-fold cover by $Z_n$.
Substituting this representation to Eq.\,\eqref{EE_Replica} gives an alternative definition of entanglement entropy in QFT,
\begin{align}\label{EE_PF}
S_A = -\lim_{n\to 1} \partial_n (\log Z_n - n \log Z) \ .
\end{align}
The $n$-fold cover $\CM_n$ has a conical singularity along the codimension-two {\it entangling surface} $\Sigma = \partial A$ with a deficit angle $2\pi (1-n)$.
This expression will be used in later sections.

It is interesting and valuable to give the path integral representation of the modular entropy in Eq.\,\eqref{ModularEntropy} to manifest the relation to thermal entropy.
If we regard $\beta = 2\pi n$ as an inverse temperature, the thermal partition function $Z(\beta)$ defined by Eq.\,\eqref{Thermal_PF} is proportional to the partition function on the singular manifold $\CM_n$,
\begin{align}
	Z(\beta) = \frac{Z_n}{(Z)^n} \ .
\end{align}
The modular entropy $\tilde S_n$, when written in terms of the replica partition function, then becomes
\begin{align}
	\tilde S_n 	= (1- n\partial_n) \log Z_n \ ,
\end{align}
which reduces to the entanglement entropy Eq.\,\eqref{EE_PF} in the $n\to 1$ limit.
Finally performing the integral Eq.\,\eqref{Renyi_Modular} gives the R{\'e}nyi entropy
\begin{align}\label{Renyi_QFT}
	S_n = \frac{1}{1-n}\,\log \left(\frac{Z_n}{Z^n}\right) \ ,
\end{align}
which can also be obtained more directly using Eq.\,\eqref{Replica_DM_PF} and the definition of the R{\'e}nyi entropy, Eq.\,\eqref{RenyiDef}.

\subsection{Entanglement entropy across a hyperplane}
\begin{figure}[htbp]
\centering
\begin{tikzpicture}
\draw [thick] (0,0) --++ (6,0) --++ (0,4) --++ (-6,0) --++ (0,-4);
\draw [thick, dashed] (3,2) -- node [below] {$A$}++ (3,0); 
\draw [thick, ->, domain=10:350] plot ({3+0.7*cos(\x)}, {2+0.7*sin(\x)});
\node at (3,3.2) {$\tau \sim \tau +2\pi n$};
\end{tikzpicture}
\caption{\label{fig:ScalarInfinite} The $n$-fold cover of $\BR^d$ with a cut along the subregion $A$ ($x_1>0$) at $t=0$ in the polar coordinates.}
\end{figure}
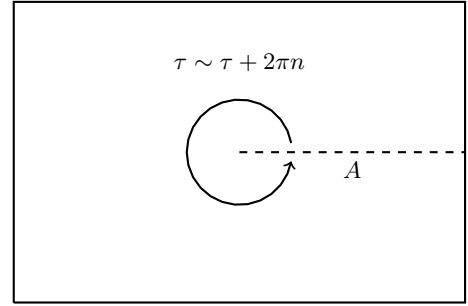

The Euclidean version of the action Eq.\,\eqref{Scalar_action_L} for a free massive scalar field is
\begin{align}
I_E = \frac{1}{2}\int \d^d x \left[ (\partial_\mu \phi)^2 + m^2 \phi^2\right] \ .
\end{align}
We denote the space and Euclidean time coordinates by $x_i\, (i=1,\cdots, d-1)$ and $x_0$.
Let $A$ and $B$ be regions in $x_1> 0$ and $x_1\le0$, respectively.
The entangling surface $\Sigma$ is chosen to be a $(d-2)$-dimensional hyperplane at $x_1=0$: $\Sigma = \{ (x_0 ,x_i)|\, x_0 = x_1 = 0\}$.
Introducing the metric in the polar coordinates for the $(x_0 ,x_1)$ plane,
\begin{align}\label{EE_Hyperplane}
\begin{aligned}
\d s^2 &= \d x_0^2 + \d x_1^2  + \sum_{i=2}^{d-1} \d x_i^2 \ ,\\
 	&= \d r^2 + r^2 \d\tau^2 + \sum_{i=2}^{d-1} \d x_i^2 \ ,
\end{aligned}
\end{align}
the $n$-fold cover $\CM_n$ with a cut along $A$ is given by the same metric with $0\le r$ and $0\le \tau \le 2\pi n$ as shown in Fig.\,\ref{fig:ScalarInfinite}.\footnote{The angular variable $\tau$ should not be confused with the Euclidean time. It is rather considered as the ``modular time" for the modular Hamiltonian.}
Hence the manifold $\CM_n$ is the direct product of the two-dimensional cone $\CC_n$ parametrized by $(r,\tau)$ and $\BR^{d-2}$.

We are concerned with the partition function $Z_n$ of the free scalar field on $\CM_n$, which is simply given by the one-loop determinant
\begin{align}\label{SchwingerRep}
\begin{aligned}
\log Z_n &= -\frac{1}{2} \log \det (-\nabla^2 + m^2) \ , \\
	&= -\frac{1}{2}\tr \log (-\nabla^2 + m^2) \ , \\
	&= \frac{1}{2}\int_{\epsilon^2}^\infty \frac{\d s}{s}\, \tr \left[ e^{-s (-\nabla^2 + m^2)} - e^{-s} \right]\ , 
\end{aligned}
\end{align}
where $\nabla$ is the covariant derivative on $\CM_n$ and we introduced the Schwinger parameter $s$ in the third equality.\footnote{
This follows from the identity for a matrix $M$,
\begin{align}
\log M =  \int_0^\infty \frac{\d s}{s}\left[ e^{-s} - e^{-s M}\right] \ .
\end{align}}
The parameter $s=\epsilon^2\ll 1$ is introduced so as to act as a regulator for the UV divergence.

In the Schwinger representation Eq.\,\eqref{SchwingerRep} the calculation of the partition function amounts to evaluating the kernel in the integrand.
This is straightforward to carry out once the eigenvalues of the Laplacian $\nabla^2$ are given, but finding the eigenfunctions is a formidable task on the $n$-fold cover $\CM_n$ of a general type of entangling surfaces.
Fortunately, the manifold $\CM_n$ for the hyperplanar entangling surface has the direct product structure $\CM_n = \CC_n \times \BR^{d-2}$ as seen from Eq.\,\eqref{EE_Hyperplane}; thus the Laplacian decomposes into the sum of those on $\CC_n$ and $\BR^{d-2}$: $\nabla^2 = \nabla^2_{\CC_n} + \nabla^2_{\BR^{d-2}}$.
We know the plane waves are the eigenfunctions of the flat space Laplacian $\nabla^2_{\BR^{d-2}}$, and we only need to solve the eigenvalue problem for the cone Laplacian $\nabla^2_{\CC_n}$.

The rotational symmetry of the cone $\CC_n$ allows the Fourier decomposition by the modes $\exp (\i\, l \tau/n)$ with integer $l$ along the angle $\tau$ of period $2\pi n$; hence the eigenfunctions $\phi_{k,l}(r,\tau)$ of the Laplacian are parametrized by two parameters $(k, l)$ satisfying
\begin{align}
\begin{aligned}
\nabla^2_{\CC_n} \phi_{k,l}(r,\tau) &= - k^2 \,\phi_{k,l}(r,\tau) \ , \qquad (k\in \BR^+, l\in \BZ) \ , \\
\phi_{k,l}(r,\tau) &= \sqrt{\frac{k}{2\pi n}} \, e^{\i l \tau/n} J_{\left| l/n\right|} (k\,r) \ , 
\end{aligned}
\end{align}
where $J_n$ is the Bessel function of the first kind.
We normalize the eigenfunctions so that they form an orthonormal basis on the cone $\CC_n$ \cite{Kabat:1995eq},
\begin{align}
\int_{\CC_n} \d^2 x\, \phi_{k,l}(x)\, \phi_{k',l'}^\ast (x) &= \delta_{l l'}\, \delta (k - k') \ .
\end{align}
The Laplacian on $\BR^{d-2}$ has the orthonormal basis of the eigenfunctions spanned by the plane waves, $\phi_{{\bm k}^\perp} (y) = \exp (\i\, {\bm k}_\perp \cdot {\bm y}) / (2\pi)^{(d-2)/2}$, with the eigenvalues $-k_\perp^2$.
Thus we can construct the orthonormal basis $\phi_{k,l,{\bm k}_\perp} (x,y) \equiv \phi_{k,l}(x)\, \phi_{{\bm k}_\perp}(y)$ of the Laplacian $\nabla^2$ with the eigenvalues $k^2 + k_\perp^2$, and evaluate the trace of the kernel $\tr\,[e^{-s (-\nabla^2 + m^2)}]$ that appeared in the third line of Eq.\,\eqref{SchwingerRep}:
\begin{align}\label{SLTrace}
\begin{aligned}
\tr &\left[e^{-s (-\nabla^2 + m^2)}\right] \\
	&= \int_{\CC_n} \d^2 x \sum_{l=-\infty}^\infty \int_0^\infty \d k \, e^{-s(k^2 + m^2)} \phi_{k,l}(x)\phi_{k,l}^\ast (x) \\	
		&\qquad \cdot \int_{\BR^{d-2}} \d^{d-2} y \int \d^{d-2}k_\perp \, e^{-s\, k_\perp^2} \, \phi_{{\bm k}^\perp} (y)\phi_{{\bm k}^\perp}^\ast (y) \ ,\\
	&= \frac{\text{Vol}(\BR^{d-2})}{12\, n} \frac{e^{-s m^2}}{(4\pi s)^{(d-2)/2}} \ .
\end{aligned}
\end{align}

There are technical subtleties in performing the integral in Eq.\,\eqref{SLTrace} associated with the UV and IR divergences, each coming from the angular modes $l$ and the volume of the spacetime, respectively. 
One way to see the IR divergence is to use the identity for the calculation that holds for $\text{Re} \,(\alpha) > -1$,
\begin{align}
\int_0^\infty \d k\, k\, e^{-s k^2} J_\alpha (kr)^2 &= \frac{e^{- r^2/(2s)}}{2s}\, I_\alpha \left( \frac{r^2}{2s}\right) \ ,\\
\int_0^\infty \d r\, r \,e^{-r^2} I_\alpha \left( r^2\right) &= - \frac{\alpha}{2} + \int_0^\infty \d r \,\frac{1}{\sqrt{2\pi}} \ ,\label{IR_Identity}
\end{align}
where $I_\alpha$ is the modified Bessel function of the first kind.
The second term on the right-hand side of Eq.\,\eqref{IR_Identity} is divergent, but independent of $\alpha\, (=|l/n|)$, thus giving rise to a term proportional to $n$ from the volume of $\CM_n$ to $\log Z_n$ and does not contribute to the entanglement entropy.
On the other hand, the UV divergence arises from the summation over the angular momentum $l$, which can be regularized, for example, by using the zeta function,
\begin{align}
\sum_{l=-\infty}^\infty |l| = 2\zeta(-1) =  - \frac{1}{6} \ .
\end{align}

Our derivation of Eq.\,\eqref{SLTrace} is slightly different from the one given by \textcite{Kabat:1995eq} that employs another regularization scheme.
An alternative way to evaluate the partition function $Z_n$ is to take a derivative with respect to $m^2$ and find the coincident Green function $G_n(x, x)$ on the $n$-fold cover space $\CM_n$ \cite{Calabrese:2004eu},
\begin{align}
\frac{1}{m^2} \log Z_n = - \frac{1}{2}\int_{\CM_n} \d^dx\, G_n(x, x) \ .
\end{align}

We are left with the other kernel $\tr \,(e^{-s})$ to evaluate in Eq.\,\eqref{SchwingerRep}, but it is easily seen to be proportional to $n$ due to the volume of the cone $\CC_n$ and does not contribute to Eq.\,\eqref{EE_PF} in contrast to Eq.\,\eqref{SLTrace}.
Thus, by putting all together, we obtain the entanglement entropy across a hyperplane $\BR^{d-2}$ for a free massive scalar field,
\begin{align}
	S_A = \frac{\pi\,\text{Vol}(\BR^{d-2})}{3} \int_{\epsilon^2}^\infty \d s \, \frac{e^{-s m^2}}{(4\pi s)^{d/2}}  \ .
\end{align}

The UV divergent terms of the entanglement entropy can be read off by expanding it around $\epsilon=0$,
\begin{align}
	S_A = \frac{\text{Vol}(\BR^{d-2})}{6(4\pi)^{d-1}}\left[ \frac{1}{(d-2)\epsilon^{d-2}} - \frac{m^2}{(d-4)\epsilon^{d-4}} + \cdots \right] \ .
\end{align}
The leading term is of order $1/\epsilon^{d-2}$ and the coefficient is
proportional to the area of the entangling surface $\Sigma = \BR^{d-2}$.
This is the manifestation of \emph{the area law of entanglement entropy}, one of the characteristics of quantum entanglement in QFTs.
The subsequent terms are less divergent of order $\epsilon^{d-2i}$ with $i=1,2,\cdots$.
We discuss these UV structures in a more general setting in Sec. \ref{ss:UVstructure} and show that they hold for any local quantum field theory on any manifold without boundary.
It, however, should be noted that the area law does not hold in $d=2$ dimensions where the entropy logarithmically diverges
\begin{align}
S_A = - \frac{1}{12}\log (m^2 \epsilon^2) \ .
\end{align}

\subsection{Rindler spacetime and thermofield double}\label{ss:RindlerTFD}

The previous example of the entanglement entropy across a hyperplane is not merely the simplest setup we can carry out for the exact calculation, but also has its physical origin in understanding a thermal aspect of black holes as an entanglement of a field across the horizon \cite{Israel:1976ur,Laflamme:1988wg}.

\begin{figure}[htbp]
\centering
\begin{tikzpicture}[%
    scale=2.7,%
    maingrid/.style={draw,very thick},%
    subgrid/.style={draw,thin},%
    tlabels/.style={pos=0.88,above,sloped,yshift=-.3ex},%
    label/.style={%
        postaction={%
            decorate,%
            transform shape,%
            decoration={%
                markings,%
                mark=at position .65 with \node #1;%
            }%
        }%
    },%
]%
    \pgfmathdeclarefunction{arcosh}{1}{\pgfmathparse{ln(#1+sqrt(#1+1)*sqrt(#1-1))}}
    \pgfmathsetmacro{\Xmax}{1.2}
    \pgfmathsetmacro{\Tmax}{1.2}
    \pgfmathsetmacro{\g}{1}
    \newcommand\mylabelstyle\footnotesize
    
    \fill[black!10] (-45:1.7)--(0,0)--(45:1.7)--cycle;
    
    \foreach \t in {-1,-0.75,...,1}{%
        \path[subgrid] (0,0) -- (\Xmax,{\Xmax*tanh(\g*\t)});
    }

    \foreach \xx in {0.3,0.6,...,\Xmax}{%
        \path[draw,thick,dotted]
            plot[domain=-{arcosh(\Xmax/\xx)/\g}:{arcosh(\Xmax/\xx)/\g}]
            ({\xx*cosh(\g*\x)},{\xx*sinh(\g*\x)});  
    }

    
    \draw[thick,-stealth] (-\Xmax,0) -- (\Xmax,0) node[right] {$x_1$}
    	node[pos=0.75,below,fill=black!10] {$A$}
	node[pos=0.25,below,fill=white] {$B$};
    \draw[thick,-stealth] (0,-\Tmax) -- (0,\Tmax) node[left] {$T$};
    \draw[dashed] (-\Xmax,-\Tmax) -- (\Xmax,\Tmax) 
        node[pos=0.67,above,sloped,yshift=-.3ex] {\mylabelstyle$r=0$}
        node[tlabels,black] {\mylabelstyle$\eta=\infty$};
    \draw[dashed] (-\Xmax,\Tmax) -- (\Xmax,-\Tmax) 	
    	node[pos=0.67,below,sloped,yshift=-.3ex] {\mylabelstyle$r=0$}
        node[tlabels,black,below] {\mylabelstyle$\eta=-\infty$};
    \node at (1.5, 0.55) {$\eta=$const};
    \node at (1.15, 0.1) {$r=$const};
\end{tikzpicture}
\caption{\label{fig:Rindler} The Rindler spacetime (shaded region) embedded in the Minkowski spacetime.}
\end{figure}
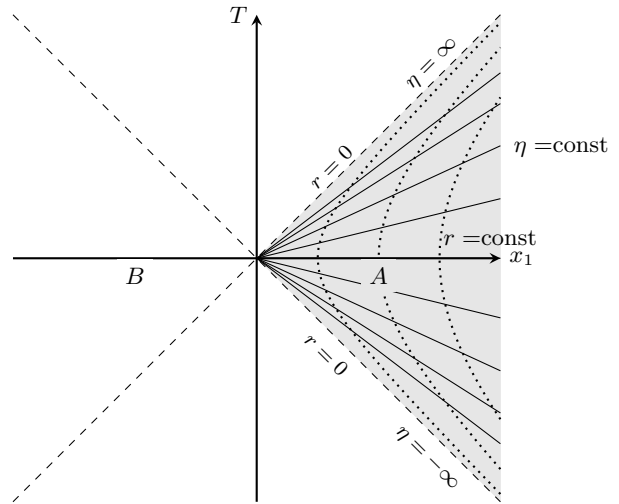

To see their relation we analytically continue the metric Eq.\,\eqref{EE_Hyperplane} by Wick rotation $\tau = \i \,\eta$ to the Rindler spacetime
\begin{align}
	\d s^{2} = - r^{2} \d\eta^{2} + \d r^{2} + \sum_{i=2}^{d-1} \d x_i^2\ ,
\end{align}
with the new variable $\eta$ ranging $-\infty < \eta < \infty$.
The Rindler spacetime covers a portion specified by $x_1 \ge |T| $ of the Minkowski spacetime $\BR^{1,d-1}$ (see Fig.\,\ref{fig:Rindler} for illustration),
\begin{align}
	\d s^{2} = -\d T^{2} + \d x_1^{2} + \sum_{i=2}^{d-1} \d x_i^2\ ,
\end{align}
as seen from the coordinate transformation
\begin{align}
	T = r \sinh \eta \ , \qquad x_1 = r \cosh \eta \ .
\end{align}
This transformation implies that an observer who is at rest in the Rindler spacetime is uniformly accelerating in the Minkowski spacetime and is confined to the wedge $x_1 \ge |T| $ separated by the ``Rindler horizon" at $r=0$ from the rest of the spacetime.
Hence at a given time slice, say, $T=0$, there is a region $x_1<0$ (denoted by $B$ in Fig.\,\ref{fig:Rindler}) that is inaccessible to the Rindler observer living on $x_1>0$ (denoted by $A$ in Fig.\,\ref{fig:Rindler}),
whose state is described by the reduced density matrix $\rho_A$ that is a mixed thermal ensemble at inverse temperature $\beta = 2\pi$ even if the total system $A\cup B$ is in a pure state.
This is precisely the same situation we considered in the previous section, and 
the entanglement entropy calculated there is equivalent to the thermal entropy that the Rindler observer measures in her/his frame.

Now we want to give a concrete expression of the modular Hamiltonian $H_A$ defined by Eq.\,\eqref{Gibbs_State} for the Rindler observer.
In order to fix its form, we consider the matrix element of the reduced density matrix $\rho_A$,
\begin{align}
	\langle \phi_A' | \rho_A | \phi_A \rangle = \frac{1}{Z}\, \langle \phi_A' | e^{-2\pi\, H_A} | \phi_A \rangle  \ ,
\end{align}
and regard the right-hand side as a transition amplitude from an initial state $|\phi_A\rangle$ to a final state $|\phi_A'\rangle$.
The (Euclidean) ``time" evolution is generated by the modular Hamiltonian $H_A$ over the period $2\pi$.
Looking back to the example in the previous subsection, we can understand that the system on the region $A$ is time evolved along the angular direction $\tau$ starting from $\tau=0$ to $2\pi$ (see Fig.\,\ref{fig:ScalarInfinite}).
We thus conclude the modular Hamiltonian in the present case is the generator of the translation along the modular time $\tau$, hence $H_A = \partial_\tau$.\footnote{In Lorentzian signature, the modular Hamiltonian generates the boost $K = \partial_\eta$ in the Rindler spacetime.}
In Euclidean QFT, the modular Hamiltonian is written explicitly by the Noether's theorem as
\begin{align}\label{Modular_Hamiltonian}
	H_A = \int_{x_0=0, x_1\ge 0} \d^{d-1}x\, x_1\, T_{00} \ ,
\end{align}
where $T_{\mu\nu}$ is the stress tensor, which will be defined in Eq.\,\eqref{STDef}.
This result is a particular case of the theorem by \textcite{Bisognano:1976za,Bisognano:1975ih} that is valid for any QFT even with interactions.

Recalling the construction of the thermofield double state Eq.\,\eqref{TFD}, we can rewrite the ground state wave function in the Minkowski spacetime in the following form,
\begin{align}
		\Psi [\phi_A, \phi_B] = \frac{1}{\sqrt{Z}} \sum_{n}\,e^{-\beta E_n/2} \varphi_n [\phi_A]\, \varphi_n [\phi_B] \ ,
	\end{align}
where $\phi_A$ and $\phi_B$ are states on the time slice $A$ and $B$ respectively, and $\varphi_n [\phi]$ is the $n^{th}$ eigenfunction of the modular Hamiltonian $H_A$ with eigenvalue $E_n$.
Namely, we are able to \emph{purify} a mixed state in the Rindler spacetime  to a pure state in the Minkowski spacetime by viewing the region inside the Rindler horizon as a fictitious system \cite{Israel:1976ur,Laflamme:1988wg}.

\subsection{Structure of UV divergences}\label{ss:UVstructure}
In the previous section, we rewrite the definition of entanglement entropy in terms of an Euclidean effective action $\log Z_n$ on an $n$-fold cover $\CM_n$ with a conical singularity along a codimension-two surface $\Sigma = \partial A$ surrounding the region $A$ of interest.
In QFT on a curved spacetime, the effective action is regarded as a function of the metric $g_{\mu\nu}$ on $\CM_n$; thus we are able to classify the UV divergent terms by diffeomorphism invariant local terms.
Here we restrict our attention to QFTs without dimensionful parameters (= CFTs) for simplicity, whose effective action may be given by\footnote{One can repeat the same argument presented here for any renormalizable QFTs with a slight modification. For example, one can add terms $\Lambda^{d-2}\,M^2, \Lambda^{d-4}\, M^2\, \CR, \cdots$ to the integrand if $M$ is a parameter of mass dimension one.
This does not change the UV structure, and we ignore them for the moment.
}
\begin{align}\label{PF_UV}
\begin{aligned}
\log Z_n [g_{\mu\nu}] =&\, \sum_{i=0}^{\left[ d/2\right]} C_{d-2i} \int_{\CM_n} \d^d x\, \sqrt{g}\, \Lambda^{d-2i}\, \CR^i \\
	& + (-1)^{\frac{d-1}{2}}F[g_{\mu\nu}] \ ,
\end{aligned}
\end{align}
where $\Lambda$ is the UV cutoff scale of mass dimension one and $\CR^i$ is a scalar polynomial of the Riemann tensors of order $i$ on $\CM_n$.
We call $F[g_{\mu\nu}]$ the renormalized free energy that is dimensionless and a scheme-independent nonlocal functional of the metric in odd dimensions.
In contrast, there is a logarithmic divergence $\sim C_0 \log \Lambda$ in addition to Eq.\,\eqref{PF_UV} in even dimensions, which makes the renormalized free energy ambiguous.
It is actually the source of the conformal anomaly and the coefficient $C_0$ depends on the central charges when the theory is conformally invariant.

We put aside the logarithmic divergence for a moment and determine the general structure of the UV divergences of entanglement entropy.
We plug the general form Eq.\,\eqref{PF_UV} into Eq.\,\eqref{EE_PF}, where there appears only the difference $\log Z_n - n\log Z$ of the effective actions for $n$ close to $1$.
The UV divergent term of order $\Lambda^{d-2i}$ is proportional to the integral\footnote{We often use the shorthand notation for an integral $\int_{\CM} \equiv \int_{\CM} \d^d x\,\sqrt{g}$ with the measure suppressed.} $\int_{\CM_n}\CR^i - n \int_{\CM}\CR^i$ that should localize on the entangling surface $\Sigma$ since the $n$-fold covering space $\CM_n$ differs from the original space $\CM$ only around $\Sigma$.
Especially, the leading divergence with $i=0$ cancels out because of $\text{Vol}(\CM_n) = n\, \text{Vol} (\CM_1)$.
We thus end up with the UV divergences in the entanglement entropy starting from $O(\Lambda^{d-2})$:
\begin{align}\label{UV_structure}
S_A = c_{d-2}\,\Lambda^{d-2} + c_{d-4}\, \Lambda^{d-4} + \cdots \ ,
\end{align}
where $c_{d-2i}$ are given by the integrals of local diffeomorphism invariants on $\Sigma$, schematically written as
\begin{align}\label{UV_coeff}
	c_{d-2i} = \sum_{l+m=i-1}\int_\Sigma \, \CR^l\, \CK^{2m} \ ,
\end{align}
where $\CK$ is the extrinsic curvature of $\Sigma$ of the order of $2m$, whose definition will be given in Eq.\,\eqref{ExtrinsicCurvature},
and $\CR^l\, \CK^{2m}$ are scalar polynomials of $\CR$ and $\CK$ of the order of $l$ and $2m$ respectively.
The reason only even powers of the extrinsic curvatures appear in Eq.\,\eqref{UV_coeff} is that the entanglement entropy for a region $A$ is equal to that of its complement $\bar A$ while their extrinsic curvatures have opposite signs, hence the odd powers of $\CK$ vanish.

We derived the UV structure Eq.\,\eqref{UV_structure} assuming the covariance and renormalizability of QFTs on a manifold without boundaries.
If we relax the assumptions, there are a variety of cases where entanglement entropy can take a different UV structure.
For instance, nonrelativistic systems with Fermi surfaces have the entanglement entropy that violates the area law logarithmically \cite{Wolf:2006zzb,Gioev:2006zz} [see also \textcite{Ogawa:2011bz,Huijse:2011ef} for the holographic descriptions].
For nonsmooth entangling surfaces such as those with corners and wedges, the entanglement entropies have additional UV divergences whose coefficients depend on the opening angles \cite{Fradkin:2006mb,Casini:2006hu,Hirata:2006jx,Casini:2008as,Bueno:2015xda,Bueno:2015rda,Elvang:2015jpa,Klebanov:2012yf,Myers:2012vs}.\footnote{See also \textcite{Ghasemi:2017pke} for a more recent work.}

The general structure Eq.\,\eqref{UV_structure} guarantees the UV finiteness of the mutual information $I(A,B)$ of two disjoint systems $A$ and $B$ defined by Eq.\,\eqref{MI} because every coefficient $c_{d-2i}$ of the UV divergent term $\Lambda^{d-2i}$ of the entanglement entropy is given by an integral on the entangling surface and they cancel in the mutual information due to the trivial identity (see Fig.\,\ref{fig:MI})
\begin{align}
	\int_{\Sigma_{A+B}} = \int_{\Sigma_A} + \int_{\Sigma_B} \ .
\end{align}

We fix the precise forms of the coefficients for free fields in the next Sec. \ref{ss:HeatKernel}.
Note that there is a logarithmic divergence in even dimensions which we separately deal with when we discuss CFT in Sec. \ref{ss:CFT}.

  \section{Heat kernel expansion}\label{ss:HeatKernel}
The effective action $\log Z_n$ is a one-loop determinant provided that a theory is noninteracting.
The computation of the effective action is carried out in a standard way using the heat kernel method \cite{Birrell:1982ix}.
Here we reexamine a free massive scalar field in $d$ dimensions on $\CM_n$ using this method.
The one-loop effective action is given by
\begin{align}\label{MassiveScalar}
\begin{aligned}
\log Z_n &= - \frac{1}{2} \log \det ( \CD + m^2) \ ,\\
	&= \frac{1}{2} \int_{\epsilon^2}^\infty \frac{\d s}{s}\,  \tr \,K_{\CM_n}(s) \,e^{-m^2 s}\ ,
\end{aligned}
\end{align}
where $\CD$ is a scalar Laplacian plus the scalar curvature $\CD = -\nabla^2 + \xi R$ 
and $m$ is the mass of the scalar field. The parameter $\xi$ is free, but becomes $(d-1)/(4d)$ when the theory is conformally invariant (see Sec. \ref{ss:CFT}).
In the second equality, we used the Schwinger representation and introduced the UV cutoff scale $\epsilon$. 
The {\it heat kernel operator}\footnote{The name follows from the fact that it satisfies the heat equation $(\partial_s + \CD) K_{\CM_n} (s) = 0$.} $K_{\CM_n}(s) = e^{-s \CD}$ has the expansion around $s=0$ of the following form:
\begin{align}\label{HKE}
\tr\, K_{\CM_n} (s) = \frac{1}{(4\pi s)^{d/2}} \sum_{i=0}^\infty a_i(\CM_n)\, s^i \ .
\end{align}
Note that the heat kernel coefficients $a_i$ do not depend on $s$.

As we see shortly, the heat kernel coefficients allow the expansion in the $n \to 1$ limit into the bulk and surface parts on a singular manifold $\CM_n$,
\begin{align}\label{HKCoefficient}
a_i = a_i^\text{bulk} + (1-n) \,a_i^\Sigma + O\left( (1-n)^2\right)\ .
\end{align}
The bulk part $a_i^\text{bulk}$ for the $n$-fold cover $\CM_n$ is $n$ times larger than the one for the original space $\CM_1$
\begin{align}\label{BulkMn}
a_i^\text{bulk}(\CM_n) = n\, a_i^\text{bulk}(\CM_1) \ .
\end{align}
The subleading terms arise from the contributions of the conical singularity at the hypersurface $\Sigma$.
Using the heat kernel expansion Eq.\,\eqref{HKE} in the effective action Eq.\,\eqref{MassiveScalar}, we obtain the entanglement entropy as a power series of the inverse of the mass with the heat kernel coefficients,
\begin{align}\label{EE_Scalar_HK}
S_A = \frac{1}{2(4\pi)^{d/2}}\sum_{i=0}^\infty \frac{a_i^\Sigma}{m^{2i - d}} \,\Gamma\left( i - \frac{d}{2}, m^2 \epsilon^2\right) \ ,
\end{align}
where $\Gamma(a,z)$ is the incomplete gamma function defined by $\Gamma (t,x) = \int_x^\infty \d s\, s^{t-1} e^{-s}$.
Hence given the surface parts of the heat kernel coefficients, the entanglement entropy is calculable at any order of the heat kernel expansion.\footnote{The heat kernel method may be adapted to the perturbative studies of R{\'e}nyi entropy near $n=1$, which requires the subleading coefficients in the expansion Eq.\,\eqref{HKCoefficient}.}

We fix the heat kernel coefficients following the strategy used by \textcite{Fursaev:1994in,Fursaev:1995ef,Fursaev:2013fta} [see also \textcite{Solodukhin:2011gn} for a review].
First, we start with the known results for the bulk heat kernel coefficients on a nonsingular manifold $\CM$ \cite{Birrell:1982ix,Vassilevich:2003xt}:
\begin{align}\label{HKbulk}
\begin{aligned}
a_0^\text{bulk} &= \int_{\CM} 1\ , \\
	a_1^\text{bulk} &= \left( \frac{1}{6} - \xi \right) \int_{\CM} \CR \ , \\
	a_2^\text{bulk} &= \int_{\CM} \left[ \frac{1}{180} \CR_{\mu\nu\rho\sigma}^2 - \frac{1}{180} \CR_{\mu\nu}^2 \right.\\
		&\qquad \qquad \left. + \frac{1}{2}
	 \left( \frac{1}{6} - \xi \right)^2 \CR^2 + \frac{1}{6} \left( \frac{1}{5} - \xi \right)\nabla^2 \CR \right]\ .
\end{aligned}
\end{align}
To obtain the surface parts, we regularize the conical singularity at $\Sigma$ and replace $\CM_n$ with a regularized manifold $\widetilde \CM_n$.
Eqs.\,\eqref{HKbulk} are valid for the smooth geometry $\widetilde \CM_n$ from which we read off the surface parts $a_i^\Sigma$.

\subsection{Regularized cone with $\U(1)$ symmetry}\label{ss:RegConeU(1)}
Before presenting general cases, we consider a two-dimensional cone $\CC_n$ with a conical singularity at the origin (we denote it by $\Sigma$),
\begin{align}
\begin{aligned}
\d s^2_{\CC_n} &= e^{\sigma (r)} (\d r^2 + r^2 \d\tau^2) \ , \\
\sigma(r) &= \sigma_1 r^2 + \sigma_2 r^4 + \cdots \ ,
\end{aligned}
\end{align}
where $0\le r \ ,  0\le \tau \le 2\pi n$, and $\sigma_1$ and $\sigma_2$ are constant.
It has a deficit angle $2\pi(1-n)$ at the tip.

\begin{figure}[htbp]
\centering
\begin{tikzpicture}[scale=1.2]
	\draw [thick] ([shift=(240:2)]1,2) --++ (1,2) -++ (1,-2);
	\draw [thick] ([shift=(240:2)]1,2) arc (240:300:2);
	\draw [thick, dashed] ([shift=(300:2)]1,2) arc (60:120:2);
	\draw [thick, ->] ([shift=(240:2)]2.2,4.2) node [above] {$r=0$} --++ (1.1,-2.2) node [below] {$r$};
	\draw [thick, ->] ([shift=(250:2)]1,1.7) arc (250:290:2);
	\node at (1,-0.6) {$\tau$};
	\node at (1,3.5) {Cone};
	
	\draw[ultra thick, ->] (3,1.5) --++ (1.5,0);
	
	\draw [thick, rounded corners=10] ([shift=(240:2)]6,2) --++ (1,2) -++ (1,-2);
		
	\draw [thick, dashed] ([shift=(240:0.5)]6,2.2) arc (240:300:0.5);
	\draw [thick, dashed] ([shift=(300:0.5)]6,2.2) arc (60:120:0.5);
	\draw [thick] ([shift=(240:2)]6,2) arc (240:300:2);
	\draw [thick, dashed] ([shift=(300:2)]6,2) arc (60:120:2);
	\node at (7,1.8) {$r=\delta$};
	\node at (6,3.5) {Regularized cone};
\end{tikzpicture}
\caption{A cone that has a singularity at $r=0$ is smoothened by replacing the tip with a disk.}
\label{fig:RegCone}
\end{figure}
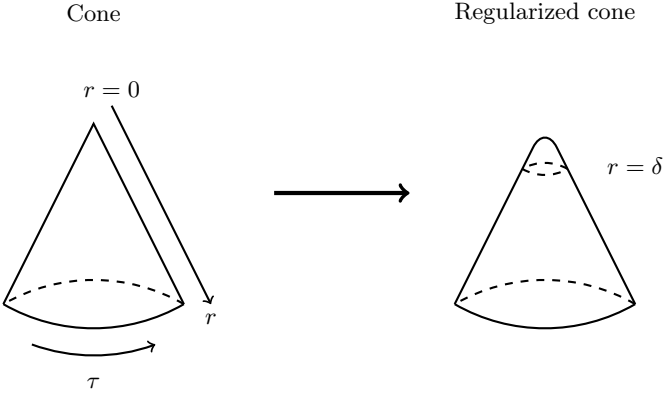

In order to properly take into account the singularity, we regularize the cone by deforming the metric so that the origin at $r=0$ becomes smooth while it has the same geometry as $\CC_n$ away from $\Sigma$.
We call such a geometry \emph{a regularized cone} $\tilde \CC_n$, letting the metric be 
\begin{align}\label{RegCone}
\d s^2_{\tilde \CC_n} = e^{\sigma (r)} \left[ f_\delta (r)\, \d r^2 + r^2 \d \tau^2 \right] \ ,
\end{align}
where $f_\delta (r)$ is a smooth function satisfying
\begin{align}\label{SmoothingFunction}
f_\delta (r \to 0) = n^2 \ , \qquad f_\delta (r>\delta) = 1 \ ,
\end{align}
with a small parameter $\delta \ll 1$.\footnote{The small parameter $\delta$ has nothing to do with the UV cutoff $\epsilon$ in general and can be tuned arbitrarily.}
The regularized cone $\tilde\CC_n$ agrees with $\CC_n$ for $r> \delta$ and becomes flat space at $r=0$ as illustrated in Fig.\,\ref{fig:RegCone}.

Next we evaluate the Riemann tensors $\CR_{\mu\nu\rho\sigma}$ on $\tilde \CC_n$.
Since the Riemann tensor in two dimensions is fixed by the Ricci scalar,
\begin{align}
	\CR_{\mu\nu\rho\sigma} = \frac{\CR}{2}(g_{\mu\rho} g_{\nu\sigma} - g_{\mu\sigma} g_{\nu\rho}) \ ,
\end{align}
it is enough to calculate $\CR$ on $\tilde \CC_n$.
The Ricci scalar for the metric \eqref{RegCone} is given by
\begin{align}
	\CR = e^{-\sigma} \left( \frac{f_\delta'}{r f_\delta^2} - \frac{\partial_r^2 \sigma}{f_\delta} \right) \ .
\end{align}
The first term in the bracket of the right-hand side, representing a contribution to the Ricci curvature from the singularity, naively vanishes when $\delta = 0$.
However, it contributes to the integral as a surface term even in $\delta \to 0$ limit as seen from the integral of the Ricci scalar,
\begin{align}
\begin{aligned}
\int_{\tilde\CC_n} \CR &= 2\pi n\int_0^\infty \d r \left[ -2 \partial_r f^{-1/2}_\delta(r) - r f_\delta^{-1/2}(r) \, \partial_r^2 \sigma \right] \ ,\\
	&= 4\pi (1-n) - 2\pi n \int_0^\infty \d r \,r f_\delta^{-1/2}(r) \, \partial_r^2 \sigma \ .
\end{aligned}
\end{align}
The first term proportional to $n-1$ is the contribution of the singularity to the curvature
while the second term is the bulk contribution accounting for the curvature of $\CC_n$ away from the singularity.
Taking the $\delta \to 0$ limit, the second term approaches the integral of the Ricci scalar on the cone $\CC_n$ with the origin $\Sigma$ removed:
\begin{align}
	\lim_{\delta \to 0} \int_{\tilde\CC_n}  \CR = \int_{\CC_n/\Sigma} \CR + 4\pi (1-n)\ .
\end{align}
The final result does not depend on the choice of the smooth function $f_\delta(r)$, thus is independent of the regularization scheme.

This argument can be equally generalized to higher-dimensional theories if the $n$-fold cover $\CM_n$ has a $\U(1)$ symmetry around $\Sigma$, namely if the metric is of the form,
\begin{align}\label{U(1)metric}
	\d s^2_{\CM_n} = e^{\sigma (r)} \left[ \d r^2 + r^2 \d\tau^2 + \gamma_{ij}(r, y)\, \d y^i \d y^j \right] \ ,
\end{align}
where $y^i~(i=1,\cdots, d-2)$ are the coordinates of the entangling surface $\Sigma$ located at $r=0$.
This is actually the case when $\Sigma$ does not have an extrinsic curvature as we see in the next section.
The regularized space $\widetilde \CM_n$ is obtained by replacing the cone with $\tilde \CC_n$, being the product manifold $\tilde \CC_n \times \Sigma$ near $\Sigma$.
We do not reproduce the derivation, but
the results are conveniently summarized in the following relations \cite{Fursaev:1995ef}:
\begin{align}\label{Formula}
\begin{aligned}
	\CR\big|_{\widetilde\CM_n} &= \CR\big|_{\CM_1} + 4\pi (1-n)\, \delta_\Sigma \ , \\
	\CR_{\mu\nu}\big|_{\widetilde\CM_n} &= \CR_{\mu\nu}\big|_{\CM_1} + 2\pi (1-n) N_{\mu\nu}\, \delta_\Sigma \ ,\\
	\CR_{\mu\nu\rho\sigma}\big|_{\widetilde\CM_n} &= \CR_{\mu\nu\rho\sigma}\big|_{\CM_1} \\
		&\qquad + 2\pi (1-n) (N_{\mu\rho}N_{\nu\sigma}-
	N_{\mu\sigma}N_{\nu\rho})\, \delta_\Sigma \ ,
\end{aligned}
\end{align}
where $\delta_\Sigma$ is the delta function for the entangling surface $\Sigma$ and $N_{\mu\nu}$ is defined by  $N_{\mu\nu}= \sum_{a=1}^2 n^a_\mu n^a_\nu$ with the normal vectors $n^a_\mu~(a=1,2)$ to the surface.

As an example, let us apply Eq.\,\eqref{Formula} to the Euler density  defined in $2m$ dimensions by
\begin{align}\label{EulerDensity}
\begin{aligned}
	E_{2m} = \frac{1}{2^{2(m+1)}\pi^m\, m!}\, &\epsilon_{\mu_1\mu_2\cdots \mu_{2m-1}\mu_{2m}} \epsilon^{\nu_1\nu_2\cdots \nu_{2m-1}\nu_{2m}}\\
		&\cdot \CR^{\mu_1\mu_2}_{~~~~~~\nu_1\nu_2} \cdots \CR^{\mu_{2m-1}\mu_{2m}}_{~~~~~~~~~~\nu_{2m-1}\nu_{2m}}  \ .
\end{aligned}
\end{align}
The integral of the Euler density over a manifold $\CM$ is called the Euler invariant $\chi[\CM] \equiv \int_{\CM} E_{2m}$ that is a topological invariant.
It is normalized so that $\chi[\BS^{2m}] = 2$ for an even-dimensional sphere.
Then the application of Eq.\,\eqref{Formula} to Eq.\,\eqref{EulerDensity} yields an interesting relation between the Euler densities of the $n$-fold cover $\widetilde \CM_n$ and the entangling surface $\Sigma$ \cite{Fursaev:1995ef}:
\begin{align}\label{EulerFormula}
	E_{2m}\big|_{\widetilde\CM_n} &= E_{2m}\big|_{\CM_1} + (1-n)\, E_{2m-2}\big|_{\Sigma} + O\left((1-n)^2\right)\ .
\end{align}
Although these relations are likely to hold for the case even without $\U(1)$ symmetry, to our best knowledge, there is no general proof so far.

\subsection{Regularized cone without $\U(1)$ symmetry}\label{ss:Reg_Cone_no_U(1)}
Without a $\U(1)$ symmetry around the entangling surface, the metric can take a more general form than Eq.\,\eqref{U(1)metric}.
If we parametrize the two-dimensional transverse directions to $\Sigma$ by $x_a$ $(a=1,2)$, the manifold $\CM$ looks like a product form $\BR^2 \times \Sigma$ near the origin $x_1=x_2 = 0$.
Even away from the origin, we can still control the metric in the so-called Riemann normal coordinates \cite{Lewkowycz:2013nqa,Fursaev:2013fta,Rosenhaus:2014woa},
\begin{align}
\begin{aligned}
	\d s^2_{\CM} &=  \delta_{ab}\, \d x^a \d x^b + \left(\gamma_{ij}(y) + 2\CK^a_{ij} \, x^a \right) \d y^i \d y^j + O(x^2) \ .
\end{aligned}
\end{align}
We introduced the extrinsic curvature $\CK^a_{\mu\nu}$ by
\begin{align}\label{ExtrinsicCurvature}
\CK^a_{\mu\nu} = h_\mu^{~\rho} h_\nu^{~\sigma} \nabla_\rho n^a_\sigma \big|_{\Sigma}\ , 
\end{align}
where $h$ is the induced metric on $\Sigma$, $h_{\mu\nu} = g_{\mu\nu} - \sum_{a=1}^2 n^a_\mu n^a_\nu$ with the unit normal vector $n^a_\mu = \delta^a_\mu$ to $\Sigma$.
Moving to the polar coordinates $x_1 = r\sin\tau, x_2 = r\cos\tau$, the metric becomes 
\begin{align}
\begin{aligned}
	\d s^2_{\CM} &= \d r^2 + r^2 \d\tau^2 \\
		&\qquad + \left( \gamma_{ij}(y) + 2r \cos\tau\, \CK^1_{ij} + 2r \sin\tau\,\CK^2_{ij} \right) \d y^i \d y^j\\
		&\qquad\quad + O(r^2)\ .
\end{aligned}
\end{align}
The $\U(1)$ symmetry is broken due to the explicit dependence on the angular coordinate $\tau$.
Compared with Eq.\,\eqref{U(1)metric} for $n=1$, the extrinsic curvature $\CK^a_{ij}$ has to vanish when the $\U(1)$ symmetry is present.

We want to calculate the integrals of geometric invariants on the $n$-fold cover as in the $\U(1)$ symmetric case.
To this end, we use the regularized $n$-fold cover $\widetilde\CM_n$ proposed by \textcite{Fursaev:2013fta},
\begin{align}
\begin{aligned}
	&\d s^2_{\widetilde\CM_n^\text{(FPS)}} \\
	&\quad = f_\delta (r) \d r^2 + r^2 \d \tau^2 \\
		&\qquad + \left[ \gamma_{ij}(y) + 2r^n c^{1-n}\left( \cos\tau\, \CK^1_{ij} + \sin\tau\,\CK^2_{ij}\right) \right] \d y^i \d y^j \\
		&\qquad \quad  + O(r^2)\ ,
\end{aligned}
\end{align}
where $\tau\sim\tau + 2\pi n$, $c$ is a constant of dimension of length, and $f_\delta(r)$ is the smoothing function Eq.\,\eqref{SmoothingFunction}.
Using the regularized metric, the following formulas for the integrals of the Riemann curvatures are argued to hold based on several explicit calculations \cite{Fursaev:2013fta}:
\begin{align}
\int_{\widetilde\CM_n} \CR &= n \int_{\CM} \CR + 4\pi (1-n) \int_\Sigma 1 \ ,\label{SingularFormulae1}\\
\int_{\widetilde\CM_n} \CR^2 &= n \int_{\CM} \CR^2 + 8\pi (1-n) \int_\Sigma 
\CR \ , \label{SingularFormulae2}\\
\int_{\widetilde\CM_n} \CR_{\mu\nu}^2 &= n \int_{\CM} \CR_{\mu\nu}^2 \nonumber\\
	&\qquad + 4\pi (1-n) \int_\Sigma \left[\CR_{aa} - \frac{1}{2}(\CK^{a\,\mu}_{\,\mu})^2\right] \ , \label{SingularFormulae3}\\
\int_{\widetilde\CM_n} \CR_{\mu\nu\rho\sigma}^2 &= n \int_{\CM} \CR_{\mu\nu\rho\sigma}^2 \nonumber\\
	&\qquad + 8\pi (1-n) \int_\Sigma \left[ \CR_{abab} - \CK^a_{\mu\nu} \CK^{a\, \mu\nu} \right] \ , \label{SingularFormulae4}
\end{align}
where Eq.\,\eqref{SingularFormulae1} is exact while only the terms up to $O(1-n)$ are shown in the rest.
The subleading integral are performed on the codimension-two surface $\Sigma$ and $\CR_\Sigma$, $\CR_{aa}$ and $\CR_{abab}$ are the Ricci scalar for the induced metric $h$, the induced Ricci tensor $\CR_{aa} = \sum_{a=1,2} n^a_\mu n^a_\nu \CR^{\mu\nu}$ and the induced Riemann tensor $\CR_{abab} = \sum_{a,b=1,2} n^a_\mu n^b_\mu n^a_\rho n^b_\sigma \CR^{\mu\nu\rho\sigma}$, respectively.
These are enough to fix the heat kernel coefficients $a_i$ in Eq.\,\eqref{HKbulk} up to second order.

The results given in Eq.\,\eqref{SingularFormulae1}-\eqref{SingularFormulae4} can be used to derive a similar relation to Eq.\,\eqref{EulerFormula}, but in the integrated forms in two and four dimensions:
\begin{align}\label{EulerFormula2}
	\chi [\widetilde \CM_n] = \chi [\CM] + (1-n) \chi[\Sigma] + O\left((1-n)^2\right)\ ,
\end{align}
where we used the fact that the regularized $n$-fold cover $\widetilde \CM_n$ is topologically the same as the original manifold $\CM$.

\subsection{Heat kernel coefficients}
We move onto the determination of the heat kernel coefficients for the surface parts.
Substituting Eqs.\,\eqref{SingularFormulae1}-\eqref{SingularFormulae4} into the coefficients \eqref{HKbulk} on the smooth geometry $\widetilde\CM_n$,
one finds the bulk parts satisfy Eq.\,\eqref{BulkMn} and the surface parts become \cite{Fursaev:2013fta}
\begin{align}\label{HKsurf}
\begin{aligned}
a_0^\Sigma &= 0\ , \\
	a_1^\Sigma &=  \frac{2\pi (1-6\xi) }{3} \int_\Sigma 1 \ , \\
	a_2^\Sigma &= \frac{\pi}{9} \int_\Sigma \Bigg[ \frac{1}{5} \left( 2 \CR_{abab} - \CR_{aa}  -2 \CK^a_{\mu\nu} \CK^{a\, \mu\nu} + \frac{1}{2}(\CK^{a\,\mu}_{\,\mu})^2\right) \\
		&\qquad \qquad\qquad + (1-6\xi)^2 \CR \Bigg]\ ,
\end{aligned}
\end{align}
where we used the Gauss-Codazzi equation to simplify the expression,
\begin{align}\label{GaussCodazzi}
\CR =  \CR_\Sigma + 2\CR_{aa} - \CR_{abab} - (\CK^{a\,\mu}_{\,\mu})^2 + \CK^a_{\mu\nu} \CK^{a\, \mu\nu} \ .
\end{align}

It follows that the entanglement entropy Eq.\,\eqref{EE_Scalar_HK} for a free massive scalar field has the UV divergences:
\begin{align}\label{EAfinite}
S_A &= \frac{(1-6\xi)}{6(d-2) (4\pi)^{d/2 -1}} \,\frac{A_\Sigma}{\epsilon^{d-2}} + \frac{c_{d-4}}{\epsilon^{d-4}}  + \cdots +\gamma(m) \ ,
\end{align}
where $\gamma(m)$ includes all the mass dependences.
The area law divergence can be found in the leading term that comes from the coefficient $a_1^\Sigma$ proportional to the area $A_\Sigma$ of the surface $\Sigma$. 
In addition, there is a universal term proportional to the area in $\gamma(m)$ \cite{Hertzberg:2010uv}:
\begin{align}\label{gamma_odd}
	\gamma(m) = (1-6\xi)\,\gamma_d \, A_\Sigma\, m^{d-2} + \cdots \ ,\quad \gamma_d = 	\frac{\Gamma(1-d/2)}{12(4\pi)^{d/2-1}} \ .
\end{align}
The coefficient $\gamma_d$ is finite for odd $d$, but is divergent due to the pole of the gamma function for even $d$.
This divergence signals the conformal anomaly in even dimensions and modifies the mass expansion of $\gamma(m)$  for even $d_e$,
\begin{align}
\begin{aligned}
	\gamma(m) =\,& (1-6\xi)\,\gamma_{d_e}^{(\text{even})}\, A_\Sigma\, m^{d_e-2}\log (m\,\epsilon) + \cdots \ , \\
	&\gamma_{d_e}^{(\text{even})} = \frac{(-1)^{d_e/2 -1}}{6(4\pi)^{d_e/2 -1} \Gamma (d_e/2)}  \ .
\end{aligned}
\end{align}
This may be inferred from Eq.\,\eqref{gamma_odd} by expanding it around $d = d_e - \epsilon$  and using the replacement $1/\epsilon \to \log (m\, \epsilon)$ for the pole in the dimensional regularization.

  \section{Conformal field theory}\label{ss:CFT}

In this section, we consider QFTs invariant under the conformal transformation
\begin{align}\label{ConfTra}
\bar g_{\mu\nu}(x') = \Omega^2 (x)\, g_{\mu\nu}(x) \ ,
\end{align}
where $\bar g_{\mu\nu}$ is the transformed metric.
In Euclidean space $\BR^{d}$, the group of the conformal transformation is $\SO (1,d+1)$ while in Minkowski space $\BR^{1,d-1}$ the group is $\SO (2,d)$.
The conformal group is an extension of the Poincar{\'e} group, including the dilatation $x'^\mu = \lambda\, x^\mu$ for a constant $\lambda$ and the special conformal transformation $x'^\mu = (x^\mu + b^\mu x^2)/(1+ 2b_\mu x^\mu + b^2 x^2)$ for a vector $b^\mu$, hence preserving the angle between two curves infinitesimally \cite{DiFrancesco:1997nk}.

Let $I_E[g_{\mu\nu}, \phi ]$ be the Euclidean action for a theory with a field $\phi(x)$ and a metric $g_{\mu\nu}$.
This theory is called classical conformal field theory if the action does not change under the following transformation:
\begin{align}
	I_E[\bar g_{\mu\nu}, \bar \phi ] =  I_E[ g_{\mu\nu}, \phi ] \ ,\qquad \bar \phi(x') = \Omega^{-\Delta} (x)\, \phi(x) \ ,
\end{align}
for some constant $\Delta$ being the conformal dimension of the field $\phi(x)$.
For example, a scalar field theory coupled to a curved space 
\begin{align}\label{ConfScalar}
	I_E[g_{\mu\nu}, \phi ] = \frac{1}{2} \int \d^d x \sqrt{g}\, \left[ \partial_\mu\phi\, \partial^\mu \phi + \frac{d-2}{4(d-1)} \CR\, \phi^2 \right] 
\end{align}
is conformally invariant with $\Delta = d/2 -1$.
It is straightforward to check that Eq.\,\eqref{ConfScalar} is conformal invariant  using the transformation law of the Ricci scalar \cite{Birrell:1982ix}:
\begin{align}\label{RicciConfTr}
\begin{aligned}
	\CR [\bar g_{\mu\nu}] &= \Omega^{-2} \CR[g_{\mu\nu}] - 2(d-1)\Omega^{-3}\,\nabla^2\Omega \\
		&\qquad - (d-1)(d-4)\Omega^{-4} \partial_\mu\Omega\, \partial^\mu\Omega \ ,
\end{aligned}
\end{align}
where the covariant derivatives on the right-hand side are taken with respect to the metric $g_{\mu\nu}$ before the conformal transformation.
An equivalent, but a slightly simpler way to derive Eq.\,\eqref{ConfScalar} is to construct a conformal invariant $\phi^{d/\Delta} \sqrt{g} \,\CR[\phi^{2/\Delta}g_{\mu\nu}]$ out of the metric and the scalar field and use Eq.\,\eqref{RicciConfTr} with $\Omega = \phi^{1/\Delta}$.
To make it quadratic in $\phi$, one needs to set $\Delta = d/2 - 1$ for the conformal dimension \cite{Brown:1976wc} and finds
\begin{align}
\begin{aligned}
	\phi^{d/\Delta} \,\CR[\phi^{2/\Delta}g_{\mu\nu}] &= \phi^2\, \CR[g_{\mu\nu}] \\
		&\qquad + \frac{4(d-1)}{d-2}\left[ \partial_\mu \phi \partial^\mu\phi - \nabla^\mu (\phi \partial_\mu \phi)\right] \ .
\end{aligned}
\end{align}
This gives the conformal invariant action \eqref{ConfScalar} up to the total derivative and normalization.

Next, consider an action $I_E[g_{\mu\nu}]$ of the metric, which may be obtained by integrating out all fields on a curved space. 
The variation of the action under the infinitesimal conformal transformation $\delta g_{\mu\nu} = 2\,\delta \Omega (x) g_{\mu\nu}$ becomes
\begin{align}
\delta I_E[g_{\mu\nu}] = \int \d^d x\,\delta g_{\mu\nu} \frac{\delta I_E[g_{\mu\nu}]}{\delta g_{\mu\nu}} =\int \d^d x \sqrt{g}\, T_{\mu}^{~\mu}\, \delta \Omega(x) \ ,
\end{align}
where the stress-energy tensor is defined by
\begin{align}\label{STDef}
T^{\mu\nu} = \frac{2}{\sqrt{g}} \frac{\delta I_E}{\delta g_{\mu\nu}} \ .
\end{align}
If the theory is conformally invariant, the action has to be invariant under any deformation $\delta \Omega(x)$; hence the trace of the stress-energy tensor has to vanish, $T_\mu^{~\mu} = 0$.

The traceless property of the stress tensor at the classical level is violated at the quantum mechanical level due to the conformal anomaly in even dimensions \cite{Bonora:1985cq,Duff:1993wm,Deser:1993yx}, which can be put in the form
\begin{align}\label{WeylAnomaly}
\langle T_\mu^{~\mu} \rangle = \frac{(-1)^{d/2}}{2} A\, E_{d} - \sum_i B_i \,I_i + \nabla_\mu J^\mu\ ,
\end{align}
where $E_{d}$ is the Euler density, normalized so that $\int_{\BS^{d}} E_{d} = 2$, and $I_i$ are a set of the independent Weyl invariants built from the Weyl tensor in $d$ dimensions.
The coefficients $A$ and $B_i$ are referred to as type $A$ and $B$ central charges, respectively.
The total derivative term $\nabla_\mu J^\mu$ is scheme dependent as it can be affected by adding local counterterms to the effective action.\footnote{For example, there is a total derivative term $\nabla_\mu J^\mu \propto \Box\, \CR$ in four dimensions, which can be removed by a local counterterm $\CR^2$ \cite{Birrell:1982ix}.}
Thus we ignore the total derivative term contribution to the conformal anomaly in the following discussion.

\subsection{Correlation functions}
The conformal symmetry restricts the form of correlation functions more severely than the Poincar{\'e} symmetry.
We start with a primary scalar operator $\CO(x)$ of conformal dimension $\Delta$  transforming under the conformal transformation Eq.\,\eqref{ConfTra} as
\begin{align}\label{ScaleTr}
	\bar \CO (x') = \Omega^{-\Delta}(x)\, \CO(x) \ .
\end{align}
The translation invariance imposes the constancy of the one-point function $\langle \CO(x)\rangle = c$, but the scaling law \eqref{ScaleTr} implies $c = \Omega^{-\Delta}(x)\, c$ for any $\Omega(x)$, thus the one-point function vanishes:
\begin{align}
	\langle \CO(x)\rangle = 0 \ .
\end{align}

Similarly the two-point function $\langle \CO_1(x)\,\CO_2(y)\rangle$ of primary operators $\CO_1$ and $\CO_2$ of conformal dimensions $\Delta_1$ and $\Delta_2$ depends only on the distance $|x-y|$ between the two points by the translation invariance. 
To incorporate the dilatation and the special conformal invariances, the two conformal dimensions should be the same. 
Then the two-point function of appropriately normalized operators takes the form
\begin{align}\label{TwoPnt}
	\langle \CO_1(x)\,\CO_2(y)\rangle = \frac{\delta_{\Delta_1\, \Delta_2}}{|x-y|^{2\Delta_1}}  \ .
\end{align}

The three-point function of primary operators $\CO_1, \CO_2$ and $\CO_3$ of conformal dimensions $\Delta_1, \Delta_2$ and $\Delta_3$ are similarly fixed by the conformal symmetry,
\begin{align}
\begin{aligned}
	\langle &\CO_1(x)\,\CO_2(y) \, \CO_3(z)\rangle \\
	&= \frac{C_{123}}{|x-y|^{\Delta_1 + \Delta_2 - \Delta_3}|y-z|^{\Delta_2 + \Delta_3 - \Delta_1}|z-x|^{\Delta_3 + \Delta_1 - \Delta_2}} \ ,
\end{aligned}
\end{align}
where $C_{123}$ is an undetermined constant.
Correlation functions of more than four operators cannot be completely fixed by only the conformal invariance in this way.

For conserved currents $j_\mu(x)$ of dimension $\Delta=d-1$ and the stress tensor $T_{\mu\nu}(x)$ of dimension $d$, their one-point functions vanish and the two-point functions are fixed to be \cite{Osborn:1993cr,Erdmenger:1996yc}
\begin{align}\label{jjCor}
	\langle j_\mu(x)\, j_\nu(y) \rangle = C_J\, \frac{I_{\mu\nu}(x-y)}{(x-y)^{2(d-1)}} \ ,
\end{align}
and
\begin{align}\label{TTCor}
\langle T_{\mu\nu}(x)\,T_{\rho\sigma}(y)\rangle
	= C_T\, \frac{I_{\mu\nu,\rho\sigma}(x-y)}{(x-y)^{2d}} \ ,
\end{align}
with the conformally invariant functions,
\begin{align}
\begin{aligned}
I_{\mu\nu}(x) &= \delta_{\mu\nu} - \frac{2x_\mu x_\nu}{x^2} \ , \\
I_{\mu\nu,\rho\sigma}(x) &= \frac{1}{2}\left[ I_{\mu\rho}(x)I_{\nu\sigma}(x) +I_{\mu\sigma}(x)I_{\nu\rho}(x) \right] -\frac{1}{d}\delta_{\mu\nu}\delta_{\rho\sigma} \ .
\end{aligned}
\end{align}
The two-point functions of a scalar primary field $\CO(x)$ and the conserved current or the stress tensor vanish,
\begin{align}\label{TO_CFT}
	\langle j(x)\, \CO(y)\rangle = \langle T_{\mu\nu} (x)\, \CO(y) \rangle = 0 \ ,
\end{align}
due to the conformal invariance and the conservation laws $\partial_\mu j^\mu = 0$ and $\partial_\mu T^{\mu\nu} = 0$.

\subsection{Conformal anomaly in entanglement entropy}
We saw in the previous section that entanglement entropy has the UV divergences whose leading part is always proportional to the area of the entangling surface $\Sigma$ in Eq.\,\eqref{EAfinite}.
There are also universal logarithmic divergences depending on whether $d$ is even or not.
We show this logarithmic divergence is induced by the conformal anomaly Eq.\,\eqref{WeylAnomaly} of the stress-energy tensor in even $d$ dimensions.

Let us consider the scale transformation of the entanglement entropy of an entangling surface $\Sigma$ of size $l$.
We need to know how the partition function $Z_n$ on the $n$-fold cover $\CM_n$ transforms under the Weyl scaling $l \to e^{\sigma} l$.
Equivalently we act the scaling of the metric $g_{\mu\nu} \to e^{2\sigma}g_{\mu\nu}$ on the effective action $\log Z_n$ and read off the variation locally given by the expectation value of the stress tensor on $\CM_n$:
\begin{align}
l \frac{\d}{\d l}\, \log Z_n = -\int_{\CM_n} \d^d x \sqrt{g}\, \langle T_\mu^{~\mu} \rangle \ .
\end{align} 
It follows with the QFT definition Eq.\,\eqref{EE_PF} that the entanglement entropy varies under the scale transformation as
\begin{align}\label{EE_Weyl}
l \frac{\d}{\d l}\, S_A = -\int_{\CM_1} \d^d x \sqrt{g}\, \langle T_\mu^{~\mu} \rangle + \lim_{n\to 1} \partial_n \int_{\CM_n} \d^d x \sqrt{g}\, \langle T_\mu^{~\mu} \rangle \ .
\end{align}
In odd dimensions, $\langle T_\mu^{~\mu} \rangle = 0$ for CFTs and there are no logarithmic divergences in the entropy.
On the other hand, the right hand side can be nonvanishing due to the conformal anomaly, Eq.\,\eqref{WeylAnomaly}, in even dimensions.
If a theory is defined on a flat space ($\CM_1 = \BR^d$), the first term always vanishes while the second term has a nontrivial contribution from the conical singularity at the entangling surface $\Sigma$.
In what follows, we estimate the second term explicitly using Eqs.\,\eqref{SingularFormulae1}-\eqref{SingularFormulae4} in two and four dimensions.

In two dimensions, the Euler density is $E_2 = \CR/(4\pi)$ while there are no Weyl invariants.
We rescale the type $A$ central charge, $A = c/3$, so as to follow the canonical convention in CFT$_2$.
Then substituting the trace of the stress tensor
\begin{align}
\langle T_\mu^{~\mu} \rangle = - \frac{c}{24\pi}\CR \ ,
\end{align}
into Eq.\,\eqref{EE_Weyl} and applying Eq.\,\eqref{SingularFormulae1}, we find the entanglement entropy for an interval of width $l$,
\begin{align}
l \frac{\d}{\d l}\, S_A = \frac{c}{6}\int_\Sigma 1 = \frac{c}{3} \ .
\end{align}
This result yields a well-known behavior of entanglement entropy of CFT$_2$ \cite{Holzhey:1994we,jin2004quantum,Calabrese:2004eu}
\begin{align}\label{EE_CFT2}
S_A = \frac{c}{3}\log \frac{l}{\epsilon} + \cdots \ ,
\end{align}
where $\epsilon$ is a UV cutoff and $\cdots$ are finite parts.

In four dimensions, there is one Euler density and one Weyl invariant,
\begin{align}
\begin{aligned}
E_4 &= \frac{1}{32\pi^2} \left( \CR_{\mu\nu\rho\sigma}^2 - 4 \CR_{\mu\nu}^2 + \CR^2 \right) \ , \\
I_4 &= \frac{1}{16\pi^2} \left( \CR_{\mu\nu\rho\sigma}^2 - 2 \CR_{\mu\nu}^2 + \frac{1}{3}\CR^2 \right) \ ,
\end{aligned}
\end{align}
whose coefficients we conventionally take to be $A = a$ and $B=c$, respectively.
Again, Eqs.\,\eqref{SingularFormulae1}-\eqref{SingularFormulae4} and the identities \eqref{WeylAnomaly} and \eqref{EE_Weyl} lead to the logarithmic divergence of entanglement entropy in CFT$_4$:
\begin{align}\label{EE_4d}
S_A = \frac{c_2}{\epsilon^2} + c_0\log \frac{l}{\epsilon}  + \cdots \ ,
\end{align}
where $c_2$ is a constant proportional to the area $A_\Sigma$ of the entangling surface $\Sigma$ and $c_0$ is fixed to be \cite{Solodukhin:2008dh,Fursaev:2013fta}
\begin{align}\label{c04d}
\begin{aligned}
c_0 &=-\frac{a}{2}\, \chi [\Sigma ]\\
	&\quad  + \frac{c}{2\pi} \int_\Sigma \left[ \CR_{abab} - \CR_{aa} + \frac{1}{3}\CR \right.\\
		&\qquad\qquad\qquad \left. + \frac{1}{2} (\CK^{a\,\mu}_{\,\mu})^2 - \CK^a_{\mu\nu} \CK^{a\,\mu\nu}\right] \ .
\end{aligned}
\end{align}
Here we used the Gauss-Codazzi equation \eqref{GaussCodazzi} relating the Ricci curvature on $\Sigma$ to the intrinsic and extrinsic curvatures,
which gives the first term proportional to the topological invariant (the Euler characteristic)
\begin{align}
	\chi[\Sigma]= \int_\Sigma E_2 = \frac{1}{4\pi}\int_\Sigma \CR_\Sigma \ ,
\end{align}
of the entangling surface $\Sigma$.
If the entangling surface is spherical, $\Sigma = \BS^2$, only the $a$-anomaly contributes to Eq.\,\eqref{c04d}, while if it is cylindrical $\Sigma = \BS^1 \times \BR$, only the $c$-anomaly survives.
The same expression \eqref{c04d} for the coefficient $c_0$ can be obtained by a holographic calculation in higher derivative gravities \cite{Hung:2011xb}, where the logarithmic divergences arise from a particular class of anomalies for a submanifold of even dimensions in CFT, known as the Graham-Witten anomalies \cite{Graham:1999pm}.

In higher dimensions, there is one Euler density and several independent Weyl invariants.
Entanglement entropy in even $d$-dimensional CFT allows the structure,
\begin{align}\label{EE_Even}
S_A = \frac{c_{d-2}}{\epsilon^{d-2}} + \frac{c_{d-4}}{\epsilon^{d-4}} + \cdots + \frac{c_2}{\epsilon^2} +  c_0\log \frac{l}{\epsilon}  + \cdots \ ,
\end{align}
with the logarithmic coefficient given by
\begin{align}\label{Log_EE_C0}
	c_0 = \frac{(-1)^{d/2+1}}{2}\, A \, \chi [\Sigma] - \sum_i\, B_i\,\left( \partial_n \, \int_{\CM_n} I_i\right)\bigg|_{n\to 1} \ .
\end{align}
The finite terms represented by the last dots depend on the regularization scheme and do not have any physical meaning.
On the other hand, entanglement entropy in odd $d$-dimensional CFT has similar expansions without the logarithmic term,
\begin{align}\label{EE_Odd}
S_A = \frac{c_{d-2}}{\epsilon^{d-2}} + \frac{c_{d-4}}{\epsilon^{d-4}} + \cdots + \frac{c_1}{\epsilon}  + (-1)^{(d-1)/2}F \ .
\end{align}
The finite part $F$ is {\it scheme independent} and physical observable.
The sign factor $(-1)^{(d-1)/2}$ is introduced for $F$ being positive in $d$ dimensions \cite{Klebanov:2011gs}, but there are no proofs for the positivity except for topological field theory where $F$ is the topological entanglement entropy that can be shown to be nonnegative \cite{Kitaev:2005dm,Levin:2006zz}.
We investigate the role of the finite part of entanglement entropy in odd dimensions in later sections.

\subsubsection{One interval in CFT$_2$}
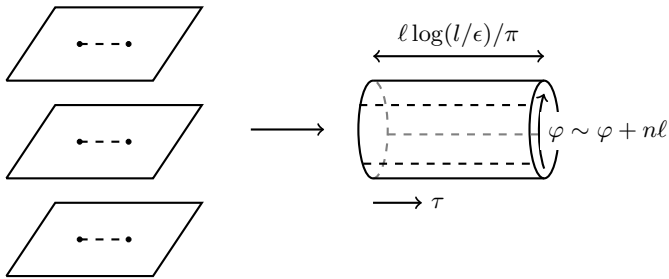
\begin{figure}[h]
\centering
\begin{tikzpicture}[scale=0.65]
	\draw[thick] (2, 3) --++ (3, 0) --++ (1, 1.5) --++ (-3, 0) --++ (-1, -1.5);
	\draw[thick, dashed] (3.5, 3.75) --++(1,0);
	\filldraw [fill=black] (3.5, 3.75) circle (0.05);
	\filldraw [fill=black] (4.5, 3.75) circle (0.05);
	\draw[thick] (2, 1) --++ (3, 0) --++ (1, 1.5) --++ (-3, 0) --++ (-1, -1.5);
	\draw[thick, dashed] (3.5, 1.75) --++(1,0);
	\filldraw [fill=black] (3.5, 1.75) circle (0.05);
	\filldraw [fill=black] (4.5, 1.75) circle (0.05);
	\draw[thick] (2, -1) --++ (3, 0) --++ (1, 1.5) --++ (-3, 0) --++ (-1, -1.5);
	\draw[thick, dashed] (3.5, -0.25) --++(1,0);
	\filldraw [fill=black] (3.5, -0.25) circle (0.05);
	\filldraw [fill=black] (4.5, -0.25) circle (0.05);

	\draw[thick, ->] (7,2) --++ (1.5,0);

	\draw[thick] (9.5,1) arc (270:90:0.3 and 1);
	\draw[thick, dashed, color=gray] (9.5,1) arc (-90:90:0.3 and 1);
	\draw[thick] (13,2) ellipse (0.3 and 1);
	\draw[thick] (9.5,3) -- (13,3);
	\draw[thick] (9.5,1) -- (13,1);
	\draw[thick, dashed] (9.3, 1.3) --++ (3.5,0);
	\draw[thick, dashed] (9.22, 2.5) --++ (3.5,0);
	\draw[thick, dashed, color=gray] (9.8, 1.9) --++ (3.5,0);
	\draw[thick, ->] (9.5,0.5) --++ (1,0) node[right] {$\tau$};
	\draw[thick, ->] (13,1.2) arc (230:130:0.3 and 1) node[fill=white, right, midway] {$\varphi \sim \varphi + n\ell$};
	\draw[thick, <->] (9.5,3.5) -- node[above] {$\ell \log (l/\epsilon)/\pi$} (13,3.5);
\end{tikzpicture}
\caption{\label{fig:CFT2EE} The $n$-fold cover $\CM_n$ of $\BC$ with a cut between $z=-l/2$ and $l/2$ is conformally mapped to a cylinder.
The intervals are mapped to lines at $\varphi = k \ell$ $(k=0,1,\cdots, n-1)$ where the $n$ copies are glued together.
This shows the conformal transformation for $n=3$.
}
\end{figure}

One can derive the entanglement entropy of one interval Eq.\,\eqref{EE_CFT2} in CFT$_2$ on a flat space $\BC$ using purely CFT techniques.
Suppose the interval is placed between $z = -l/2$ and $l/2$ in the complex $z$ plane $\BC$.
We first map the interval to a semi-infinite line from the origin to the infinity by the transformation 
\begin{align}
	\zeta\equiv \frac{z+l/2}{z-l/2} \ .
\end{align}
Then we make use of the conformal transformation from the complex plane to a cylinder,
\begin{align}\label{ConfMapCyl}
w\equiv \tau + \i\, \varphi = \frac{\ell}{2\pi} \log \zeta \ ,
\end{align}
where $\ell$ is the circumference of the cylinder $\varphi \sim \varphi + \ell$ as $\zeta$ is the complex coordinate of $\BC$.\footnote{Although the final result does not depend on $\ell$, we keep track of the $\ell$ dependence so as to make the argument as general as possible.}
For a later convenience, we excise two small holes of radius $\epsilon$ at $z=\pm l/2$ to regularize the UV divergence \cite{Ohmori:2014eia}.
Then the interval or equivalently the semi-infinite line is mapped to the line at $\varphi=0$ ranging from $\tau=- \ell\, \log (l/\epsilon)/(2\pi)$ to $\ell\, \log (l/\epsilon)/(2\pi)$ (see Fig.\,\ref{fig:CFT2EE}).

Now we want to construct the $n$-fold cover $\CM_n$ of $\BC$ by gluing $n$ copies of the complex plane along the interval.
This is easy to carry out in the $w$ coordinates by just gluing $n$ sheets of cylinder at $\varphi = k\ell$ $(k=0,1,\cdots, n)$, resulting in one cylinder of circumference $n\ell$ as in Fig.\,\ref{fig:CFT2EE}.
Then the partition function on $\CM_n$ is mapped to the cylinder partition function,
\begin{align}\label{CylinderPF}
	Z[\CM_n] = \langle 0| \,e^{- \beta H}\, |0 \rangle \ ,
\end{align}
where $\beta$ is the regularized length of the cylinder
\begin{align}
	\beta \equiv \frac{\ell}{\pi}\log \frac{l}{\epsilon} \ ,
\end{align}
and $H$ is the Hamiltonian along $\tau$ that will be fixed.

The stress tensor in the holomorphic sector transforms under the conformal mapping as
\begin{align}\label{HolTra}
	T(z) = \left( \frac{\d w}{\d z}\right)^2 T(w) + \frac{c}{12}\{ w, z\} \ ,
\end{align}
where $\{ w, z\} \equiv w'''(z)/w'(z) - 3[ w''(z)/w'(z)]^2/2$ is the Schwarzian derivative, and similarly for the antiholomorphic sector.\footnote{We consider a theory only  with equal left and right central charges $c_L=c_R = c$. A theory with $c_L\neq c_R$ has a gravitational anomaly and the entanglement entropy of such a theory was studied by \textcite{Castro:2014tta,Azeyanagi:2015uoa,Nishioka:2015uka,Hughes:2015ora,Iqbal:2015vka}.}
We parametrize the cylinder of circumference $n\ell$ by the complex coordinate $w$ and map it to a complex plane parametrized by $z$ using the conformal transformation of Eq.\,\eqref{ConfMapCyl} with $\ell$ and $\zeta$ replaced with $n\ell$ and $z$, respectively.
Using Eq.\,\eqref{HolTra}, we find the Hamiltonian along the $\tau$ direction on the cylinder,
\begin{align}\label{CylHam}
\begin{aligned}
	H &= \int_0^{n\ell} \frac{\d\varphi}{2\pi} \left(T(w) + \bar T(\bar w)\right) \ , \\
		&= \frac{2\pi}{n\ell} \left( L_0 + \bar L_0 - \frac{c}{12} \right) \ ,
\end{aligned}
\end{align}
where $L_0\, (\bar L_0)$ is the Virasoro generator in the (anti-)holomorphic sector on $\BC$.
 
The ground state $|0\rangle$ is annihilated by $L_0$ and $\bar L_0$, hence the partition function Eq.\,\eqref{CylinderPF} with the Hamiltonian \eqref{CylHam} resulting in
\begin{align}
	\log Z[\CM_n] = \frac{c }{6n} \log \frac{l}{\epsilon} \ .
\end{align}
The entanglement entropy follows from Eq.\,\eqref{EE_PF},
\begin{align}
	S_A = \frac{c}{3}\log \frac{l}{\epsilon} \ ,
\end{align}
which agrees with the previous result Eq.\,\eqref{EE_CFT2} derived with the conformal anomaly.

\subsection{R{\'e}nyi entropy and conformal maps}\label{ss:Renyi_CFT}

In this section, we are concerned with the R{\'e}nyi entropy $S_n$, Eq.\,\eqref{Renyi_QFT}, for a spherical entangling surface in CFT and discuss the connection to a thermal entropy on a conformally flat space.

First, let $F_n \equiv - \log Z_n$ be a free energy for the partition function $Z_n$ on the $n$-fold cover, then the R{\'e}nyi entropy takes the following form:
\begin{align}\label{RenyiNewDef}
S_n = \frac{n F_1 - F_n}{1-n} \ .
\end{align}
Calculating the free energy $F_n$ on the $n$-fold cover is far from reach even in CFT, but if we restrict our attention to a spherical entangling surface in CFT, we are able to find universal results for the R{\'e}nyi entropy.

We locate a spherical entangling surface $\BS^{d-2}$ at $\rho=R$ (and $t=0$) in the Euclidean polar coordinates,
\begin{align}\label{CFT_Flat}
\d s^2_\text{Flat} = \d t^2 + \d\rho^2 + \rho^2 \d\Omega_{d-2}^2 \ .
\end{align}
The conformal symmetry allows one to connect various geometries that are  conformally equivalent to the metric Eq.\,\eqref{CFT_Flat}, which simplifies the calculation of $F_n$ as we will see below.

\subsubsection{Conformal map to hyperbolic coordinates}\label{ss:Rel_to_Therm}
\begin{figure*}[htbp]
\centering
\begin{tikzpicture}[scale = 1.3, baseline={([yshift=-.5ex]current bounding box.center)}, scale=0.7]
\draw [->, thick] (-0.5,0) -- (-0.5,3) node [above] {$t$};
\draw [thick] (0,0.5) --++ (3,0) --++ (1,2) --++ (-3,0) -- (0,0.5);
\draw [thick, orange] (1.14,1.65) arc [start angle = 150, end angle = 280, radius = 0.3];
\draw [thick, dashed, tsuyukusairo, fill=white] (2,1.5) circle [x radius = 0.7, y radius = 0.4, rotate = 360];
\node [fill=white]  at (3.1, 1){$\BS^{d-2}$};
\draw [thick, ->] (2,1.5) --++ (45:0.8) node [right] {$\rho$};
\draw [thick, orange, fill=white] (1.7,1.5) arc [start angle = 0, end angle = 150, radius = 0.3];
\draw [thick, <-, orange] (1.7,1.5) arc [start angle = 360, end angle = 310, radius = 0.3];
\node at (2,-1) {$\BR^{d}$};
\node at (2,-2) {\footnotesize $\d s_\text{Flat}^2 = \d t^2 + \d\rho^2 + \rho^2 \,\d\Omega_{d-2}^2$};
\end{tikzpicture}
\begin{tikzpicture}[baseline={([yshift=-.ex]current bounding box.center)}, scale=0.5]
\draw[thick, ->] (1, 0) --++(2,0);
\end{tikzpicture}
\begin{tikzpicture}[scale=1.2, baseline={([yshift=1.2ex]current bounding box.center)}, scale=0.7]
\draw[thick] (0,3) arc [start angle = 90, end angle = 270, x radius = 0.2, y radius = 0.5];
\draw[thick, dashed] (0,2) arc [start angle = 270, end angle = 450, x radius = 0.2, y radius = 0.5];
\draw [thick] (4,2.5) circle [x radius = 0.2, y radius = 0.5, rotate = 0];
\node at (4.6,2.5) {$\tau$};
\draw [thick] (0,3) -- (4,3);
\draw [thick] (0,2) -- (4,2);
\draw[thick] (0,1.6) arc [start angle = 90, end angle = 270, x radius = 0.2, y radius = 0.8];
\draw[thick, dashed] (0,0) arc [start angle = 270, end angle = 450, x radius = 0.1, y radius = 0.8];
\draw[thick, orange,
	decoration={markings, mark=at position 0.5 with {\arrow{<}}},
        postaction={decorate}] (2,3) arc [start angle = 90, end angle = 270, x radius = 0.2, y radius = 0.5];
\draw[thick, dashed, orange] (2,2) arc [start angle = 270, end angle = 450, x radius = 0.1, y radius = 0.5];
\draw [thick, tsuyukusairo] (4,0.8) circle [x radius = 0.1, y radius = 0.8, rotate = 0];
\draw [thick] (0.05,1.55) arc [start angle = 190, end angle = 345, x radius = 2, y radius = 0.92];
\draw [thick] (3.94,0.1) arc [start angle = 15, end angle = 170, x radius = 2, y radius = 0.92];
\node at (4.9,0.8) {$\BS^{d-2}$};
\draw [->, thick] (2,-0.5) -- (4,-0.5) node [right] {$u$};
\node at (2,-1.5) {$\BS^{1} \times \BH^{d-1}$};
\node at (2,-2.5) {\footnotesize $\d s^2_\text{Hyp} = \d\tau^2 + \d u^2 + \sinh^2 u\, \d\Omega_{d-2}^2$};
\end{tikzpicture}
\begin{tikzpicture}[baseline={([yshift=-.5ex]current bounding box.center)}, scale=0.5]
\draw[thick, ->] (1, 0) --++(2,0);
\end{tikzpicture}
\begin{tikzpicture}[scale=1.2, baseline={([yshift=3ex]current bounding box.center)}, scale=0.7]
\draw [thick] (0,0) -- (3,0);
\draw [thick, tsuyukusairo] (0,1) circle [x radius = 0.3, y radius = 0.5, rotate = 360];
\node at (-0.9,1) {$\BS^{d-2}$};
\draw [thick] (0,1.5) -- (3,1) -- (0, 0.5);
\draw [thick, orange,
	decoration={markings, mark=at position 0.5 with {\arrow{<}}},
        postaction={decorate}] (3,2) circle [x radius = 0.3, y radius = 0.5, rotate = 360];
\node at (3.6,2) {$\tau$};
\draw [thick] (3, 2.5) -- (0,2) -- (3, 1.5);
\filldraw [fill=black] (0,0) circle (0.1) node [below] {$\theta=0$};
\filldraw [fill=black] (3,0) circle (0.1) node [below] {$\pi/2$};
\node at (1.5,-1.5) {$\BS^{d}$};
\node at (1.5,-2.5) {\footnotesize $\d s^2_\text{Sp} = \d\theta^2 + \sin^2\theta\, \d\tau^2 + \cos^2\theta\, \d\Omega_{d-2}^2$};
\end{tikzpicture}
\caption{\label{fig:ConformalMap}  The coordinate transformation Eq.\,\eqref{BoundaryCM} induces the conformal map from a flat space [Left] to a hyperbolic space times a circle [Middle].
The latter is further conformally mapped to a sphere [Right] by the coordinate transformation, $\sinh u = \cot\theta$.
The blue circles represent the entangling surface $\BS^{d-2}$, and the orange circles with arrows stand for the modular time, respectively.
}
\end{figure*}
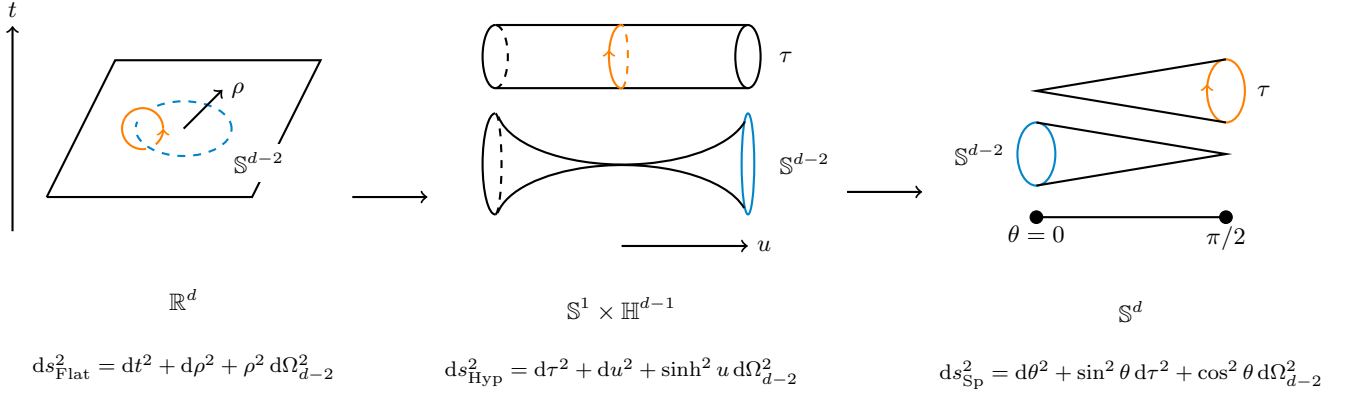

We first map the flat space $\BR^d$ to a hyperbolic space $\BS^1 \times \BH^{d-1}$ with the metric
\begin{align}\label{CFT_Hyp}
\d s^2_\text{Hyp} = \d\tau^2 + \d u^2 + \sinh^2 u \,\d\Omega_{d-2}^2 
\end{align}
with the ranges $0\le \tau \le 2\pi$ and $0\le u <\infty$, by the coordinate transformation 
\begin{align}\label{BoundaryCM}
\begin{aligned}
t &= R \frac{ \sin\tau}{\cosh u + \cos\tau} \ , \qquad
\rho = R \frac{\sinh u}{\cosh u + \cos\tau} \ .
\end{aligned}
\end{align}
The two metrics are conformally equivalent:
\begin{align}
	\d s^2_\text{Flat} = \Omega^2_\text{Hyp}\, \d s^2_\text{Hyp} \ ,	
\end{align}
with the conformal factor
\begin{align}
	\Omega_\text{Hyp} = \frac{R}{\cosh u + \cos \tau} \ .
\end{align}
Under this coordinate transformation, the entangling surface at $\rho=R$ is mapped to a sphere $\BS^{d-2}$ at $u=\infty$ in the hyperbolic coordinates.
The $n$-fold cover $\CM_n$ around the entangling surface at $\rho= R$ in Eq.\,\eqref{CFT_Flat} is equivalent to extending the ranges of $\tau$ to $0\le \tau \le 2\pi n$ (see Fig.\,\ref{fig:ConformalMap}).
After this conformal map, CFT is defined on the hyperbolic space $\BH^{d-1}$ at temperature $T_n=1/(2\pi n)$.

We assume there are no conformal anomalies (i.e., $d$ is odd) for the moment. Then the partition function is invariant under the conformal map; hence on the hyperbolic coordinates Eq.\,\eqref{CFT_Hyp}, $Z_n$ is the thermal partition function
\begin{align}\label{PFHyp}
Z_n = \tr \left(e^{-H_\text{Hyp} /T_n}\right) \ ,
\end{align}
where $H_\text{Hyp}$ is the Hamiltonian generating the translation along $\tau$ on $\BH^{d-1},$\footnote{The Hamiltonian or the energy is given by the generator of $\tau$ translation $T_{\mu\nu}\xi^\mu$ with $\xi^\mu\partial_\mu = \partial_\tau$ integrated over a spacelike surface
\begin{align}
H  =\int_{\BH^{d-1}}\d^{d-1}x\, \sqrt{g} \, T_{\mu\nu}(x)\,\xi^\mu n^\nu \ ,
\end{align}
where $n^\mu\partial_\mu = \sqrt{g^{\tau\tau}} \partial_\tau$ is the unit normal vector.
}
\begin{align}\label{ModularHamiltonian}
H_\text{Hyp}  =  \int_{\BH^{d-1}}\d^{d-1}x\, \sqrt{g} \, T_{\tau\tau}(x) \ .
\end{align}
The free energy $F_n = -\log Z_n$ and the thermal free energy $F(T)$ on $\BH^{d-1}$ at temperature $T=T_n$ are related by $F_n = F(T_n)/T_n$. 
Using the thermodynamic identity $S_\text{therm}(T) = - \partial F(T)/\partial T$, the R{\'e}nyi entropy Eq.\,\eqref{RenyiNewDef} is given by the integral of the thermal entropy \cite{Hung:2011nu},
\begin{align}\label{RenyiTherm}
\begin{aligned}
S_n &= \frac{n}{1-n} \frac{F(T_1) - F(T_n)}{T_1} \ , \\
	&= \frac{n}{n-1}\frac{1}{T_1} \int_{T_n}^{T_1} \d T \, S_\text{therm}(T) \ .
\end{aligned}
\end{align}
This is an intriguing connection between the R{\'e}nyi entropy of a spherical entangling surface on the flat space and the thermal entropy on $\BH^{d-1}$ at finite temperature, which will be useful for the holographic studies in the following sections.
Moreover, it follows from the $n\to 1$ limit of Eq.\,\eqref{RenyiTherm} that the entanglement entropy equals the thermal entropy at temperature $T= T_1$,
\begin{align}\label{RenyiAndTherm}
S_1 = \lim_{n\to 1} S_n = S_\text{therm}(T_1) \ .
\end{align}
There is a subtle issue in this relation associated with the boundary of the hyperbolic space in even dimension \cite{Casini:2011kv}.
The universal logarithmic divergence proportional to the type $A$ anomaly in the entanglement entropy \eqref{EE_Even} is attributed to a boundary effect \cite{Herzog:2015ioa,Herzog:2016kno}.

\subsubsection{Conformal map to spherical coordinates}
We can make a further conformal transformation from the hyperbolic coordinates
 Eq.\,\eqref{CFT_Hyp} to the spherical coordinates $\BS^{d}$ of unit radius,
\begin{align}\label{CFT_Sphere}
\d s^2_\text{Sp} = \d\theta^2 + \sin^2 \theta\, \d\tau^2 + \cos^2\theta\, \d\Omega_{d-2}^2 \ ,
\end{align}
with the ranges $0\le \tau \le 2\pi$ and $0\le \theta \le \pi/2$,
by the coordinate transformation $\sinh u = \cot\theta$ (see Fig.\,\ref{fig:ConformalMap}).
It is also conformally equivalent to the flat space
\begin{align}
	 \d s^2_\text{Flat} = \Omega^2_\text{Sp}\,\d s^2_\text{Sp}
\end{align}
with the conformal factor
\begin{align}
	\Omega_\text{Sp}= \frac{R}{1 + \sin\theta\, \cos\tau} \ .
\end{align}

The $n$-fold cover $\CM_n$ of $\BR^d$ is conformally equivalent to the $n$-fold cover of the $d$-sphere $\BS_n^d$ with the extended range $0\le \tau \le 2\pi n$, and 
the free energy $F_n$ equals to the logarithm of the partition function $Z[\BS^d_n]$ for CFT,
\begin{align}\label{F_nZ}
F_n = -\log Z[\BS^d_n] \ .
\end{align}
The R{\'e}nyi entropies for free fields were computed by \textcite{Klebanov:2011uf}
based on  Eq.\,\eqref{F_nZ} and shown to agree with the calculation on the hyperbolic coordinates.

Next we look into Eq.\,\eqref{RenyiAndTherm}, the entanglement entropy to the thermal entropy.
Through the thermodynamic relation, $S_\text{therm}$ is given by the free and total energies $F(T), E(T)$ of the CFT$_d$,
\begin{align}\label{TDR_CFT}
S_\text{therm}(T) = \frac{E(T) - F(T)}{T} \ .
\end{align}
When $d$ is odd, the total energy vanishes for CFT$_d$ because the round sphere $\BS^d$ is conformally equivalent to the flat space $\BR^d$: $E(T)=0$ \cite{Casini:2011kv,Klebanov:2011uf}.
Combining Eqs.\,\eqref{RenyiAndTherm} and \eqref{TDR_CFT} with $F(T_1)/T_1 = F_1 = -\log Z[\BS^d_n]$, we obtain the equality between the entanglement entropy of a sphere $\BS^{d-2}$ and the partition function on $\BS^d$,
\begin{align}\label{EE_F}
S_1 = \log Z[\BS^d] \ .
\end{align}
Note that this equality should be taken up to UV divergences as the vanishing of the total energy holds for CFT$_d$ only after renormalizing the divergences.
If the right-hand side is the correctly renormalized effective action, it should be identified with the universal part of the entanglement entropy in Eq.\,\eqref{EE_Odd}, 
\begin{align}
F = (-1)^{(d-1)/2}\log Z[\BS^d] \ .
\end{align}
This will be the key relation in Sec. \ref{ss:RGflow} when we construct a monotonically decreasing function along RG flows in odd dimensions.

When $d$ is even, there are conformal anomalies in the partition function $Z[\BS^d]$, which gives rise to the logarithmic divergence in entanglement entropy as appeared in Eq.\,\eqref{EE_Even}.
This is most easily seen by analytically continuing Eq.\,\eqref{EE_F} from odd to even dimensions and replacing the pole with the logarithm.

\subsection{Universal behavior of R{\' e}nyi entropy in CFT}
Having established the universal relation \eqref{EE_F} for the spherical entanglement entropy, we now seek for the universal aspects of the R{\'e}nyi entropies.
We set $R=1$ since the result should not depend on $R$ because of the conformal symmetry.

We want to examine the universal part of the spherical R{\'e}nyi entropy in odd dimensions and expand it around $n=1$ where $S_1 = -F_1$ up to the UV divergences. 
From Eq.\,\eqref{RenyiNewDef} we have the expansion
\begin{align}
\begin{aligned}
S_n &= S_1 + \frac{F_n - F_1}{n-1} \ , \\
	&= S_1 + \sum_{k=1}^\infty \frac{1}{k!} \frac{\partial^k F_n}{\partial n^k}\bigg|_{n=1} (n-1)^{k-1} \ .
\end{aligned}
\end{align}
For a spherical entangling surface in CFT, the free energy is given by the partition function \eqref{PFHyp} on the hyperbolic coordinates, which fixes the derivatives of $F_n$, 
\begin{align}
\begin{aligned}
\frac{\partial^k F_n}{\partial n^k}\bigg|_{n=1} &= - (-2\pi)^k \langle H_\text{Hyp}^k\rangle  \ , 
\end{aligned}
\end{align}
where $\langle\, \cdot\, \rangle \equiv \tr (\,\cdot\, e^{-2\pi H_\text{Hyp}})/\tr (e^{-2\pi H_\text{Hyp}})$ stands for the expectation value of an operator and we used the fact that the energy vanishes for CFT in odd dimensions: $E(T) = \langle H_\text{Hyp} \rangle= 0$.
The form of the Hamiltonian \eqref{ModularHamiltonian} on the hyperbolic space allows us to obtain the expansion in terms of the correlation functions of the stress tensors,
\begin{align}\label{REexpand}
\begin{aligned}
S_n &= S_1 \\
	&~- \sum_{k=1}^\infty \frac{(-2\pi)^{k+1}}{(k+1)!} (n-1)^k \\
	&\quad \cdot \int_{\BH^{d-1}}\!\cdots \int_{\BH^{d-1}}\prod_{i=1}^{k+1} \d^{d-1}x_i\,\sqrt{g}\, \langle T_{\tau\tau}(x_1) \cdots T_{\tau\tau}(x_{k+1}) \rangle \ .
\end{aligned}
\end{align}

The leading term is fixed by the two-point function of the stress tensors, whose integral results in \cite{Perlmutter:2013gua}
\begin{align}\label{LeadingRE}
\partial_n S_n\big|_{n=1} = -C_T\,\text{Vol}\left(\BH^{d-1}\right)\,\frac{\pi^{d/2+1}(d-1)\Gamma\left(d/2\right)}{\Gamma(d+2)} \ ,
\end{align}
where $C_T$ is a constant present in Eq.\,\eqref{TTCor}.
For odd $d$, the regularized volume of $\BH^{d-1}$ is [see \textcite{Klebanov:2011uf} for $d=3$]
\begin{align}\label{HypVolume}
\text{Vol}(\BH^{d-1}) = (-1)^{(d-1)/2}\frac{\pi^{d/2}}{\Gamma \left(d/2\right)} = \pi^{d/2-1}\Gamma(1-d/2)\ ,
\end{align}
which may be obtained by putting the cutoff $\Lambda\gg 1$ at the spatial infinity of $\BH^{d-1}$ and picking up a constant term in $\Lambda \to \infty$ limit.\footnote{In the second equality, we used the reflection formula for the gamma functions, $\Gamma(z) \Gamma(1-z) = \pi/ \sin (\pi z)= \pi (-1)^{(d-1)/2}$.}

While the universal result Eq.\,\eqref{LeadingRE} was derived under the assumption that $d$ is \emph{odd}, it is shown to hold for \emph{even} $d$ \cite{Perlmutter:2013gua} where there are additional terms including the one-point function $\langle\CH\rangle$ which no longer vanishes due to the conformal anomaly.
In addition, the volume of the hyperbolic space $\text{Vol}(\BH^{d-1})$ logarithmically diverges for even $d$ after regularizing the power law divergence, which may be inferred from Eq.\,\eqref{HypVolume} as a pole of the gamma function by analytically continuing odd $d$ to even.

We finally note that the first derivative of the sphere entanglement entropy Eq.\,\eqref{LeadingRE} can be useful to read off the coefficient $C_T$ of the two-point function of the stress tensor.
A similar expansion and a universal leading coefficient will be derived for a supersymmetric generalization of the R{\'e}nyi entropy in Sec. \ref{ss:SRE}.

  \section{Holographic method}\label{ss:Holography}

The holographic principle, stating the equivalence between gauge theory and gravity theory, is implied by string theory.
One of the most famous duality is the AdS/CFT correspondence conjectured
by \textcite{Maldacena:1997re}, which has been the main focus of research in
high energy theory for two decades.
String theory has extended objects called D-branes that allow two complementary descriptions as a black hole in general relativity and a gauge theory on the world volume.
In the former picture, one finds the AdS geometry in the near horizon region, which turns out to be equivalent to the gauge theory of the latter picture.
The AdS/CFT correspondence crystalizes the equivalence by connecting the type IIB superstring theory on the AdS space with a supersymmetric CFT
living on the boundary of the AdS space.
This is a concrete example of duality where a strongly coupled region of one theory is described by a weakly coupled region of the other theory.
Hence the direct check of this correspondence is far reaching in its nature and remains to be proved.

Apart from the string theory implication, we can take the AdS/CFT correspondence 
as a gauge/gravity duality in its own right simply based on the symmetry argument that the isometry $\SO(2,d)$ of the AdS$_{d+1}$ space agrees with 
the conformal group of CFT$_d$.

In this section, we give a quick review of the AdS/CFT
correspondence. 
This subject is comprehensively covered by a review paper by
\textcite{Aharony:1999ti} and many others in the literature including \textcite{Klebanov:2000me,McGreevy:2009xe} from different points of view.
We then introduce the holographic formula of entanglement entropy proposed by \textcite{Ryu:2006bv,Ryu:2006ef} and discuss the implications.
The reviews on entanglement entropy from the holographic point of view can be found in \textcite{Nishioka:2009un,Takayanagi:2012kg,Rangamani:2016dms}.

\subsection{The AdS geometries}

We briefly sketch several types of the useful coordinates for the AdS geometry that we adopt in the following sections.
Table \ref{tab:AdS_Coord_Summary} is the summary of the relations between the AdS coordinates and the corresponding conformally flat spaces on their boundaries.

To begin with, consider the flat $(d+2)$-dimensional pseudo Euclidean space defined by
\begin{align}\label{PseudEuclid}
 \d s^2&= - \d y_{-1}^2 - \d y_0^2 + \d y_1^2 + \cdots + \d y_d^2 \ .
\end{align}
The AdS$_{d+1}$ space with radius $L$ is an embedding hypersurface
satisfying 
\begin{align}
 -y_{-1}^2 - y_0^2 + y_1^2 + \cdots + y_d^2&= -L^2 \ .
\end{align}
This construction manifests the isometry $\SO(2,d)$ of the AdS$_{d+1}$ space.
Among many coordinates in the AdS space we focus on the typical examples,
the global, Poincar{\'e} and hyperbolic coordinates.

\subsubsection{Global coordinates}
The global coordinates are introduced by the coordinate transformations:
\begin{align}
\begin{aligned}
 y_{-1}&= L\,\cosh\rho\, \sin t \ , \\
 y_0&=L\,\cosh\rho\,\cos t \ , \\
 y_i &= L\,\sinh \rho\, \ e^i \ , \quad (i=1,\dots,d)\ ,
\end{aligned}
\end{align}
where $e^i$ satisfy the relation $\sum_{i=1}^d (e^i)^2=1$ and 
span a $d$-dimensional sphere.
The metric Eq.\,\eqref{PseudEuclid} then becomes
\begin{align}
 \d s^2&= L^2\left[ -\cosh^2\rho\, \d t^2 + \d\rho^2 + \sinh^2\rho\,
 \d\Omega_{d-1}^2 \right] \ .
\end{align}
By the coordinate transformation $r = \sinh\rho$, we find
another form of the global coordinates often used in literature,
\begin{align}\label{GlobalAdS}
 \d s^2&= L^2\left[ -\left( r^2+1\right) \d t^2 + \frac{\d r^2}{r^2 + 1
 } + r^2 \,\d\Omega_{d-1}^2 \right]\ .
\end{align}
These coordinates cover the whole AdS$_{d+1}$ space whose boundary at $r=\infty$ is $\BR \times \BS^{d-1}$ space.

The global AdS$_{d+1}$ space can have de Sitter space (dS$_{d}$) as its boundary,
\begin{align}
\d s^2 = L^2 \left[ \d\rho^2 + \sinh^2\rho \left( -\sin^2\theta\, \d t^2 + \d\theta^2 + \cos^2\theta \,\d\Omega_{d-2}^2 \right)\right] \ ,
\end{align}
which follows from the coordinate transformations
\begin{align}
\begin{aligned}
y_{-1} &= L\,\cosh\rho \ , \\
y_0 &= L\,\sinh\rho\, \sin\theta\, \sinh t \ , \\
y_d &= L\,\sinh\rho\, \sin\theta\, \cosh t \ , \\
y_i &= L\,\sinh \rho\, \cos\theta\,  \tilde e^i \ , \qquad (i=1,\cdots, d-1) \ , 
\end{aligned}
\end{align}
where $\sum_{i=1}^{d-1}(\tilde e^i)^2 = 1$.
This is also useful when we consider a holographic dual of a Euclidean QFT on $\BS^d$ after the Wick rotation.

\subsubsection{Poincar{\'e} coordinates}
The Poincar{\'e} coordinates are introduced by the coordinate transformations:
\begin{align}
\begin{aligned}
y_{-1} &= \frac{1}{2r}\left[ 1 + r^2\left(L^2 -t^2 + \sum_{i=1}^{d-1}x_i^2\right)\right] \ , \\
y_0 &= L \,r\, t\ ,\\
y_d &= \frac{1}{2r}\left[ 1 + r^2\left(-L^2 -t^2 + \sum_{i=1}^{d-1}x_i^2\right)\right]  \ , \\
y_i &= L\, r\, x_i\ , \qquad (i=1,\cdots d-1) \ ,
\end{aligned}
\end{align}
whose metric is given by
\begin{align}
 \d s^2&=L^2\left[ \frac{\d r^2}{r^2} + r^2\left(-\d t^2 + \sum_{i=1}^{d-1} \d x_i^2 \right) \right] \ .
\end{align}
These coordinates cover a half portion of the whole AdS$_{d+1}$ space and the boundary at $r=\infty$ is the Minkowski space $\BR^{d-1,1}$.

Inverting the radial coordinate $z\equiv 1/r$ yields another version of the Poincar{\'e} coordinates,
\begin{align}\label{AdS_Poincare_z}
 \d s^2&= L^2\,\frac{\d z^2 - \d t^2 + \sum_{i=1}^{d-1} \d x_i^2}{z^2} \ .
\end{align}

\subsubsection{Hyperbolic coordinates}
The hyperbolic coordinates of the AdS$_{d+1}$ space are obtained by the coordinate transformations:
\begin{align}
\begin{aligned}
y_{-1} &= L\, r\, \cosh u \ , \\ 
y_0 &= L\,\sqrt{r^2 - 1}\, \sinh t \ , \\
y_d &= L\,\sqrt{r^2 - 1}\, \cosh t \ ,  \\
y_i &= L\,r\,\sinh u \,\tilde e^i \ , \qquad (i=1,\cdots,d-1) \ .
\end{aligned}
\end{align}
The resulting metric becomes
\begin{align}\label{AdS_Hyp}
\begin{aligned}
\d s^2 &=L^2\left[  -\left(r^2 - 1 \right) \d t^2 + \frac{\d r^2}{r^2 - 1}\right.\\ 
	&\qquad \left. + r^2 \left( \d u^2 + \sinh^2 u\, \d\Omega_{d-2}^2\right) \right]\ .
\end{aligned}
\end{align}
These coordinates cover a half portion of the whole AdS$_{d+1}$ space whose boundary at $r=\infty$ is $\BR \times \BH^{d-1}$ space.
While these coordinates are locally equivalent to the other coordinates, there is a coordinate singularity at the event horizon, $r=1$.
In Euclidean signature, the Euclidean time $\tau = - \i\, t$ becomes a circle of period  $\beta = 2\pi$ to avoid the conical singularity at the horizon.
The period is identified with an inverse temperature of the dual CFT.

Table \ref{tab:AdS_Coord_Summary} summarizes the relations of the conformally flat (Euclidean) spaces to the corresponding (Lorentzian) AdS metrics.
\begin{table}[htbp]
\caption{\label{tab:AdS_Coord_Summary}Conformally flat spaces and the corresponding Lorentzian AdS metrics.}
\begin{ruledtabular}
\begin{tabular}{llll}
\multicolumn{2}{c}{ Conformally flat space} & \multicolumn{2}{c}{AdS metric}\\ \hline
 $\BR^d$ & Eq.\,\eqref{CFT_Flat} & Poincar{\'e} coordinates & Eq.\,\eqref{AdS_Poincare_z} \\
 $\BS^1 \times \BH^{d-1}$ & Eq.\,\eqref{CFT_Hyp} & Hyperbolic coordinates & Eq.\,\eqref{AdS_Hyp} \\
 $\BS^d$ & Eq.\,\eqref{CFT_Sphere} & Global coordinates & Eq.\,\eqref{GlobalAdS}
\end{tabular}
\end{ruledtabular}
\end{table}

\subsection{The GKP-W relation}
In what follows we deal with the Euclidean case.
The Euclidean AdS$_{d+1}$ spacetime is a solution to the Einstein equation of the Einstein-Hilbert action
\begin{align}\label{EH_action}
\begin{aligned}
I_\text{bulk} [\CB] &= - \frac{1}{16\pi G_N} \int_{\CB} \d^{d+1} x \, \sqrt{g}\, \left(\CR + \frac{d(d-1)}{L^2} \right) \\
	&\qquad - \frac{1}{8\pi G_N} \int_{\partial\CB} \d^d x\, \sqrt{h}\, \CK\ ,
\end{aligned}
\end{align}
where the cosmological constant is chosen so that $L$ is the radius of the AdS$_{d+1}$ space when $\CB = \text{AdS}_{d+1}$.
The second integral is the Gibbons-Hawking term that ensures the variational principle with the induced metric $h_{\alpha\beta}$ fixed on the boundary $\partial \CB$.
$\CK$ is the trace of the extrinsic curvature $\CK = \nabla_\alpha n^\alpha$ for the vector $n^\alpha$ normal to the boundary $\partial\CB$.

The most fundamental relation in the AdS/CFT correspondence is the equality between the partition functions of the gravity on an asymptotically AdS$_{d+1}$ space $\CB$ and the dual CFT$_d$ living on the boundary $\CM =\partial \CB$,
\begin{align}\label{GKPW}
	 e^{- I_\text{bulk}[\CB]} = Z_\text{CFT}[\CM] \ .
\end{align}
This is called the GKP-W relation named after \textcite{Gubser:1998bc,Witten:1998qj}.
When a bulk scalar field $\phi$ is present in the bulk $\CB$, we need to add to the left-hand side the matter action $I_\text{matter}[\phi]$ and replace the right-hand side with the generating functional of correlation functions,
\begin{align}
	e^{- I_\text{bulk+matter} [\CB, \phi |_{\CM} = \phi_0 ] } = \langle e^{\int_{\CM} \phi_0 \CO }\rangle_\text{CFT} \ ,
\end{align}
where $\phi_0 (x)$ is the boundary value of the bulk field $\phi$ that couples to an operator $\CO(x)$ as the external source in CFT.

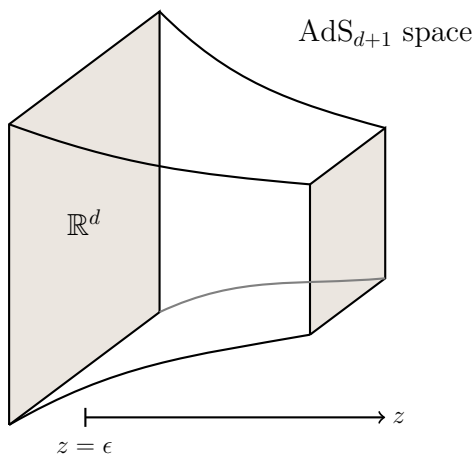
\begin{figure}[htbp]
\centering
                \begin{tikzpicture}[thick,scale=1, every node/.style={scale=.8}]
			   \tikzstyle{ann} = [fill=white,inner sep=2pt]
			   \draw[fill=verylightgray] (-1,-1.2)--(1,0.3)--(1,4.3)--(-1,2.8)--(-1,-1.2);
			   \draw[fill=verylightgray] (3, 0)--(4,0.75)--(4,2.75)--(3,2)--(3,0);
			   \draw (-1,-1.2) to [out=30, in=190] (3,0);
			   \draw[gray] (1,0.3) to [out=25, in=185] (4,0.75);
			   \draw (1,4.3) to [out=-45, in=165] (4,2.75);
			   \draw (-1, 2.8) to [out=-20, in=175] (3,2);
      			   \draw (4,4) node {\Large \textrm{AdS}$_{d+1}$ space};
			   \draw (0,1.5) node {\Large $\BR^d$};
			   \draw[|->] (0,-1.1) -- (4,-1.1) node[right] {\large $z$};
			   \draw (0,-1.5) node[ann] {\large $z=\epsilon$};
		\end{tikzpicture}
	\caption{The AdS/CFT correspondence in the Poincar{\'e} coordinates. The boundary of AdS$_{d+1}$ is the flat space $\BR^d$ on which CFT$_d$ lives.} 
	\label{fig:AdSCFT}
\end{figure}

Figure\,\ref{fig:AdSCFT} shows a schematic picture of the AdS/CFT correspondence when $\CB$ is the $\text{AdS}_{d+1}$ space in the Poincar{\'e} coordinates Eq.\,\eqref{AdS_Poincare_z}.
In this case, the boundary of AdS$_{d+1}$ at $z=0$ is a flat space $\CM = \BR^d$ where the dual CFT$_d$ is supposed to live, but we introduced a cutoff at $z=\epsilon \ll 1$ to regularize the volume of the AdS space.
This cutoff in the bulk corresponds to the UV cutoff $\Lambda$ in CFT as $\Lambda \sim 1/\epsilon$, while the large $z$ region corresponds to the IR region of the CFT.

\subsection{Holographic entanglement entropy}\label{ss:HEE}

The AdS/CFT correspondence connects, through the GKP-W relation \eqref{GKPW}, the partition function of QFT$_d$ on a manifold $\CM$ to a classical gravity action on an asymptotically AdS$_{d+1}$ space $\CB$ that asymptotes to $\CM$ on its boundary. 
Assuming that the bulk is described by the Einstein gravity Eq.\,\eqref{EH_action}, we present the holographic formula of entanglement entropy and its derivation.

\begin{figure}[htbp]
\centering
                \begin{tikzpicture}[thick,scale=1, every node/.style={scale=1}]
			   \tikzstyle{ann} = [fill=white,inner sep=2pt]
			   \draw (-1,-1.2)--(1,0.3)--(1,4.3)--(-1,2.8)--(-1,-1.2);
			   \draw[orange,fill=orange!20, opacity=0.6] (0,0) arc (-90:90:2.2 and 1.5);
			   \draw[dashed,aomidori,fill=cyan!50] (0,0) arc (-90:90:0.5 and 1.5);
			   \draw[aomidori,fill=cyan!50] (0,0) arc (270:90:0.5 and 1.5);
			   \draw (0,1.5) node {$\textcolor{aomidori} A$};
   			   \draw (3,3) node {\large\textrm{AdS space}};
			   \draw[|->] (0,-0.3) -- (3,-0.3) node[right] {$z$};
			   \draw (0,-0.7) node[ann] {$z=\epsilon$};		   
			   \draw[dashed,aomidori,fill=orange!40, opacity=0.5] (0,0) arc (-90:270:0.5 and 1.5);
			   \draw (2.6,1.7) node {\large\textcolor{orange}{$\gamma_A$}};
		\end{tikzpicture}
	\caption{The minimal surface $\gamma_A$ in orange anchored on the boundary of the region $A$ in blue located at the AdS boundary $z=\epsilon$.
	The holographic entanglement entropy of the region $A$ is given by the area of the surface $\gamma_A$ divided by $4$ times the Newton constant $G_N$ as in Eq.\,\eqref{ARE}.
	}
	\label{fig:HEE}
\end{figure}
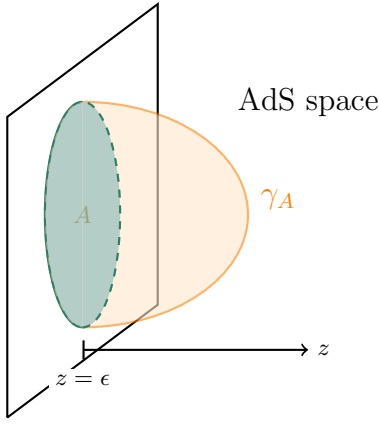

Given the GKP-W relation \eqref{GKPW}, we can, in principle, rewrite the replica trick formula \eqref{EE_PF} of entanglement entropy in terms of the bulk action.
We however need a dual gravity solution $\CB_n$ whose boundary is the $n$-fold cover $\CM_n$ with conical singularity at the entangling surface $\Sigma = \partial A$ for a given region $A$.
Assuming the existence of such a bulk space $\CB_n$ we arrive at the holographic definition of entanglement entropy
\begin{align}\label{HEE_Replica}
	S_A = \lim_{n\to 1} \partial_n \left( I_\text{bulk} [\CB_n] - n\, I_\text{bulk} [\CB]\right) \ .
\end{align}
We need a sophisticated machinery to find a bulk solution $\CB_n$ and evaluate the on-shell action $I_\text{bulk}[\CB_n]$ and we postpone it to Sec. \ref{ss:HEE_Derivation}.
\textcite{Ryu:2006bv} rather conjectured the holographic formula of entanglement entropy (also known as the Ryu-Takayanagi formula) should be 
\begin{align}\label{ARE}
S_A = \frac{\CA}{4G_N}\ ,
\end{align}
where $\CA$ is the area of a (unique) minimal surface $\gamma_A$ in $\CB$ anchored on the entangling surface $\Sigma$ (see Fig.\,\ref{fig:HEE}).
We see this geometric formula shows the same characteristics as entanglement entropy in QFT after reviewing its proof by \textcite{Lewkowycz:2013nqa}.\footnote{An earlier attempt for the proof was put forward by \textcite{Fursaev:2006ih}.}

The holographic formula for higher derivative gravity theory is derived along the same lines of arguments as \textcite{Lewkowycz:2013nqa} by \textcite{Dong:2013qoa,Camps:2013zua,Miao:2014nxa}, which correctly reproduces the Jacobson-Myers type formula \cite{Jacobson:1993xs} for the Lovelock gravity \cite{deBoer:2011wk,Hung:2011xb}.
A generalization to the covariant formula of holographic entanglement entropy in a time-dependent QFT was proposed by \textcite{Hubeny:2007xt} and later proved by \textcite{Dong:2016hjy}.
We refer the interested readers to the recent textbook by \textcite{Rangamani:2016dms} and references therein for the details.

\subsubsection{A derivation of the holographic entanglement entropy}\label{ss:HEE_Derivation}
\begin{figure*}[htbp]
\centering
	\begin{tikzpicture}[thick]
	\draw[|->] (-2,4.7) --++ (0,2) node[above] {$z$};
	\draw[fill=verylightgray, thick] (-2, 4) --++ (3, 0) --++ (1, 1.5) --++ (-3, 0) --++ (-1, -1.5);
	\node at (-1.2,4.3) {$\CM_n$};
	\node at (-1.2, 7) {$\CB_n$};
	\node[orange] at (0.6,6.2) {$\gamma_A^{(n)}$};
	\node at (0,4.4) {$\textcolor{aomidori} A$};
	\draw[very thick, aomidori] (-0.6, 4.7) --++(1.2,0);
	\draw[->] (0.3,4.6) arc (-140:70:0.3 and 0.2);
	\draw[->] (0,5.85) arc (-90:170: 0.2 and 0.3);
	\node at (0,6.7) {$\tau$};
	\node at (1,4.5) {$\tau$};
	\draw[very thick, orange, fill=magenta!30, opacity=0.6] (0.6, 4.7) arc (0:180: 0.6 and 1.5);
	\draw[fill=verylightgray, thick] (2, 4) --++ (3, 0) --++ (1, 1.5) --++ (-3, 0) --++ (-1, -1.5);
	\draw[very thick, aomidori] (3.4, 4.7) --++(1.2,0);
	\draw[->] (4.3,4.6) arc (-140:70:0.3 and 0.2);
	\draw[->] (4,5.85) arc (-90:170: 0.2 and 0.3);
	\draw[very thick, orange, fill=magenta!30, opacity=0.6] (4.6, 4.7) arc (0:180: 0.6 and 1.5);
	\node at (6.5, 4.7) {$\cdots$};
	\draw[fill=verylightgray, thick] (7, 4) --++ (3, 0) --++ (1, 1.5) --++ (-3, 0) --++ (-1, -1.5);
	\draw[very thick,  aomidori] (8.4, 4.7) --++(1.2,0);
	\draw[->] (9.3,4.6) arc (-140:70:0.3 and 0.2);
	\draw[->] (9,5.85) arc (-90:170: 0.2 and 0.3);
	\draw[very thick, orange, fill=magenta!30, opacity=0.6] (9.6, 4.7) arc (0:180: 0.6 and 1.5);
	\end{tikzpicture}
\caption{The smooth bulk manifold $\CB_n$ whose boundary is the $n$-fold cover $\partial \CB_n = \CM_n$ used for the boundary QFT in the replica trick by gluing $n$ copies of a manifold along the entangling region $A$ (the blue lines).
The entangling surface $\Sigma = \partial A$ is the fixed point of the $\BZ_n$ symmetry acting as the shift $\tau \to \tau + 2\pi$, which extends to the bulk as a codimension-two hypersurface $\gamma_A^{(n)}$ (the orange curves).
}
\label{fig:Bulk_Replica}
\end{figure*}
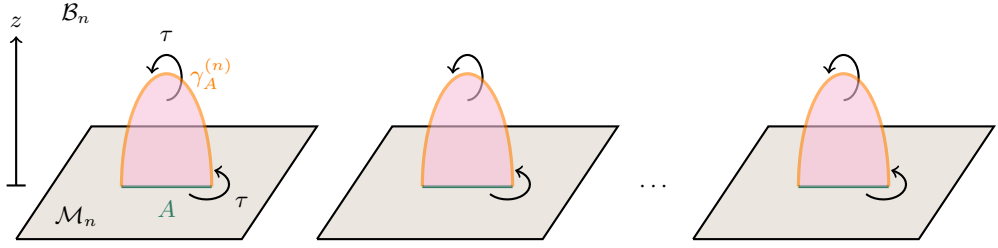

In evaluating the right-hand side of Eq.\,\eqref{HEE_Replica} we need to analytically continue the bulk solution $\CB_n$ from an integer $n$ to a real value.
The boundary condition for $\CB_n$ is by the $n$-fold cover $\CM_n$ that has periodicity $2\pi n$ along the modular time $\tau$.
$\CM_n$ is invariant under the $\BZ_n$ symmetry that shifts $\tau$ by $2\pi$ when $n$ is an integer.
The bulk geometry $\CB_n$ is a \emph{smooth} solution to the Einstein equation derived from the action Eq.\,\eqref{EH_action} with the prescribed boundary $\partial\CB_n = \CM_n$, and 
the modular time $\tau$ naturally extends into the bulk (see Fig.\,\ref{fig:Bulk_Replica}).
In order to guarantee the uniqueness of the analytic continuation of $\CB_n$ to a noninteger $n$, we assume that \emph{the bulk solution $\CB_n$ is invariant under the replica $\BZ_n$ symmetry} that also acts as a shift in $\tau$ by $2\pi$.
The entangling surface $\Sigma$, which is the fixed locus of the $\BZ_n$ action on $\CM_n$, should extend to a codimension-two hypersurface $\gamma_A^{(n)}$ anchored on $\Sigma$ at the boundary of $\CB_n$ as in
Fig.\,\ref{fig:Bulk_Replica}.
After rescaling the modular time by $\hat \tau \equiv \tau/n$, both the bulk and boundary manifolds are periodic in $\hat \tau$ of period $2\pi$.
We emphasize that there is no deficit angle around $\gamma_A^{(n)}$ in the smooth bulk geometry $\CB_n$; the $\hat\tau$ circle with period $2\pi$ shrinks smoothly at $\gamma_A^{(n)}$ due to the regularity of the bulk space $\CB_n$.

The replica $\BZ_n$ symmetry in the bulk allows one to define the orbifold
\begin{align}
	\hat \CB_n \equiv \CB_n / \BZ_n \ ,
\end{align}
which has a conical singularity along the hypersurface $\gamma_A^{(n)}$ with the deficit angle
\begin{align}
	\delta_n  = 2\pi \left(1-\frac{1}{n}\right) \ ,
\end{align}
when measured in the rescaled modular time $\hat \tau$.
While the original bulk solution $\CB_n$ has the singular boundary $\CM_n$,
the boundary of the orbifold $\hat \CB_n$ is regular,
\begin{align}
	\partial \hat \CB_n = \partial \CB = \CM .
\end{align}
We find it convenient to introduce the bulk-per-replica action $\hat I [\hat \CB_n]$ for the orbifold $\hat \CB_n$ by 
\begin{align}\label{Bulk_Per_Replica}
	\hat I [\hat \CB_n] \equiv I_\text{bulk}[\CB_n]/n \ ,
\end{align}
which simplifies Eq.\,\eqref{HEE_Replica} to
\begin{align}\label{ARE_II}
\begin{aligned}
	S_A &= \lim_{n\to 1} \partial_n \left[ n \left( \hat I [\hat \CB_n] - I_\text{bulk} [\CB]\right) \right] \ ,\\
	&= \partial_n \hat I [\hat \CB_n]\big|_{n=1}\ .
\end{aligned}
\end{align}
With this representation of the holographic entanglement entropy, 
we derive the Ryu-Takayanagi formula \eqref{ARE} by showing the bulk-per-replica action $\hat I [\hat \CB_n]$ contains a term proportional to the area of the hypersurface $\gamma_A^{(n)}$.

There is an important difference between the bulk-per-replica action $\hat I [\hat \CB_n]$ and the on-shell action $I_\text{bulk}[\hat \CB_n]$ of the orbifold $\hat \CB_n$.
The former has an additional contribution from the conical singularity at $\gamma_A^{(n)}$ compared to the latter.
To see this explicitly, we adapt the argument in Sec. \ref{ss:Reg_Cone_no_U(1)} to the singular bulk manifold by replacing $\CM_n \to \hat \CB_n$, $\Sigma \to \gamma_A^{(n)}$ and $n \to 1/n$ as the orbifold $\hat \CB_n$ has a deficit angle not $2\pi (1-n)$ but $2\pi(1-1/n)$.
With this in mind in applying Eq.\,\eqref{SingularFormulae1} to the present case, 
the bulk-per-replica action in the Einstein gravity Eq.\,\eqref{EH_action} is shown to be \cite{Dong:2016fnf,Nakaguchi:2016zqi},
\begin{align}\label{BPR_action2}
	\hat I [\hat \CB_n] = I_\text{bulk}[\hat \CB_n] + \frac{\delta_n}{8\pi G_N}\, \CA_n \ ,
\end{align}
where $\CA_n$ is the area of the hypersurface $\gamma_A^{(n)}$
\begin{align}
	\CA_n \equiv \int_{\gamma_A^{(n)}} \d^{d-1} y\,\sqrt{h} \ ,
\end{align}
with the worldvolume coordinates $y^a~(a=1,\cdots, d-1)$ and the induced metric $h_{ab}$.
The right-hand side of Eq.\,\eqref{BPR_action2} may be interpreted as the action of the Einstein gravity coupled to a codimension-two cosmic brane on $\gamma_A^{(n)}$ with tension $\delta_n/(8\pi G_N)$.
We denote collectively by $\Phi$ all the fields such as the metric and matter fields in the action, whose configurations are to be determined by solving the equations of motion $\delta \hat I [\hat \CB_n]/\delta \Phi = 0$ with fixing the replica parameter $n$.
Varying the action Eq.\,\eqref{BPR_action2} with respect to $n$ one finds
\begin{align}\label{BPR_action_Der}
	\partial_n \hat I [\hat \CB_n] = \frac{\delta \Phi}{\delta n} \frac{\delta \hat I [\hat \CB_n]}{\delta \Phi} + \frac{\CA_n}{4G_N n^2} \ .
\end{align}
Then imposing the equations of motion leaves the second term proportional to the area, which results in the Ryu-Takayanagi formula \eqref{ARE} with $\CA = \CA_1$ when plugged into Eq.\,\eqref{ARE_II}.

The minimality of the surface $\gamma_A = \gamma_A^{(1)}$ in Eq.\,\eqref{ARE} is guaranteed in this derivation as follows.
We can use the probe approximation for the cosmic brane $\gamma_A^{(n)}$ in the $n \to 1$ limit where the tension proportional to $\delta_n$ vanishes and fix the position without taking into account the backreaction to the background bulk geometry $\hat \CB_n$.
In other word, the surface $\gamma_A^{(1)}$ is just a solution to the equation of motion of the area functional $\delta \CA_1 = 0$ in the background $\CB$.
If there exist multiple solutions, we pick up a solution with the least area that dominates in the gravity partition function.

\subsubsection{Inequalities satisfied by the holographic formula}\label{ss:HEE_Ineq}
Provided the proof of the holographic formula \eqref{ARE}, the next thing to be done is to check if it satisfies the basic properties of entanglement entropy given in Sec. \ref{ss:prop EE}.
The equality $S_A=S_B$ is somewhat trivial in the holographic picture.
Since the boundary of the region $A$ is equal to the boundary of its compliment $B$, 
the minimal surface $\gamma_A$ is the same as $\gamma_B$ as long as there is no obstruction in the bulk AdS space.\footnote{Such an obstruction shows up at finite temperature as the horizon of an AdS black hole.}

\begin{figure}[htbp]
\centering
	\begin{tikzpicture}[baseline={([yshift=-.5ex]current bounding box.center)}, scale=.8]
			\draw[thick] (0,-1) --++ (0,2.3);
			\draw[thick,aomidori] (0,-0.7) arc (-90:90:0.5 and 0.5);
			\draw[thick,niiro] (0,0) arc (-90:90:0.5 and 0.5);
			\node at (-0.3,0.8) {$A$};
			\node at (-0.3,0.2) {$B$};
			\node at (-0.3,-0.4) {$C$};
		\end{tikzpicture}
		$=$ 
		\begin{tikzpicture}[baseline={([yshift=2ex]current bounding box.center)}, scale=.8]
			\draw[thick] (0,-1) --++ (0,2.3);
			\draw[thick,aomidori] (0,-0.7) arc (-90:45:0.5 and 0.5);
			\draw[thick,aomidori] (0,1) arc (90:-45:0.5 and 0.5);
			\draw[thick,niiro] (0,0) arc (-90:-45:0.5 and 0.5);
			\draw[thick,niiro] (0,0.3) arc (90:45:0.5 and 0.5);
			\node at (-0.3,0.8) {$A$};
			\node at (-0.3,0.2) {$B$};
			\node at (-0.3,-0.4) {$C$};
			\node at (0,-1.5) {$(a)$};
		\end{tikzpicture}
		$\ge$
		\begin{tikzpicture}[baseline={([yshift=-.5ex]current bounding box.center)}, scale=.8]
			\draw[thick] (0,-1) --++ (0,2.3);
			\draw[thick,aomidori] (0,-0.7) arc (-90:90:0.8 and 0.8);
			\draw[thick,niiro] (0,0) arc (-90:90:0.3 and 0.15);
			\node at (-0.3,0.8) {$A$};
			\node at (-0.3,0.2) {$B$};
			\node at (-0.3,-0.4) {$C$};
		\end{tikzpicture}
		~
		\begin{tikzpicture}[baseline={([yshift=-.5ex]current bounding box.center)}, scale=.8]
			\draw[thick] (0,-1) --++ (0,2.3);
			\draw[thick,aomidori] (0,-0.7) arc (-90:90:0.5 and 0.5);
			\draw[thick,niiro] (0,0) arc (-90:90:0.5 and 0.5);
			\node at (-0.3,0.8) {$A$};
			\node at (-0.3,0.2) {$B$};
			\node at (-0.3,-0.4) {$C$};
		\end{tikzpicture}
		$=$
		\begin{tikzpicture}[baseline={([yshift=2ex]current bounding box.center)}, scale=.8]
			\draw[thick] (0,-1) --++ (0,2.3);
			\draw[thick,aomidori] (0,-0.7) arc (-90:45:0.5 and 0.5);
			\draw[thick,niiro] (0,1) arc (90:-45:0.5 and 0.5);
			\draw[thick,aomidori] (0,0) arc (-90:-45:0.5 and 0.5);
			\draw[thick,niiro] (0,0.3) arc (90:45:0.5 and 0.5);
			\node at (-0.3,0.8) {$A$};
			\node at (-0.3,0.2) {$B$};
			\node at (-0.3,-0.4) {$C$};
			\node at (0,-1.5) {$(b)$};
		\end{tikzpicture}
		$\ge$
		\begin{tikzpicture}[baseline={([yshift=-.5ex]current bounding box.center)}, scale=.8]
			\draw[thick] (0,-1) --++ (0,2.3);
			\draw[thick,aomidori] (0,-0.7) arc (-90:90:0.35 and 0.35);
			\draw[thick,niiro] (0,0.3) arc (-90:90:0.35 and 0.35);
			\node at (-0.3,0.8) {$A$};
			\node at (-0.3,0.2) {$B$};
			\node at (-0.3,-0.4) {$C$};
		\end{tikzpicture}
	\caption{The holographic proof of the strong subadditivity. 
	$(a)$ The minimal surfaces $\gamma_{A\cup B}$ and $\gamma_{B\cup C}$ are orange and blue on the leftmost side, which can be reconnected to surfaces $\gamma_B'$ (orange) and $\gamma_{A\cup B\cup C}'$ (blue) bounding the regions $B$ and $A\cup B \cup C$ in the middle.
	The total area is minimized by the minimal surfaces $\gamma_B$ (orange) and $\gamma_{A\cup B\cup C}$ (blue) on the rightmost side, proving the strong subadditivity $S_{A\cup B\cup C} + S_{B} \le S_{A\cup B}+S_{B\cup C} $.
	$(b)$ A similar setup for proving the inequality $S_{A} + S_{C} \le S_{A\cup B}+S_{B\cup C}$.
	}
	\label{fig:SSA}
\end{figure}
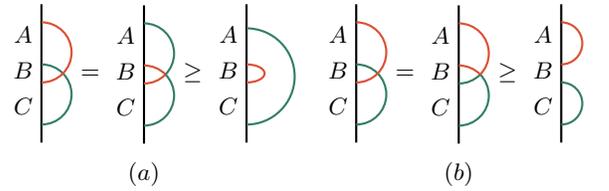

First, consider the strong subadditivity Eq.\,\eqref{SSA}, whose proof relies on an ingenious inequality for Hermitian operators in quantum mechanics.
We show Eq.\,\eqref{SSA} simply follows from the minimality of the Ryu-Takayanagi surface in the holographic formula \cite{Headrick:2007km,Headrick:2013zda}.
For three adjacent subsystems $A,B$ and $C$, the holographic entanglement entropies $S_{A\cup B}$ and $S_{B\cup C}$ are given by the areas of the minimal surfaces $\gamma_{A\cup B}$ and $\gamma_{B\cup C}$ colored in red and blue, respectively, on the leftmost side of Fig.\,\ref{fig:SSA}($a$).
By decomposing each minimal surface into two pieces and reconnecting them appropriately as in the middle of Fig.\,\ref{fig:SSA}($a$), 
we obtain the other set of surfaces $\gamma_B'$ and $\gamma_{A\cup B\cup C}'$ colored in orange and green
whose total area is equal to that of the original minimal surfaces.
Although the new surfaces $\gamma_B'$ and $\gamma_{A\cup B\cup C}'$ are anchored on $B$ and $A\cup B\cup C$ respectively, they are not necessarily the minimal surfaces $\gamma_B$ and $\gamma_{A\cup B\cup C}$ [see the rightmost side of Fig.\,\ref{fig:SSA}($a$)].
Since the minimal surfaces have less areas than the surfaces $\gamma_B'$ and $\gamma_{A\cup B\cup C}'$ in the middle of Fig.\,\ref{fig:SSA}($a$), we obtain the inequality
\begin{align}\label{Hol_SSA_1}
\CA_{A\cup B} + \CA_{B\cup C} \ge \CA_{A\cup B\cup C} + \CA_{B} \ .
\end{align}
This is equivalent to the first line of the strong subadditivity Eq.\,\eqref{SSA}.
The other inequality can be proved similarly by taking a different choice of reconnection of the surfaces as in Fig.\,\ref{fig:SSA}($b$).

\begin{figure}[htbp]
\centering
		\begin{tikzpicture}[baseline={([yshift=-.5ex]current bounding box.center)}]
			\draw[thick] (0,-1) --++ (0,3);
			\draw[thick,koinezu] (0,-0.8) arc (-90:90:1 and 1.3);
			\draw[thick,niiro] (0,0.2) arc (-90:90:0.5 and 0.8);
			\draw[thick,koinezu] (0,0.2) arc (-90:90:0.2 and 0.2);
			\draw[thick,aomidori] (0,-0.8) arc (-90:90:0.5 and 0.7);
			\node at (-0.3,1.2) {$A$};
			\node at (-0.3,0.4) {$B$};
			\node at (-0.3,-0.4) {$C$};
		\end{tikzpicture}
		$=$ 
		\begin{tikzpicture}[baseline={([yshift=-.5ex]current bounding box.center)}]
			\draw[thick] (0,-1) --++ (0,3);
			\draw[thick,koinezu] (0,-0.8) arc (-90:90:1 and 1.3);
			\draw[thick,niiro] (0,1.8) arc (90:-50:0.5 and 0.8);
			\draw[thick,aomidori] (0,0.2) arc (-90:-50:0.5 and 0.8);
			\draw[thick,koinezu] (0,0.2) arc (-90:90:0.2 and 0.2);
			\draw[thick,aomidori] (0,-0.8) arc (-90:50:0.5 and 0.7);
			\draw[thick,niiro] (0,0.6) arc (90:50:0.5 and 0.7);
			\node at (-0.3,1.2) {$A$};
			\node at (-0.3,0.4) {$B$};
			\node at (-0.3,-0.4) {$C$};
		\end{tikzpicture}
		$\ge$
		\begin{tikzpicture}[baseline={([yshift=-.5ex]current bounding box.center)}]
			\draw[thick] (0,-1) --++ (0,3);
			\draw[thick,koinezu] (0,-0.8) arc (-90:90:1 and 1.3);
			\draw[thick,niiro] (0,0.6) arc (-90:90:0.35 and 0.6);
			\draw[thick,koinezu] (0,0.2) arc (-90:90:0.2 and 0.2);
			\draw[thick,aomidori] (0,-0.8) arc (-90:90:0.35 and 0.5);
			\node at (-0.3,1.2) {$A$};
			\node at (-0.3,0.4) {$B$};
			\node at (-0.3,-0.4) {$C$};
		\end{tikzpicture}
	\caption{A schematic proof of the monogamy of mutual information.}
	\label{fig:MMI}
\end{figure}
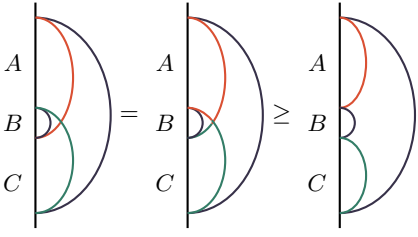

In the holographic setup, one can prove another nontrivial inequality \cite{Hayden:2011ag},
\begin{align}\label{MMI}
	S_{A\cup B} + S_{B\cup C} + S_{C\cup A} \ge S_A + S_B + S_C + S_{A\cup B\cup C} \ .
\end{align}
It takes a simpler form when written in the mutual information, 
\begin{align}
	I(A, B+C) \ge I(A, B) + I(A,C) \ ,
\end{align}
which is called a \emph{monogamy of mutual information}.
The proof proceeds in a very similar way to the strong subadditivity as in Fig.\,\ref{fig:MMI}.\footnote{One should not resort to the pictorial proof too much as the derived inequalities may fail \cite{Hayden:2011ag}.}

Actually there are an infinite set of entanglement inequalities generalizing the monogamy of mutual information that must be satisfied for a holographic system \cite{Bao:2015bfa}.

An important caution is that the inequality \eqref{MMI} holds in any holographic system, but does not in general quantum systems.
For example, a three qubit state with the density matrix
\begin{align}
	\rho_{A\cup B\cup C} = \frac{1}{2}\left( |000\rangle \langle 000 | + |111\rangle \langle 111|\right) 
\end{align}
violates the monogamy inequality because the entropies in the subsystems are \cite{Hayden:2011ag}
\begin{align}
\begin{aligned}
	&S_{A\cup  B\cup C}\\
		 &\quad = S_{A\cup B} = S_{B\cup C} = S_{C\cup A} \ ,\\
		 &\quad = S_A = S_B = S_C\ ,\\
		 &\quad = \log 2 \ .
\end{aligned}
\end{align}
Turning it around, the monogamy of mutual information is requisite for a quantum system having a holographic description in gravity theory on an asymptotically AdS space.

Although these geometric proofs are easy to understand, they are not enough to take account of subtle situations such that the subregions share their boundaries in an intricate way.
The rigorous proofs in more general settings in the Einstein gravity can be found in \textcite{Headrick:2013zda} for a time-independent case where only the minimality condition of the Ryu-Takayanagi surface is assumed with the null energy condition to exclude unphysical bulk configurations.\footnote{For a time-dependent holographic system where entanglement entropy is given by the Hubeny-Rangamani-Takayanagi prescription \cite{Hubeny:2007xt}, the null curvature condition $\CR_{\mu\nu}\xi^\mu\xi^\nu\ge 0,~\xi^2 = 0$ is shown to be sufficient for the entanglement inequalities to hold \cite{Wall:2012uf}.
In general higher derivative gravity theories, however, there are no known proofs for the inequalities as the null curvature condition would be invalidated and a minimization procedure of the surface may fail \cite{Wall:2012uf,Dong:2013qoa}.}

\subsection{Spherical entangling surface}
\begin{figure}[htbp]
\centering
	\begin{tikzpicture}[thick, scale=.8, every node/.style={scale=1}]
			   \tikzstyle{ann} = [fill=white,inner sep=2pt]
			   \draw (-1,-1.2)--(1,0.3)--(1,4.3)--(-1,2.8)--(-1,-1.2);
			   \draw[orange,fill=orange!20, opacity=0.6] (0,0) arc (-90:90:2.2 and 1.5);
			   \draw[dashed,aomidori] (0,0) arc (-90:90:0.5 and 1.5);
			   \draw[aomidori, fill=orange!20, opacity=0.6] (0,0) arc (270:90:0.5 and 1.5);
			   \draw[->] (0,1.5)--++(120:1.1) node[left] {$\rho$};
			   \draw[|->] (0,-0.3) -- (3,-0.3) node[right] {$z$};
			   \draw (0,-0.7) node[ann] {$z=\epsilon$};
			   \draw (2.6,1.7) node {\textcolor{orange}{$\gamma_A$}};
		\end{tikzpicture}
\quad
	\begin{tikzpicture}[thick, scale=.9, every node/.style={scale=.8}]
	\draw[->] (0,0)--(3,0) node[right] {\large $\rho$};
	\draw[->] (0,0)--(0,3) node[above] {\large $z$};
	\draw[ultra thick, orange] (2,0) node[below, black] {\large $R$} arc (0:90:2) node[left, black] {\large $R$};
	\node at (-0.2,-0.2) {\large 0};
	\end{tikzpicture}
\caption{The minimal surface of a spherical entangling surface in the Poincar{\'e} coordinates of AdS$_{d+1}$.}
\label{fig:Sphere}
\end{figure}
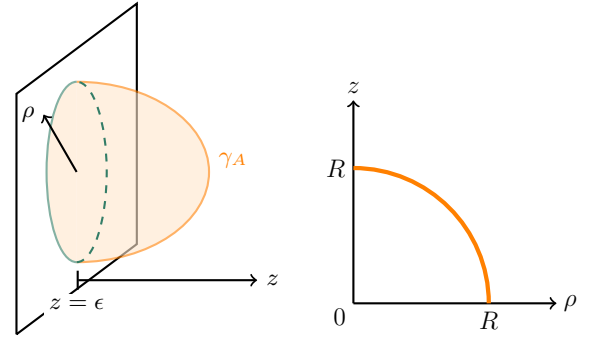

We illustrate the calculation of the holographic entanglement entropy for
the entanglement entropy across
a sphere $\BS^{d-2}$ of radius $R$ located at $\rho=R$ in the polar coordinates.
It is easiest to work in the (Euclidean) Poincar{\'e} AdS$_{d+1}$ space,
\begin{align}\label{SpherePoincare}
\d s^2 = L^2\, \frac{\d z^2 + \d t^2 + \d\rho^2 + \rho^2\, \d\Omega_{d-2}^2}{z^2} \ .
\end{align}
We let $z=z(\rho)$ to respect the spherical symmetry of the entangling surface (see Fig.\,\ref{fig:Sphere}).

The minimal surface $\gamma_A$ is the solution of the variational problem for the area functional with respect to $z=z(\rho)$,
\begin{align}\label{areaSp}
\CA = L^{d-1} \text{Vol}(\BS^{d-2}) \int_0^R \d\rho\, \frac{\rho^{d-2}}{z^{d-1}(\rho)} \sqrt{1+ (\partial_\rho z)^2} \ .
\end{align}
Solving the equation of motion with the boundary condition $\rho=R$ at the boundary $z=0$,\footnote{The equation of motion is a second-order differential equation, but can be reduced to a first-order one \cite{Bakhmatov:2017ihw,Colgain:2018tzo}.}
the solution turns out to be a hemisphere in any $d$ dimensions \cite{Ryu:2006bv} (see Fig.\,\ref{fig:Sphere})
\begin{align}\label{Sp_sol}
z(\rho) = \sqrt{R^2 - \rho^2} \ .
\end{align}
It follows that the holographic entanglement entropy of a sphere of radius $R$ is
\begin{align}\label{EE_Sphere}
\begin{aligned}
S_A = \frac{L^{d-1}\text{Vol}(\BS^{d-2})}{4G_N} & \int_{\epsilon/R}^1 \d y\, \frac{(1-y^2)^{(d-3)/2}}{y^{d-1}} \ , 
\end{aligned}
\end{align}
where we introduced the UV cutoff $\epsilon \ll 1$ at $z=\epsilon$.

Expanding the integral around $y=0$ and performing the integration near $y=\epsilon /R$, one finds the UV divergent terms consistent with the generic structure in Eq.\,\eqref{EE_Even} for even $d$ and in Eq.\,\eqref{EE_Odd} for odd $d$.
The leading divergent part is proportional to the area of the entangling surface: 
\begin{align}
	\text{Area}(\partial A) = R^{d-2}\,\text{Vol}(\BS^{d-2}) \ ,
\end{align}
which is known to be the area law of entanglement.

\subsubsection{Universal terms}

There is the logarithmic divergence of the entanglement entropy for even $d$, whose coefficient $c_0$ is proportional to the type $A$ anomaly for a spherical entangling surface [see Eq.\,\eqref{Log_EE_C0}],
\begin{align}
	c_0 = (-1)^{d/2+1} A \ .
\end{align}
Compared with Eq.\,\eqref{EE_Sphere}, the type $A$ central charge in a large-$N$ CFT$_d$, described by the Einstein gravity on the AdS$_{d+1}$ spacetime, is determined to be \cite{Myers:2010xs,Myers:2010tj}
\begin{align}
	A = \frac{L^{d-2}}{2G_N} \frac{\pi^{d/2 -1}}{\Gamma (d/2)} \ .
\end{align}

The universal finite term $F$ for odd $d$ in Eq.\,\eqref{EE_Odd} is similarly read off from Eq.\,\eqref{EE_Sphere}.
By analytically continuing $d$ and letting $\epsilon$ be zero before carrying out the integral, we find \cite{Kawano:2014moa}
\begin{align}\label{CFT_F_value}
F = \frac{L^{d-1}}{4G_N} \frac{\pi^{d/2}}{\Gamma(d/2)} \ .
\end{align}
This is always positive thanks to the sign factor put in front of $F$ in Eq.\,\eqref{EE_Odd}.
This universal term is conjectured to decrease monotonically under any renormalization group flow \cite{Myers:2010xs,Myers:2010tj,Klebanov:2011gs}, and known as the holographic $\CC$-theorem \cite{Girardello:1998pd,Freedman:1999gp,Myers:2010xs,Myers:2010tj} that will be discussed in Sec. \ref{ss:RGflow}.

\subsubsection{Relation to thermal entropy on $\BH^{d-1}$}\label{ss:HEE_thermal}
We revisit Eq.\,\eqref{RenyiAndTherm} between the entanglement entropy across a sphere and the thermal entropy on a hyperbolic space derived in Sec. \ref{ss:Rel_to_Therm} from the holographic point of view.
We show, in the hyperbolic coordinates Eq.\,\eqref{AdS_Hyp} of the AdS space, that the minimal surface coincides with the black hole horizon of the AdS topological black hole, and thus the entanglement entropy agrees with the thermal entropy of the black hole.

\begin{figure}[htbp]
\centering
\begin{tikzpicture}[thick]
	\draw[ultra thick, cyan!70, fill=blue!12] (0,2) arc (90:270:2);
	\draw (0,-2) arc (-90:90:2);
	\draw[ultra thick, orange] (0,-2)--(0,2);
	\draw[orange!50] (0,-2) to [out=110, in=250] (0,2);
	\draw[orange!50] (0,-2) to [out=160, in=200] (0,2);
	\node[orange] at (0.7,-1) {$r=R$};
	\node[orange, fill=white] at (2,-0.5) {\large minimal surface};
	\node[cyan!60!blue] at (-2.5,-1) {$r=\infty$};
	\node[cyan!60!blue] at (-3.2,-0.5) {\large $\BS^1 \times \BH^{d-1}$};
	\node[fill=white] at (0,-2.5) {\large Global AdS$_{d+1}$};
\end{tikzpicture}

\caption{The global AdS space obtained by the conformal map Eq.\,\eqref{BulkCM} from the Poincar{\'e} coordinates. The outside of the AdS topological black hole horizon covers half of the global coordinates. The minimal surface (the thick orange curve) in Fig.\,\ref{fig:Sphere} is on the horizon of the AdS topological black hole.
}
\label{fig:TopAdS}
\end{figure}
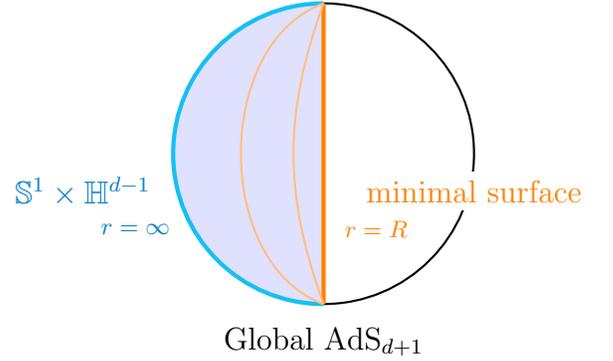

The Euclidean Poincar{\'e} coordinates Eq.\,\eqref{SpherePoincare} can be mapped by the coordinate transformations
\begin{align}\label{BulkCM}
\begin{aligned}
z &= R\,\frac{1}{r\cosh u + \sqrt{r^2 - 1} \cos \tau} \ , \\
t &= R\, \frac{\sqrt{r^2 -1} \sin(\tau)}{r\cosh u + \sqrt{r^2 -1} \cos \tau} \ , \\
\rho &= R\, \frac{ r \sinh u}{r\cosh u + \sqrt{r^2 -1} \cos \tau} \ ,
\end{aligned}
\end{align}
to the new coordinates 
\begin{align}\label{AdS_Hyp2}
\begin{aligned}
\d s^2 = L^2 \bigg[& \left( r^2 -1\right) \d\tau^2 + \frac{\d r^2}{r^2 -1} \\
	&\qquad + r^2(\d u^2 + \sinh^2 u \,\d\Omega_{d-2}^2)\bigg] \ .
\end{aligned}
\end{align}
This is the Euclidean AdS topological black hole given by Eq.\,\eqref{AdS_Hyp} whose boundary is $\BS^1\times \BH^{d-1}$ (see Fig.\,\ref{fig:TopAdS}).
The $\tau$ coordinate has period $\tau\sim \tau + 1/T$ with the temperature $T=1/(2\pi)$ fixed to avoid the conical singularity at $r=0$.

The entangling surface located at $\rho = R, t=0, z=0$ in the original coordinates Eq.\,\eqref{SpherePoincare} is conformally mapped to a hypersurface at $r=\infty, \tau=0$ and $u=\infty$ in the hyperbolic coordinates Eq.\,\eqref{AdS_Hyp} under the transformations Eq.\,\eqref{BulkCM}.
The minimal surface in the latter coordinates anchored on constant time $\tau=0$ and $r=\infty, u=\infty$ is obtained by extremizing the area functional with respect to $r=r(u)$,
\begin{align}
\begin{aligned}
\CA = L^{d-1} \int_0^\infty \d u\, (r \sinh u)^{d-2} \sqrt{r^2 + \frac{(\partial_u r)^2}{r^2 -1}} \ .
\end{aligned}
\end{align}
It is easy to check $r=1$ is the solution to the equation of motion, which
is nothing but the horizon of the topological AdS black hole.
We thus find the holographic entanglement entropy is equal to the black hole entropy \cite{Casini:2011kv},
\begin{align}
S_A(R) = S_\text{BH}(T) = S_\text{therm}(T)\ ,
\end{align}
where we used one more relation between the AdS$_{d+1}$ black hole entropy $S_\text{BH}$ and the thermal entropy $S_\text{therm}$ of the dual CFT$_d$ on $\BH^{d-1}$ at temperature $T$.
This completes the holographic derivation of Eq.\,\eqref{RenyiAndTherm}.

\subsection{Holographic R{\'e}nyi entropy}\label{ss:HRE}
Before concluding this section, we make a few comments on the holographic calculation of the R{\'e}nyi entropy.
We first derive the holographic formula of the modular entropy in a similar manner to the holographic entanglement entropy in Sec. \ref{ss:HEE} and show it is given by the area of a cosmic brane.
The R{\'e}nyi entropy is constructed out of the modular entropy by Eq.\,\eqref{Renyi_Modular}.
As an example, we calculate the holographic R{\'e}nyi entropy across a spherical entangling surface in CFT .

\subsubsection{A derivation of the holographic formula}
The argument for deriving the Ryu-Takayanagi formula in Sec. \ref{ss:HEE_Derivation} has more implications for the holographic description of quantum entanglement than it appears.
The modular entropy Eq.\,\eqref{ModularEntropy} written in terms of the partition function
\begin{align}
	\tilde S_n &= (1 - n\partial_n) (\log Z_n - n\log Z)
\end{align}
can be turned, with the GKP-W relation and the bulk-per-replica action Eq.\,\eqref{Bulk_Per_Replica}, into
\begin{align}
	\begin{aligned}
		\tilde S_n 
		&= (n\partial_n -1 ) \left[ n (\hat I [\hat \CB_n ] - I_\text{bulk} [ \CB ] )\right] \  , \\
		&=  n^2 \partial_n \hat I [\hat \CB_n] \ .
	\end{aligned}
\end{align}
This reminds us of the holographic definition of entanglement entropy Eq.\,\eqref{ARE_II}, and one can recycle the resulting relation \eqref{BPR_action_Der} to find \emph{the holographic formula of the modular entropy} \cite{Dong:2016fnf},
\begin{align}\label{HRE_formula}
	\tilde S_n = \frac{\CA_n}{4G_N}\bigg|_{\delta \hat I = 0, \,\partial \gamma^{(n)}_A = \Sigma} \ .
\end{align}
It resembles the Ryu-Takayanagi formula \eqref{ARE} and actually reproduces it when $n=1$.
This formula, however, contains the area of a codimension-two \emph{cosmic brane} $\gamma_A^{(n)}$ with tension $\delta_n/(8\pi G_N)$ that backreacts to the bulk geometry.
Hence it is more intricate in practice than the case of the Ryu-Takayanagi formula \eqref{ARE} as we have to fix the location of the cosmic brane by solving the equations of motion of the action $\hat I$ describing the Einstein gravity coupled to the cosmic brane and possibly matter fields as indicated by the subscript $\delta \hat I = 0$ in Eq.\,\eqref{HRE_formula} [see Eq.\,\eqref{BPR_action_Der} for the definition of $\hat I$].
We are also able to build the \emph{holographic R{\'e}nyi entropy} by combining Eq.\,\eqref{HRE_formula} with the relation \eqref{Renyi_Modular}.

Given the holographic formula of the R{\'e}nyi entropy, one may ask whether it satisfies the inequalities Eqs.\,\eqref{RenyiIneq1}-\eqref{RenyiIneq4} characterizing the R{\'e}nyi entropy.
To this end it is enough to check the three inequalities Eq.\,\eqref{Modular_Ineq} implying the non-negativities of the modular energy, entropy and capacity.
The easiest one to show is the inequality $\tilde S_n \ge 0$ that follows from the non-negativity of the area of a cosmic brane in Eq.\,\eqref{HRE_formula}.
The non-negativity of the modular energy $E\ge 0$ becomes clear by expressing $E$ in the form
\begin{align}
	E  = I_\text{bulk}[\hat \CB_n] - I_\text{bulk}[\CB] + \frac{\CA_n}{4G_N} \ ,
\end{align}
and invoking $\CB$ the on-shell solution with the least action while $\hat \CB_n$ is not necessarily so.
The last inequality $C\ge 0$ for the modular capacity is most nontrivial and turns out to require the stability of the bulk geometry $\CB_n$ \cite{Nakaguchi:2016zqi}.
It remains an open issue whether and when such a stability condition is guaranteed to hold in the holographic setup.

\subsubsection{Spherical entangling surface in CFT}

We have seen in Sec. \ref{ss:HEE_thermal} that the entanglement entropy across a spherical entangling surface in CFT is equal to the thermal entropy on $\BH^{d-1}$ at inverse temperature $2\pi R$ and can be calculated holographically as the black hole entropy in the dual AdS spacetime.
Here we see the holographic formula of the modular entropy \eqref{HRE_formula} also has an elegant interpretation as a black hole entropy when the entangling surface is spherical in accordance with the CFT story in Sec. \ref{ss:Renyi_CFT} \cite{Hung:2011nu}.

The AdS topological black hole given by Eq.\,\eqref{AdS_Hyp2} is a particular case of the following AdS black hole,
\begin{align}\label{AdS_Hyp_Renyi}
\d s^2 = L^2\left[ f(r)\, \d\tau^2 + \frac{\d r^2}{f(r)}+ r^2(\d u^2 + \sinh^2 u \,\d\Omega_{d-2}^2)\right] \ ,
\end{align}
where the function $f(r)$ vanishes at $r = r_H$,
\begin{align}
	f(r) =  r^2 -1 - \frac{r_H^{d-2}}{r^{d-2}}\left( r_H^2 - 1\right) \ .
\end{align}
This black hole has a temperature parametrized by $r_H$,
\begin{align}\label{RenyiBHTemp}
	T(r_H) = \frac{1}{4\pi} \left[ d\, r_H - \frac{d-2}{r_H} \right] \ ,
\end{align}
and the dual CFT lives on the boundary that is a hyperbolic space $\BH^{d-1}$ at finite temperature $T(r_H)$.

In this setup, the codimension-two cosmic brane in \eqref{HRE_formula} is at the black hole horizon and the modular entropy can be identified with the thermal entropy by matching the black hole temperature Eq.\,\eqref{RenyiBHTemp} with $T_n = 1/(2\pi n)$.
By setting $r_H$ to
\begin{align}
	r_n = \frac{1+\sqrt{d\,n^2(d - 2) + 1}}{d\, n} \ ,
\end{align}
the black hole entropy at temperature $T_n$ becomes 
\begin{align}
	\tilde S_n = S_\text{therm}\left(T_n\right) = \frac{\text{Vol}(\BH^{d-1})L^{d-1}}{4G_N} r_n^{d-1} \ .
\end{align}
Performing the integration in Eq.\,\eqref{Renyi_Modular} or Eq.\,\eqref{RenyiTherm} leads to the R{\'e}nyi entropy,
\begin{align}
\begin{aligned}
	S_n 	&= \frac{\text{Vol}(\BH^{d-1})L^{d-1}}{4G_N}\frac{n}{2(n-1)}\left(2 - r_n^{d-2} - r_n^d\right) \ .
\end{aligned}
\end{align}

  \section{Renormalization group flows}\label{ss:RGflow}

In this section, we turn our attention to the dynamical aspects of entanglement entropy under an RG flow.
We first overview the idea of RG flows as a course graining of microscopic degrees of freedom and detail the motivation and implication of the so-called $\CC$-theorem realizing our intuition that the effective degrees of freedom monotonically decrease along RG flows in the theory space of QFT.
The current status of the $\CC$-theorems in various dimensions will be recapitulated for the readers' convenience.
We then proceed to the proofs for the entropic version of the $c$-theorem in two dimensions and the $F$-theorem in three dimensions.
The monotonicities of $\CC$-functions built from entanglement entropy will be shown as a consequence of the strong subadditivity and Lorentz invariance of QFT.
As a specific example, we consider a free massive scalar field and examine the $F$-theorem by introducing a systematic method of the large mass expansion.
After sketching the attempts to formulate the $\CC$-theorems in higher dimensions, we move onto the holographic descriptions of entanglement under RG flows and reveal the role of entanglement entropy as an order parameter of phase transitions.
We also make a small test of the $F$-theorem in holographic models of RG flows and conclude the section with comments on some exact results for entanglement in supersymmetric field theories.

\subsection{Ordering theories along RG flows}\label{ss:OrderingRG}
In the Wilsonian picture of QFT, the renormalization group transformation takes one theory  to another effective theory by coarse graining the microscopic degrees of freedom heavier than the energy scale of our interest, resulting in a trajectory called an RG flow in the space of QFTs parametrized by coupling constants \cite{Wilson:1973jj,Polchinski:1983gv}.

To be more concrete, let us denote the theory space of QFTs by $\CT$ whose coordinates are given by a set of coupling constants $\{ g_i\}$, and the RG transformation by $\mathfrak{R}_t$ that induces a flow from one theory $T \in\CT$ to another theory $\mathfrak{R}_t\, T$.
The parameter $t \ge 0$ counts how many times the coarse-graining is processed, hence playing a role of time for RG flows.
It will be convenient to relate the RG time $t$ with the energy scale $\mu$ by $t = - \log (\mu/\Lambda)$ where $\Lambda$ is the UV cutoff.
One can follow the trajectory of the RG flow of a given theory $T$ by changing $t$ from the UV ($t=0$) to the IR limit ($\infty$).

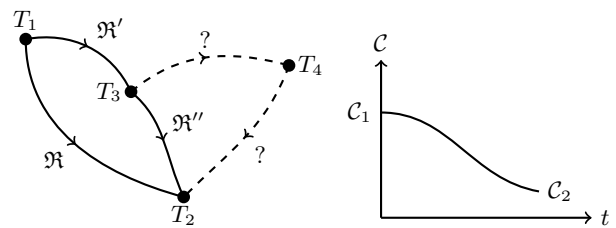
\begin{figure}[ht]
\centering
		\begin{tikzpicture}[scale=0.7, thick,decoration={
    markings,
    mark=at position 0.5 with {\arrow{>}}}]
		\draw[postaction={decorate}] (0,3)  to [out = -90, in = 165] node[below left] {$\mathfrak R$} (3,0);
		\draw[postaction={decorate}] (0,3)  to [out = 10, in = 115] node[above right] {$\mathfrak R'$} (2,2);
		\draw[postaction={decorate}] (2,2)  to [out = -40, in = 115] node[above right] {$\mathfrak R''$} (3,0);
		\draw[postaction={decorate},dashed] (2,2)  to [out = 40, in = 165] node[above] {$?$}(5,2.5);
		\draw[postaction={decorate},dashed] (5,2.5)  to [out = 250, in = 45] node[below right] {$?$}(3,0);
		\draw[fill=black] (0,3) node[above] {$T_1$} circle (0.1);
		\draw[fill=black] (2,2) node[left] {$T_3$} circle (0.1);
		\draw[fill=black] (5,2.5) node[right] {$T_4$} circle (0.1);
		\draw[fill=black] (3,0) node[below] {$T_2$} circle (0.1);
	\end{tikzpicture}
	\begin{tikzpicture}[thick,scale=0.7]
		\draw[->] (0,0) -- (4,0) node[right] {$t$};
		\draw[->] (0,0) -- (0,3) node[above] {$\CC$};
		\draw (0,2) node[left] {$\CC_\text{1}$} to [out=0, in=170] (3,0.5) node[right] {$\CC_\text{2}$};
	\end{tikzpicture}
\caption{[Left] An example of RG flows in the theory space $\CT$. The black dots stand for the fixed points of RG flows.
[Right] A $\CC$-function monotonically decreases as the RG time $t$ is increased.
}
\label{fig:c-function}
\end{figure}

The fixed points of RG flows are by definition scale invariant field theories, which are  expected to be CFTs under the assumptions of unitarity and Poincar{\'e} invariance \cite{Nakayama:2013is}.
We thus associate to an RG flow the UV and IR CFTs denoted by $T_\text{UV}$ and $T_\text{IR}$ at $t = 0$ and $t=\infty$ respectively.
There can be multiple RG flows attached to one fixed point depending on the types of perturbation added to the corresponding CFT.
Figure \ref{fig:c-function} illustrates a situation of the theory space with four fixed points $T_1,T_2,T_3$, and $T_4$ represented by the black dots, some of which are connected by RG flows in a nontrivial way.

Intuitively, physical degrees of freedom must decrease monotonically under any RG flow because massive degrees of freedom are integrated out once the energy scale of the flow becomes below the scale set by the masses.
In other words, we hope to find a function $\CC: \CT \to \BR_{\ge 0}$ that quantifies the effective degrees of freedom of a given theory $T\in \CT$ by a non-negative real number and decreases monotonically along any RG flow on $\CT$:\footnote{The number of degrees of freedom in QFT is conventionally measured by a $\CC$-function in unit of a simplest quantum field such as a free scalar theory.}
\begin{align}\label{c_monotonic_decreasing}
	\frac{\d\, \CC (\mathfrak{R}_t\, T)}{\d t} \le 0 \ .
\end{align}
This type of a function, broadly termed a \emph{$\CC$-function}, brings us a natural interpretation that every RG flow goes downward when its value is regarded as a height on the theory space $\CT$ (Fig.\,\ref{fig:c-function}).
In particular, the UV fixed point is higher than (or at the same altitude as) the IR fixed point:
\begin{align}\label{c_monotonic}
	\CC_\text{UV} \ge \CC_\text{IR} \ ,
\end{align}
where $\CC_i \equiv \CC (T_i)$ are the fixed point values.
As we will see, the fixed point value of a $\CC$-function is often calculable as a scheme-independent quantity in the corresponding CFT without any information around the point.
If this is the case, the inequality \eqref{c_monotonic} constrains the dynamics of RG flows as follows.
Suppose there are two CFTs $T_3$ and $T_4$ in the theory space as in Fig.\,\ref{fig:c-function}, and we are interested in whether there exists an RG flow interpolating them.
In principle, we could perturb one of the theories, e.g., $T_3$, by relevant operators and see if it triggers an RG flow toward the other $T_4$ by perturbative calculations.
On the other hand, if the fixed point values $\CC_3$ and $\CC_4$ are available to us and it turns out that $\CC_3 < \CC_4$, then we can conclude, without any perturbative calculation, that there are no RG flows from $T_3$ to $T_4$ just by resorting to the inequality \eqref{c_monotonic}.

We are thus able to characterize a $\CC$-function by the following conditions:
\begin{itemize}
\item $\CC$ is a function of a renormalization scale $\mu = \Lambda\, e^{-t}$ that we can tune by hand.
\item $\CC$ is well-defined for any QFT, independent of the types of field contents and their interactions.
\item $\CC$ is a monotonic function of the renormalization time $t$ for any unitary and Lorentz invariant QFT in one of the following senses \cite{Barnes:2004jj,Gukov:2015qea}:
	\begin{itemize}
	\item[(i)] [Weak version] The fixed point value of $\CC$ should decrease along RG flows as in Eq.\, \eqref{c_monotonic}.
	\item[(ii)] [Stronger version] $\CC$ monotonically decreases along the entire RG flow as in Eq.\,\eqref{c_monotonic_decreasing}.
	\item[(iii)] [Strongest version] $\CC$ is a potential that generates RG flows:
		\begin{align}
			\beta_i (g) = G_{ij}(g)\,\frac{\partial \CC}{\partial g_j} \ ,
		\end{align}
		where $\beta_i (g) \equiv - \d g_i /\d t$ and $G_{ij}(g)$ is a positive-definite metric on the theory space $\CT$.
	\end{itemize}
\end{itemize}
Here we distinguish the three versions of a $\CC$-function in increasing order of strength for completeness.
The weak version follows from the stronger one by integrating Eq.\,\eqref{c_monotonic_decreasing} along the flow.
The stronger version is implied by the strongest one too,
\begin{align}
	\frac{\d\,\CC}{\d t} = - \beta_i \,\frac{\partial\,\CC}{\partial g_i} = - G_{ij}\,\frac{\partial\,\CC}{\partial g_i} \frac{\partial\,\CC}{\partial g_j}  \le 0 \ .
\end{align}

While the existence of a $\CC$-function may be taken for granted from the physical point of view, the problem is how to construct such a well-behaved function satisfying the aforementioned properties.
Indeed the construction of a $\CC$-function for QFT in dimensions higher than $2$ has been a long-standing problem for three decades since the initiating work by \textcite{Zamolodchikov:1986gt}, and was successfully settled down recently in only three and four dimensions.

One possible approach to this problem is to formulate a set of axioms that guarantee the existence of a $\CC$-function in any dimension without any explicit construction.
The power of an axiomatic approach is well demonstrated in axiomatic thermodynamics whose most important consequence is the very existence of a unique entropy function that never decreases under adiabatic processes and characterizes thermal equilibrium states when represented as a function of extensive variables such as energy and volume [see, e.g., \textcite{Lieb:1997mi} for a review of the subject].
In fact, there are considerable similarities between the theory space of QFT and the state space of thermodynamics.
For instance, the irreversibility of an RG flow between two fixed points $T_a,T_b \in \CT$ corresponds to an irreversible adiabatic process $a \to b$ that takes one equilibrium state $a$ to the other $b$.
A partial list of the analogies between thermodynamics and QFT is shown in Table\,\ref{tab:Analogy}.
It would be of great interest to translate the set of axioms for equilibrium thermodynamics into the language of QFT and examine to what extent the formulation of an entropy function carries over to the construction of a $\CC$-function.
We will not address this possibility any further in this review and leave it to future investigations.
Instead, we will take a constructive approach for a $\CC$-function in the subsequent sections.

\begin{table}
\caption{\label{tab:Analogy}An analogy between thermodynamics and QFT.}
\begin{ruledtabular}
\begin{tabular}{cc}
	Thermodynamics & QFT \\
	\hline
	Equilibrium states  & Conformal fixed points\\
	$a,b$ & $T_a, T_b$ \\
	Union of two states & Coupling of two theories\\
	$a+b$ & $T_a$ and $T_b$ \\
	Irreversible adiabatic process & RG flow \\
		$a\rightarrow b$ & from $T_a$ to $T_b$\\
	Entropy function & $\CC$-function
\end{tabular}
\end{ruledtabular}
\end{table}

\subsection{List of $\CC$-theorems}
We now provide a list of the known $\CC$-theorems and related conjectures in various dimensions along with comments on their historical backgrounds.

\subsubsection{Two dimensions}
\textcite{Zamolodchikov:1986gt} proved his celebrated $c$-theorem in the following form.

\begin{theorem*}[Zamolodchikov's $c$-theorem]
In two-dimensional renormalizable QFTs, there exists Zamolodchikov's $c$-function $c(g_i, \mu)$ that depends on a set of dimensionless coupling constants $\{ g_i\}$ and the energy scale $\mu (= \Lambda\,e^{-t})$, satisfying the following properties:
\begin{enumerate}
\item It takes the same value as the central charge $c$ of the CFT corresponding to each fixed point of RG flows:
	\begin{align}
		c(g_i, \mu)|_\mathrm{CFT} = c \ .
	\end{align}
\item It monotonically decreases along any RG flow:
	\begin{align}
		\frac{\d\, c(g_i, \mu)}{\d t} = - \mu \frac{\d\, c(g_i, \mu)}{\d\mu} \le 0 \ .
	\end{align}
\item It is stationary only at the fixed points:
	\begin{align}
		\frac{\partial\, c(g_i, \mu)}{\partial g_i}\bigg|_\mathrm{CFT} = 0 \ .
	\end{align}
\end{enumerate}
\end{theorem*}
Hence, Zamolodchikov's $c$-theorem falls into the strongest version in the terminology introduced in the previous section \cite{Friedan:2009ik}.
The Zamolodchikov's $c$-function can be built explicitly from the two-point correlation functions of the stress-energy tensor in two dimensions, and the monotonicity follows from the reflection positivity of the correlator.

\subsubsection{Three dimensions}

Given the fact that Zamolodchikov's $c$-function gives the central charge at the fixed points, it is tempting to look for counterparts of the two-dimensional central charge in $d>2$ dimensions.
Let us consider a few candidates for central charges that can be defined irrespective of dimensionality.
\begin{enumerate}
\item[$(a)$]
A ``thermal central charge" $C_\text{Therm}$ as a coefficient of the thermal free energy density $f_\text{Therm}$ at finite temperature $T$ defined by
\begin{align}\label{Thermal_C}
	f_\text{Therm} \sim C_\text{Therm}\, T^d  \ .
\end{align}
\item[$(b)$]
The coefficient $C_T$ for the two-point function of the stress-energy tensor in Eq.\,\eqref{TTCor}.
\item[$(c)$]
The central charges $A$, $B_i$ for the conformal anomalies defined by Eq.\,\eqref{WeylAnomaly} that exist only in even $d$ dimensions.
\end{enumerate}
In $d=2$ dimensions, all of these are known to be proportional to the central charge $C_\text{Therm} \propto C_T \propto A \propto c$ for CFT, but they do not generally have linear relations in higher dimensions.

The possibility $(a)$ of adopting the thermal central charge $C_\text{Therm}$ as a measure of degrees of freedom was studied by \textcite{Sachdev:1993pr} for the three-dimensional $\O(N)$ vector model in the large-$N$ limit.
This model has the RG flow from the Gaussian fixed point $T_\text{Gaussian}$, where all the $N$ scalar fields are free and massless, to the critical $\O(N)$ (interacting) fixed point $T_\text{critical}$ by keeping the renormalized masses vanishing.
Perturbing the critical $\O(N)$ fixed point by giving the scalars negative mass squared terms triggers the RG flow to the $\O(N-1)$ symmetric (Goldstone) fixed point $T_\text{Goldstone}$ with $N-1$ massless free scalar fields (Goldstone modes).\footnote{There is one massive scalar field that decouples at the Goldstone fixed point.}
If the thermal central charge measures the degrees of freedom correctly, we expect them to be ordered along the RG flow $C_\text{Therm} (T_\text{Gaussian}) > C_\text{Therm} (T_\text{critical}) > C_\text{Therm} (T_\text{Goldstone})$.
The thermal central charges at the Gaussian and Goldstone fixed points are simply given by $C_\text{Therm} (T_\text{Gaussian}) = N\,C_\text{Therm}(T_\text{scalar})$ and $C_\text{Therm} (T_\text{Goldstone}) = (N -1)\,C_\text{Therm}(T_\text{scalar})$ as consistent with the RG flow, where  $T_\text{scalar}$ is a real massless free scalar field theory.
On the other hand, the large-$N$ analysis showed $C_\text{Therm} (T_\text{critical}) = \left(4N/5 + O(1)\right)\,C_\text{Therm}(T_\text{scalar})$, which can be less than the IR value $C_\text{Therm} (T_\text{Goldstone})$ for large $N$ and ruled out the possibility of $C_\text{Therm}$ being the fixed point value of a $\CC$-function \cite{Sachdev:1993pr}.

The option $(b)$ employing the coefficient $C_T$ as a fixed point value of a $\CC$-function was suggested by \textcite{Petkou:1995vu} based on the result $C_T (T_\text{critical}) = \left( N - 40/(9\pi^2) \right)\, C_T (T_\text{scalar})$ consistent with the RG flows described for the $\O(N)$ vector model.
This possibility was explored by \textcite{Nishioka:2013gza} more systematically for $\CN =2$ supersymmetric field theories in three dimensions that allows the exact calculation of $C_T$ due to the supersymmetric localization technique.
In that work it was shown that a supersymmetric analog of the $\O(N)$ vector model ($\CN=2$ Wess-Zumino model) can be a counterexample for the ``$C_T$"-theorem.
Another counterexample was also found by \textcite{Fei:2014yja} by examining the RG flows in the five-dimensional $\O(N)$ symmetric scalar field theory with the $\O(N)$ symmetric interacting fixed point.

In three dimensions, there exists no conformal anomaly; hence the third option $(c)$ is unavailable on its own.
Alternatively, a proposal was made by \textcite{Jafferis:2011zi} [and independently by \textcite{Myers:2010xs,Myers:2010tj} in a different form from the holographic viewpoint] based on the extremization principle of the supersymmetric partition function \cite{Jafferis:2010un}, which is now known as the $F$-theorem,
\begin{theorem*}[$F$-theorem]
In three-dimensional QFTs, there exists a function $\CF(g_i, \mu)$ on the theory space satisfying the following properties:
\begin{enumerate}
\item It takes the same value as the sphere free energy $F$ of the CFT corresponding to each fixed point of RG flows,
	\begin{align}\label{REE_F}
		\CF(g_i, \mu) |_\mathrm{CFT} = F \ ,
	\end{align}
	where $F=-\log |Z[\BS^3]|$ and $Z[\BS^3]$ is the (renormalized) Euclidean partition function on $\BS^3$.
\item It is a monotonically decreasing function under any RG flow,
	\begin{align}\label{REEMonotonicity}
		\frac{d\,\CF}{d\,t}  = - \mu \frac{\d\,\CF}{\d\mu} \le 0 \ .
	\end{align}
\end{enumerate}
\end{theorem*}
This statement can be regarded as a variant of the third option $(c)$ since we can extract the type $A$ central charge of the conformal anomaly from the sphere partition function $A \propto \log Z[\BS^d]$ in even dimensions.

While the $F$-theorem was originally stated in the weak form $F_\mathrm{UV} \ge F_\mathrm{IR}$, a perturbative argument \cite{Klebanov:2011gs,Yonekura:2012kb} suggests the sphere free energy is the strongest version of a $\CC$-function.
Strong evidence supporting the $F$-theorem was given for $\CN=2$ supersymmetric theories with $\U(1)_R$ symmetry, where
the sphere free energy $F$ is shown to be locally maximized at RG fixed points \cite{Jafferis:2010un,Closset:2012vg} unless there is no accidental symmetry at the IR fixed point [see also the review by \textcite{Pufu:2016zxm} for details].
A nonperturbative proof of the strong version of the $F$-theorem was given by \textcite{Casini:2012ei} by exploiting the unexpected relation between the free energy and  the entanglement entropy of a disk we derived in Eq.\,\eqref{EE_F}. 
We outline the proof and the implications in Sec. \ref{ss:Ftheorem}.

\subsubsection{Four dimensions}
Seeking for a $\CC$-theorem in four dimensions has a long history starting from the conjecture [of the type $(c)$ in Sec. \ref{ss:OrderingRG}] by \textcite{Cardy:1988cwa} for the $a$-theorem.
\begin{theorem*}[$a$-theorem]
In four-dimensional QFTs, the type $A$ central charge $a$ of the conformal anomaly decreases along any RG flow
\begin{align}\label{a_theorem_claim}
a_\mathrm{UV} \ge a_\mathrm{IR} \ .
\end{align}
\end{theorem*}
This weak version of the $a$-theorem has passed a number of tests over the years in perturbative QFTs \cite{Osborn:1989td,Jack:1990eb} and was given an evidence for $\CN=1$ supersymmetric theories by the $a$-maximization method that implies the strongest $a$-theorem \cite{Intriligator:2003jj,Barnes:2004jj}.
A proof of the $a$-theorem without relying on supersymmetry was presented more recently by \textcite{Komargodski:2011vj} who connect the difference of the $a$-coefficient with the scattering amplitude of four dilaton fields that naturally couple to the stress-energy tensor [see also \cite{Komargodski:2011xv}].
Interestingly the unitarity of the scattering amplitude turns out to guarantee the inequality \eqref{a_theorem_claim} in their proof.

Recently, an alternative proof of the $a$-theorem was given by \textcite{Casini:2017vbe} who resort to the monotonicity of entanglement as in the case of the $F$-theorem.
We touch on the proof in Sec. \ref{ss:C_theorem_Higher}.

Finally, a few comments on the other possibilities are in order:
\begin{enumerate}
\item[$(a)$] The thermal central charge $C_\text{Therm}$ does not satisfy a $\CC$-theorem in a class of four-dimensional gauge theories that are not asymptotically free at UV \cite{Appelquist:1999hr}.
\item[$(b,c)$] The $C_T$ coefficient of the stress-tensor two-point function is proportional to the type $B$ central charge $c$, $C_T \propto c$, in four dimensions \cite{Osborn:1993cr}.
The latter was extensively studied in supersymmetric and nonsupersymmetric theories and many counterexamples are known in a typical class of gauge theories  \cite{Cappelli:1990yc,Anselmi:1997am}.
\end{enumerate}

\subsubsection{Higher dimensions}
Given the success in $d\le 4$ dimensions, one may be tempted to extend a $\CC$-theorem to higher dimensions $d>4$.
This problem was less explored than the lower-dimensional cases, but once formulated it will provide us a unifying picture of RG flows in QFT irrespective of dimensionality.
It is too early to survey the whole subject as it is still under active investigation.
We thus comment only on the two following promising conjectures\footnote{See also \textcite{Gukov:2015qea,Gukov:2016tnp} for a recent discussion on  constraining RG flows by topological data of the theory space.
}:
\begin{itemize}
\item The $a$-theorem in higher even dimensions postulates the monotonicity of the type $A$ central charge for the conformal anomaly under RG flows.
There is supporting evidence for the $a$-theorem from holography \cite{Myers:2010xs,Myers:2010tj} and even proofs for a certain class of supersymmetric field theories in six dimensions \cite{Cordova:2015fha,Cordova:2015vwa}.
\item The $F$-theorem in higher odd dimensions was suggested by \textcite{Klebanov:2011gs,Myers:2010xs,Myers:2010tj} and further incorporated into the generalized $F$-theorem proposal \cite{Giombi:2014xxa} that interpolates between the $F$-theorem and the $a$-theorem in noninteger dimensions.
A few examples supporting the $F$-theorem in five dimensions are presented by \textcite{Jafferis:2012iv,Fei:2014yja}.
\end{itemize}

\subsection{The entropic $c$-theorem in $(1+1)$ dimensions}

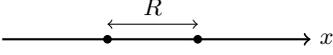
\begin{figure}[h]
\centering
\begin{tikzpicture}
\filldraw [fill=black] (-0.6,-1.8) circle (0.05) ;
\filldraw [fill=black] (0.6,-1.8) circle (0.05);
\draw[<->] (-0.6,-1.6)  -- node[above] {$R$}(0.6,-1.6);
\draw[thick, ->] (-2, -1.8) -- (2.1, -1.8) node[right] {$x$};
\end{tikzpicture}
\caption{An interval of width $R$ as an entangling region.}
\label{fig:2dCFT_EE2}
\end{figure}

We shall prove the strong version of the $\CC$-theorem in two dimensions
by constructing a $\CC$-function built from entanglement entropy.
A $\CC$-function should be well-defined for any QFT and monotonically decreases as  the energy scale $\mu$ is lowered while fixing the size of a system (e.g., an entangling region).
One good candidate for it is the entanglement entropy $S(R)$ of an interval of width $R$ in $(1+1)$ dimensions (see Fig.\,\ref{fig:2dCFT_EE2}).

Instead of tuning the energy scale $\mu$ with the system size fixed, 
we would rather change the size $R$ to trigger an RG flow while fixing $\mu$.
In this picture, we can probe the physics at the UV and IR scales by small and large intervals respectively.
The conditions for a function $c_E(R)$ being a $\CC$-function are restated in the following form:
\begin{enumerate}
\item $c_E(R)$ coincides with the central charge $c$ of CFT$_2$ at a fixed point of an RG flow,
	\begin{align}
		c_E(R) |_\text{CFT} = c \ .
	\end{align}
\item $c_E(R)$ is a monotonically decreasing function under any RG flow,
	\begin{align}\label{EntropicMonotonicity}
		c_E'(R)\le 0 \ .
	\end{align}
\end{enumerate}
Recalling from Eq.\,\eqref{EE_CFT2}, the entanglement entropy of the interval in CFT$_2$ becomes\footnote{We can redefine the UV cutoff $\epsilon$ so as to remove a finite term in Eq.\,\eqref{EE_CFT2}.}
\begin{align}
S(R)|_\text{CFT} = \frac{c}{3}\log \frac{R}{\epsilon} \ ,
\end{align}
where the central charge $c$ appears as a coefficient of the logarithmic divergence.
It follows that we can introduce the \emph{entropic $c$-function} \cite{Casini:2004bw} satisfying the first condition,
\begin{align}\label{Entropic_C}
c_E(R) \equiv 3R\, S'(R) \ .
\end{align}
Note that the entropic $c$-function is well-defined for \emph{any} QFT as it is built from the entanglement entropy of an interval.
It should also be stressed that the entropic $c$-function is free from the UV divergence as there is only logarithmic divergence in two dimensions, and it always takes a finite value in contrast to the entanglement entropy itself.

\begin{figure}[htbp]
\centering
	\begin{tikzpicture}[thick]
		\draw[->] (0,0) --++ (5,0) node[right] {$x$};
		\draw[->] (-1,0) --++ (0,4) node[above] {$t$};
		\draw (2,1)-- node[below] {$B$} ++(1,0);
		\draw (0,3)-- node[above] {$A\cup B \cup C$} ++(5,0);
		\draw[dashed] (2,1) -- node[below, sloped] {$A$}++(-2,2);
		\draw[dashed] (3,1) -- node[below, sloped] {$C$} ++(2,2);
		\draw (2,1) -- node[above,sloped] {$B\cup C$} (5,3);
		\draw (3,1) -- node[above,sloped] {$A\cup B$} (0,3);
	\end{tikzpicture}
\caption{A proof of the entropic $c$-theorem. Two intervals $A$ and $C$ are on the light rays $t=\pm x$ while an interval $B$ is on a time slice $t=0$.}
\label{fig:EntropicC}
\end{figure}
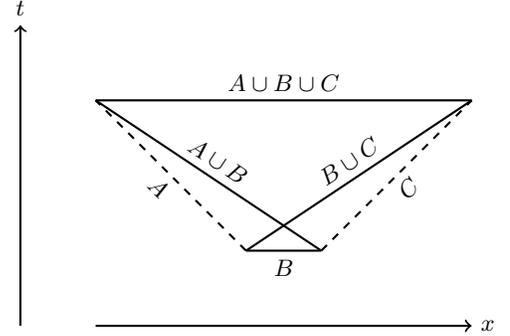

In order to prove that the monotonicity of the entropic $c$-function, Eq.\,\eqref{EntropicMonotonicity}, we consider two intervals $A$ and $C$ on the light rays $t=\pm x$ (the dashed lines) and an interval $B$ of width $r$ on a time slice $t=0$ as in Fig.\,\ref{fig:EntropicC}.
More precisely we choose the three regions as follows:
\begin{align}
	\begin{aligned}
		A &= \left\{ t = -x, \, -\frac{R}{2} \le x \le - \frac{r}{2} \right\} \ , \\
		B&= \left\{ t = 0, \, - \frac{r}{2} \le x \le \frac{r}{2} \right\} \ , \\
		C&= \left\{ t = x, \,  \frac{r}{2} \le x \le \frac{R}{2} \right\} \ .
	\end{aligned}
\end{align}
Entanglement entropy is invariant under unitary time evolution and hence is Lorentz invariant in QFT.
We thus are able to use the boosted interval connecting the left end point of $A$ and the right end point of $B$ to calculate the entanglement entropy of the union $A\cup B$.
Similarly the entanglement entropy of the union $B\cup C$ is equal to that of the boosted interval connecting the left end point of $B$ and the right end point of $C$,\footnote{The diffeomorphism invariant length $\Delta R$ between two points $(t,x)$ and $(t+\Delta t, x+\Delta x)$ is defined by
\begin{align}
	\D R= \sqrt{- (\D t)^2 + (\D x)^2} \ .
\end{align}
}
\begin{align}
	S_{A\cup B} = S_{B\cup C} = S\left(\sqrt{rR}\right) \ .
\end{align}
The union of the three regions $A\cup B\cup C$ can also be time-evolved into the interval of width $R$ at $t=(R-r)/2$.
Hence the entanglement entropies of $A\cup B\cup C$ and $B$ are
\begin{align}
	S_{A\cup B\cup C} = S(R) \ , \qquad S_B = S(r) \ .
\end{align}

Now we apply the strong subadditivity Eq.\,\eqref{SSA} and find the inequality,
\begin{align}\label{ECSSA}
 S(R) + S(r) \le 2 S\left(\sqrt{rR}\right)\ .
\end{align}
We can reduce it in the limit $r\to R$ to \cite{Casini:2004bw}\footnote{See also \textcite{Bhattacharya:2014vja} for the generalization of this argument to prove the positivity of the entanglement density.}
\begin{align}\label{ECmonotonic}
\frac{c_E'(R)}{3} = S'(R) + RS''(R) \le 0 \ ,
\end{align}
which is actually what we wanted to prove, Eq.\,\eqref{EntropicMonotonicity}.
It is worthwhile to emphasize that conformal symmetry plays no roles in the proof of the inequality \eqref{ECmonotonic}, and hence it is valid for \emph{any} QFT as it follows from the strong subadditivity of entanglement entropy and Lorentz invariance in QFT.

We note that the entropic $c$-function Eq.\,\eqref{Entropic_C} is not stationary at the UV fixed point for a free massive scalar theory \cite{Casini:2005zv}; hence it is an example of the stronger $\CC$-theorem in the terminology of Sec. \ref{ss:OrderingRG}.
It behaves quite differently from Zamolodchikov $c$-function under RG flows although they coincide at conformal fixed points.
It remains open whether there exists a strongest $\CC$-function built from entanglement entropy.

\subsection{The $F$-theorem in $(2+1)$ dimensions}\label{ss:Ftheorem}
We adapt the argument for the entropic $c$-theorem to the proof of the strong version of the $F$-theorem in a $(2+1)$-dimensional QFT.
A counterpart of the entropic $c$-function will be introduced as the renormalized entanglement entropy $\CF$ \cite{Liu:2012eea}.
We content ourselves with outlining the proof of the monotonicity of $\CF$ and leave the technical details to the original paper \cite{Casini:2012ei}.
After examining the UV and IR behaviors of $\CF$, we discuss the possibility of the strongest version of the $F$-theorem.

\subsubsection{A sketch of the proof}
The nonperturbative proof of the $F$-theorem goes in parallel with that of the entropic $c$-theorem in the previous section.
To this end, we construct a function $\CF(R)$ from the entanglement entropy of a disk satisfying the following:
\begin{enumerate}
\item $\CF(R)$ coincides with the sphere free energy $F$ of CFT$_3$ at a fixed point of an RG flow,
	\begin{align}\label{REE_F_proof}
		\CF(R) |_\text{CFT} = F \ .
	\end{align}
\item $\CF(R)$ is a monotonically decreasing function under any RG flow,
	\begin{align}\label{REEMonotonicity_proof}
		\CF'(R)\le 0 \ .
	\end{align}
\end{enumerate}
The key in finding such a function is the relation \eqref{EE_F} between the sphere free energy $F$ and the entanglement entropy $S(R)$ of a disk of radius $R$ for CFT$_3$,
\begin{align}\label{S_F}
S(R)|_\text{CFT} = \alpha \frac{2\pi R}{\epsilon} - F \ ,
\end{align}
where $\alpha$ is a theory-dependent constant.
In three dimensions, the UV divergence of entanglement entropy $S(R)$ is of order $O(1/\epsilon)$ for any QFT, and it is always proportional to the radius $R$ of the disk due to the area law.
We want to regularize the divergence as in the entropic $c$-function Eq.\,\eqref{Entropic_C}, thus
introducing the \emph{renormalized entanglement entropy} or the $\CF$-function  \cite{Liu:2012eea}
\begin{align}\label{REE}
\CF (R) \equiv (R\, \partial_R - 1) S(R) \ .
\end{align}
Then we see the first condition Eq.\,\eqref{REE_F_proof} for $\CF$ being a $\CC$-function immediately follows from Eq.\,\eqref{S_F}.

Furthermore, \textcite{Casini:2012ei} completed a proof of the $F$-theorem by showing the monotonicity of $\CF$, Eq.\,\eqref{REEMonotonicity_proof}, with the strong subadditivity for any Lorentz invariant field theory.
We only sketch the proof of the inequality Eq.\,\eqref{REEMonotonicity_proof}, leaving the technical details to the original paper \cite{Casini:2012ei} [see also \textcite{Casini:2015woa} for a proof using the mutual information].

\begin{figure}[htbp]
\includegraphics[width=7cm]{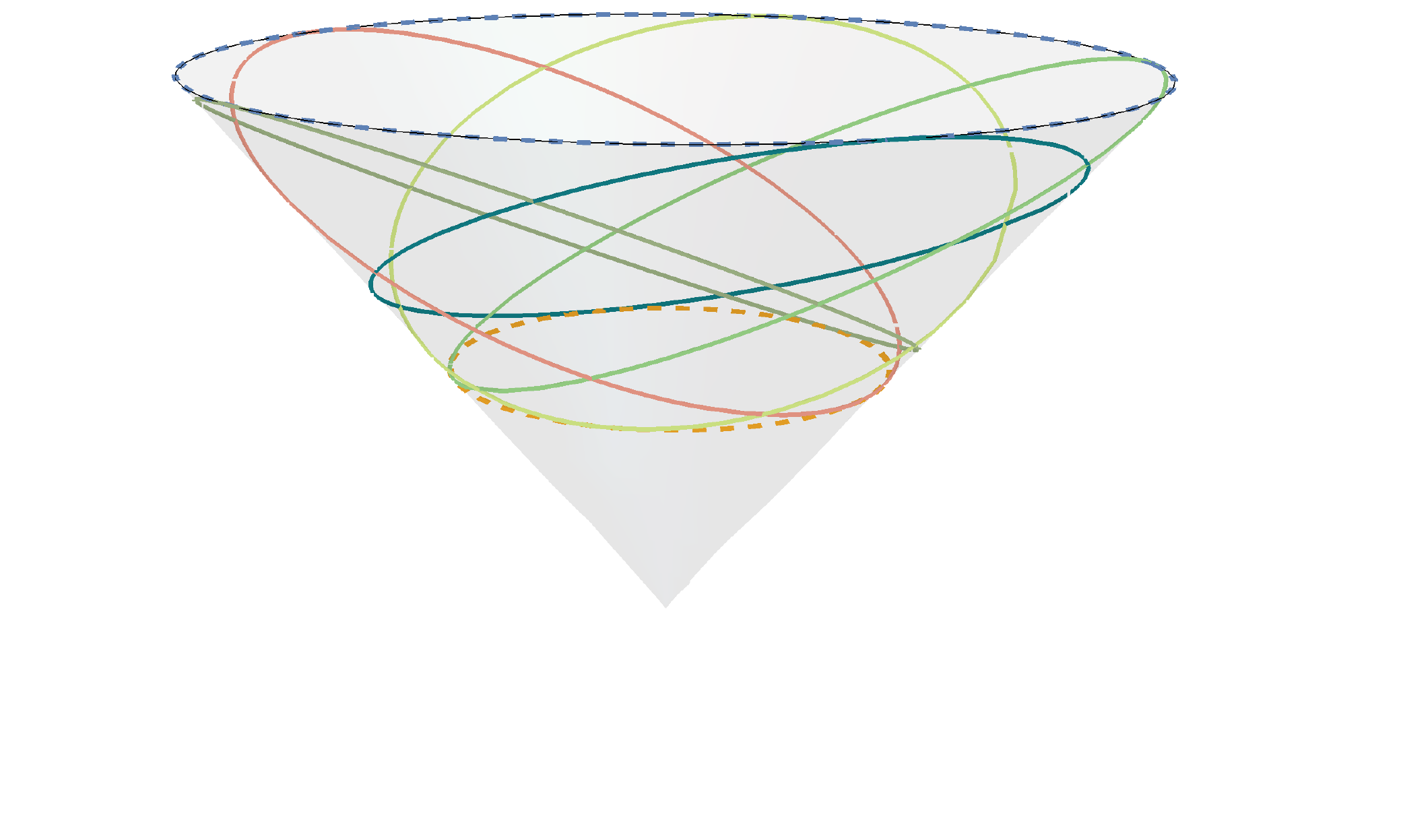}
\caption{\label{fig:Fproof} $N=5$ boosted disks of radii $\sqrt{rR}$ touching on the null cone in gray color.
The dashed curves are circles of radii $R$ and $r$ at constant time slices $t=R$ and $r$, respectively.
}
\end{figure}

The point for proving the entropic $c$-theorem was the choice of the boosted intervals whose union and intersection are intervals of different widths at different time slices as in Fig.\,\ref{fig:EntropicC}.
In order to extend the previous argument to the present case, we have to choose a set of boosted disks touching on the null cone, whose union and intersection approximate disks of different radii, say, $R$ and $r\, (\le R)$ (see Fig.\,\ref{fig:Fproof}).
Clearly an infinite number of boosted disks are needed to realize such a situation.
Hence, we begin with $N$ boosted disks labeled by $X_i~(i=1,\cdots, N)$ whose diffeomorphism invariant radii are $\sqrt{r R}$ and apply to them the inequality obtained by the repeated use of the strong subadditivity 
\begin{align}\label{SSA_general}
\begin{aligned}
	S_{\cup_i X_i} &+ S_{\cup_{\{i,j\}} X_{ij}} \\
		& + S_{\cup_{\{i,j,k\}} X_{ijk}} + \cdots + S_{\cap_i X_i} \le \sum_i S_{X_i} \ ,
\end{aligned}
\end{align}
where we used the shorthand notation $X_{ij\cdots k} \equiv X_i \cap X_j \cap \cdots \cap X_k$.
There are $N$ terms on the left-hand side, each of which gives a disk entropy of a radius ranging from $r$ to $R$.
Dividing by $N$ on both sides, the sum on the left-hand side approaches an integral in the $N\to \infty$ limit,
\begin{align}
	\frac{1}{\pi} \int_0^\pi \d\theta \, S\left(\frac{2rR}{R + r - (R-r)\cos\theta} \right) \le S\left( \sqrt{rR} \right) \ .
\end{align}
Letting $R = r - \epsilon$ and expanding both  sides at $\epsilon = 0$, one finds $S''(R)\le 0$ at order $O(\epsilon^2)$,
\begin{align}\label{REEmonotonic}
\CF'(R) = R\,S''(R) \le 0 \ ,
\end{align}
which is the other necessary condition Eq.\,\eqref{REEMonotonicity_proof} for $\CF$ being a $c$-function in three dimensions.
We again stress, as in the case of the entropic $c$-theorem in two-dimensions, that the inequality \eqref{REEmonotonic} is valid for \emph{any} QFT as the proof does not rely on conformal symmetry at all.

\subsubsection{Constraints from the $F$-theorem}

As applications of the $F$-theorem, we want to constrain the phase diagram of 
noncompact QED$_3$ coupled to $2N_f$ two-component Dirac fermions of unit charges $\psi^i~(i=1,\cdots, 2N_f)$ \cite{Grover:2012sp,Giombi:2015haa}.
This theory enjoys the $\SU(2N_f)$ global symmetry and flows to the conformal phase for large $N_f$ at the IR fixed point.
On the other hand, the theory is suspected to exhibit a spontaneous symmetry breaking when the number of the fermions are smaller than a critical value $N_f\le N_\text{crit}$ due to the condensation of the operator of the form $\sum_{i=1}^{N_f}(\bar\psi_i\,\psi^i - \bar\psi_{i+N_f}\,\psi^{i+N_f})$.
The chiral symmetry broken (CSB) phase preserves the subgroup $\SU(N_f)\times \SU(N_f) \times \U(1)$ of $\SU(2N_f)$ and is described by the $2N_f^2$ Nambu-Goldstone bosons associated with the spontaneous symmetry breaking and a free Maxwell field that is dual to a free scalar field in the IR.

In order to estimate the critical value $N_\text{crit}$, we compare the sphere free energies $F_\text{conf}(N_f),~F_\text{CSB}(N_f)$ of the conformal and CSB phases.
If there exists an RG flow from the conformal phase to the CSB phase, the latter should have a lower sphere free energy than that of the former $F_\text{CSB}(N_f) < F_\text{conf}(N_f)$ as dictated by the $F$-theorem.

In the large $N_f$ expansion, the conformal phase has \cite{Klebanov:2011td},
\begin{align}
	F_\text{conf}(N_f) = 2 N_f\, F_\text{ferm} + \frac{1}{2}\log \left( \frac{\pi N_f}{4}\right) + O(1/N_f) \ ,
\end{align} 
while there are $(2N_f^2 + 1)$ massless free scalar fields in the CSB phase
\begin{align}
	F_\text{CSB} (N_f) = \left(2N_f^2 + 1\right)\, F_\text{scalar} \ ,
\end{align}
which grows quadratically in $N_f$ and exceeds $F_\text{conf}(N_f)$ for large $N_f$ as expected.
Using the values of the sphere free energies for a free scalar and a free Dirac fermion \cite{Klebanov:2011gs,Dowker:2010yj},
\begin{align}\label{F_value_scalar}
\begin{aligned}
F_\text{scalar} &= \frac{\log 2}{8} - \frac{3\zeta (3)}{16\pi^2} \approx 0.0638 \ ,\\
F_\text{fermion} &= \frac{\log 2}{4} + \frac{3\zeta (3)}{8\pi^2} \approx 0.2189\ ,
\end{aligned}
\end{align}
we can estimate the value $N_f \approx 4.4$ at which $F_\text{CSB}(N_f)  = F_\text{conf}(N_f)$.
A more precise analysis using the $\epsilon$-expansion and the Pad{\'e} approximation also yields a similar value \cite{Giombi:2015haa}.
Hence the $F$-theorem rules out any RG flow from the conformal phase to the broken symmetry phase for $N_f \ge 5$ and predicts the upper bound for the critical value $N_\text{crit}\le 4$.

\subsubsection{Large mass expansion}
 
We consider a free massive scalar field to examine the $F$-theorem by a concrete example.

In general, the entanglement entropy across an entangling surface $\Sigma$ for a theory with a mass gap of order $m$ allows an expansion in terms of large $m$ \cite{Grover:2011fa,Huerta:2011qi},
\begin{align}\label{EE_LargeMass}
S_\Sigma = \alpha \frac{\ell_\Sigma}{\epsilon} - \gamma + \sum_{i\in \BZ} c_i^\Sigma\,m^i \ ,
\end{align}
where we separate the mass effect with the coefficients $c_{i}^\Sigma$ of mass dimension $-i$ from the contributions of the UV divergence proportional to the length $\ell_\Sigma$ of the contour $\Sigma$ and the topological entanglement entropy $\gamma$.
In the presence of the mass gap, the entanglement entropy would be dominated by the local contributions within the correlation length $\xi \sim 1/m$ from the entangling region $\Sigma$, and it is reasonable to expect $c_{i}^\Sigma$ are characterized by local geometric data around $\Sigma$,
\begin{align}
c_{i}^\Sigma = \oint_\Sigma \d s \,f(\kappa, \partial_s \kappa, \partial_s^2 \kappa,\cdots) \ ,
\end{align}
where $\kappa$ is the extrinsic curvature and $s$ is a coordinate parametrizing $\Sigma$.
It follows that there are no $c_{i}^\Sigma$'s for $i \ge 2$, and the leading coefficient is given by $c_{1}^\Sigma = \beta \oint_\Sigma \d s  = \beta\,\ell_\Sigma$ for a constant $\beta$.

For a pure ground state, the entanglement entropy is symmetric under the exchange of 
the region $A$ bounded by $\Sigma$ and its complement, hence the coefficients $c_{i}^\Sigma$ must be invariant under the $\BZ_2$ symmetry acting on the integrand $f(\kappa, \partial_s \kappa, \cdots)$ with $\kappa\to -\kappa$ and $s\to -s$ \cite{Grover:2011fa}.
To illustrate it, let us work out the first few coefficients.
Since $\kappa$ and $s$ have dimensions $1$ and $-1$, the $\BZ_2$ symmetry allows
\begin{align}
	\begin{aligned}
		c_{-1}^\Sigma &=  \beta' \oint_\Sigma \d s \,\kappa^2 \ , \\
		c_{-3}^\Sigma &=  \oint_\Sigma \d s \, \left[\beta''_1\, \kappa^4 + \beta''_2\,(\partial_s\kappa)^2\right] \ ,
	\end{aligned}
\end{align}
where $\beta'$ and $\beta''_1, \beta''_2$ are some constants depending on the details of QFT, but the possibilities $c_0^\Sigma \propto \oint_\Sigma \d s\,\kappa, \, c_{-2}^\Sigma \propto \oint_\Sigma \d s\,\kappa\, (\partial_s\kappa)$ are excluded.
Extending this argument, we find $f(\kappa, \partial_s \kappa, \cdots)$ always has even dimensions, and there only remain terms with the odd $i$ in the expansion Eq.\,\eqref{EE_LargeMass},
\begin{align}\label{EE_MassScalar}
S_\Sigma = \alpha \frac{\ell_\Sigma}{\epsilon} + \beta\, m\, \ell_\Sigma - \gamma + \sum_{n=0}^\infty \frac{c_{-2n-1}^\Sigma}{m^{2n+1}} \ ,
\end{align}

\begin{figure}[htbp]
\centering
\begin{tikzpicture}
			\draw[thick, fill=verylightgray] (0,0) --++ (3,0) --++ (1,1.5) --++ (-3,0) --++ (-1, -1.5);
			\draw[thick] (0,-3) --++ (3,0) --++ (1,1.5) --++ (0,3);
			\draw[thick, dashed] (0, -3) --++ (1, 1.5) --++ (3,0);
			\draw[thick] (0, -3) --++ (0,3);
			\draw[thick] (3, -3) --++ (0,3);
			\draw[thick, dashed] (1, -1.5) --++ (0,3);
			\draw[thick, aomidori, fill=cyan!50] (2,0.8) ellipse (0.8 and 0.5);
			\draw[thick, aomidori, fill=cyan!20] (2,-2.2) ellipse (0.8 and 0.5);
			\draw[thick, dashed, aomidori] (1.2, 0.8) --++ (0,-3);
			\draw[thick, dashed, aomidori] (2.8, 0.8) --++ (0,-3);
			\draw[thick, <->] (4.4,-1.5) -- node[right] {\large $\BS^1$ of size $L$} (4.4,1.5)  ;
			\node at (2, 0.8) {$A$};
			\node[black!50] at (2, -2.2) {$A$};
			\draw[->, thick] (-0.4, 1) node[left] {\large $\BR^{2,1}$} --(0.5,0.5);
			
		\end{tikzpicture}
\caption{The dimensional reduction of a $(3+1)$-dimensional field theory to the $(2+1)$-dimensional field theory.
}
\label{fig:KKmodes}
\end{figure}
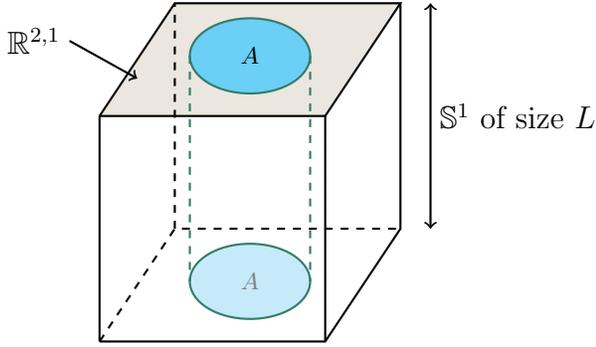

Now we want to compute the leading coefficient $c_{-1}^\Sigma$ for a free massive scalar field. 
We start with a $(3+1)$-dimensional free \emph{massless} scalar theory on $\BR^{2,1}$ times $\BS^1$ of circumference $L$ as in Fig.\,\ref{fig:KKmodes}.
By this compactification, the higher-dimensional scalar field reduces a tower of massive scalar fields in $(2+1)$ dimensions with masses,
\begin{align}
m_n^2 = \left(\frac{2\pi}{L}\right)^2 n^2 \ , \qquad n\in \BZ \ .
\end{align}
If we choose an entangling surface of the form $\Sigma_2 = \Sigma \times \BS^1$ with $\Sigma$ a smooth curve in $(3+1)$ dimensions, the entanglement entropy in four dimensions should equal the summation of the entanglement entropies across $\Sigma$ for the massive scalar fields in $(2+1)$ dimensions:
\begin{align}\label{DimRed}
\begin{aligned}
S_{\Sigma_2}^{(3+1)} &= \sum_{n\in\BZ} S_{\Sigma}^{(2+1)} (m_n) \ ,\\
	&\xrightarrow{L\to \infty} \quad \frac{L}{\pi} \int_0^{1/\epsilon} \d p\, S_{\Sigma}^{(2+1)} (m=p) \ ,
\end{aligned}
\end{align}
where we took the large $L$ limit and replaced the summation over the massive modes with the integral over a continuous parameter $p$.
We also introduced the UV cutoff $\epsilon$ for the integral in the large $L$ limit, which is to be regarded as the UV cutoff in $(3+1)$ dimensions.
As seen from Eq.\,\eqref{EE_Even}, the entanglement entropy in even-dimensional CFT has the logarithmic divergence whose coefficient is fixed by the conformal anomaly by Eq.\,\eqref{c04d}.
Then the logarithmically divergent part of the left-hand side becomes
\begin{align}\label{EE4dmassless}
S_{\Sigma_2}^{(3+1)}\big|_\text{log} = \frac{c}{2\pi} \int_{\Sigma_2}\left[\frac{1}{2}(\CK^{a~\mu}_{~\mu})^2 -  \CK^a_{\mu\nu}\CK^{a\,\mu\nu} \right] \log \frac{l}{\epsilon}\ ,
\end{align}
where we assume $\Sigma$ is topologically $\BS^1$ whose typical size is $l$.
Here we use the fact that the Euler number of $\Sigma_2 \sim \BT^2$ vanishes, and the induced curvatures $\CR_{aa}$ and $\CR_{abab}$ are zero on a flat space.
On the other hand, we found the logarithmic divergent term on the right-hand side of Eq.\,\eqref{DimRed} from the leading term of the mass expansion Eq.\,\eqref{EE_MassScalar},
\begin{align}\label{EE3dmassive}
- \frac{L}{\pi} c_{-1}^\Sigma \log\epsilon \ .
\end{align}
Comparing Eqs.\,\eqref{EE4dmassless} and \eqref{EE3dmassive}, the coefficient $c_{-1}^\Sigma$ is fixed by the central charge $c$ and the integral of the extrinsic curvatures.
If we choose $\Sigma$ to be a disk of radius $R$, for a scalar field with $c=1/120$, we obtain \cite{Huerta:2011qi,Klebanov:2012yf}
\begin{align}
c_{-1}^\Sigma = - \frac{\pi}{240R} \ .
\end{align}

This coefficient also fixes the leading form of the $\CF$-function Eq.\,\eqref{REE} in the large mass limit,\footnote{The topological entanglement entropy $\gamma$ is zero for a massive scalar field since there is nothing left in the IR fixed point.}
\begin{align}\label{LargeMassExpansionScalar}
\CF = \frac{1}{120(mR)} + O(1/(mR)^3) \ ,
\end{align}
which is indeed monotonically decreasing as consistent with the $F$-theorem.\footnote{Since $\CF$ is dimensionless, the mass $m$ and the radius $R$ are always paired in the expansion.}
One can generalize the argument and compute the higher order coefficients $c_{-2n-1}^\Sigma$ by starting from a free theory on $\BR^{2,1}\times \BT^{2n+1}$ and comparing the logarithmic divergence of the entanglement entropy of $\Sigma_{2n+2} = \Sigma\times \BT^{2n+1}$ with that of massive fields in three dimensions \cite{Klebanov:2012yf}.
The coefficients $c_{-2n-1}^\Sigma$ are generally given by the conformal anomalies in $(2n+4)$ dimensions.
The subleading coefficient $c_{-3}^\Sigma$ is calculated in this way by \textcite{Klebanov:2012yf,Safdi:2012sn}.

\begin{figure}[htbp]
\centering
\includegraphics[width=8cm]{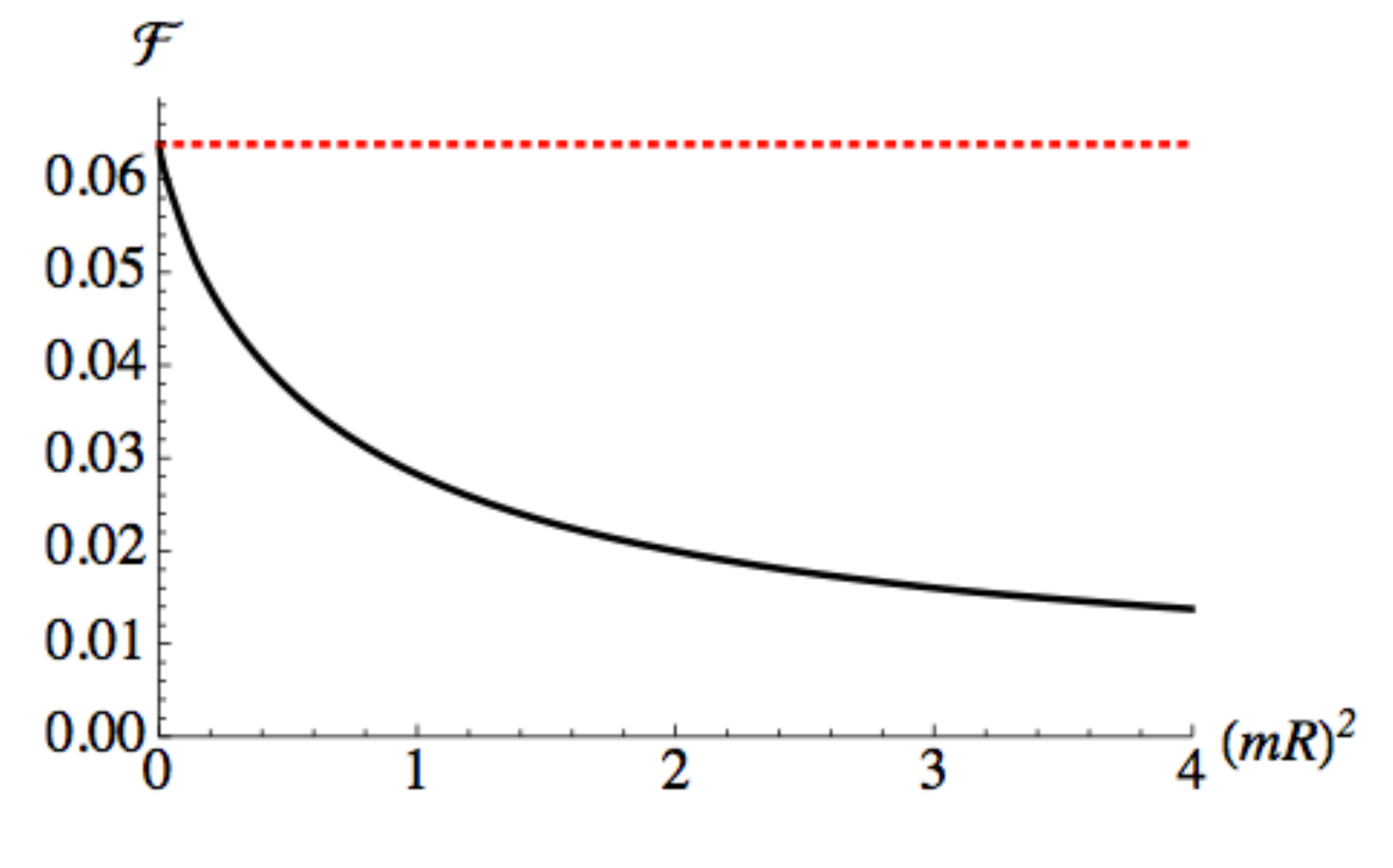}
\caption{The numerical plot of the $\CF$-function of a free massive scalar field (the black curve).
The dotted red line is the UV fixed point value $\CF_\text{UV}\approx 0.064$ of the $\CF$-function.
}
\label{fig:REEmassive}
\end{figure}

We can check the validity of the large mass expansion by comparing with the numerical estimation using the real time approach developed in Sec. \ref{ss:RTapproach}.
The numerical plot of the $\CF$-function is shown in Fig.\,\ref{fig:REEmassive} with respect to $(mR)^2$ \cite{Klebanov:2012va,Nishioka:2014kpa}.
It starts from $\CF_\text{UV}\approx 0.064$ and monotonically approaches to $\CF \sim 1/(120mR)$ in the large $mR$ limit as expected.
In addition, the UV value of the $\CF$-function agrees with the sphere free energy $F_\text{scalar}$ of a free massless scalar field given by Eq.\,\eqref{F_value_scalar}.
Hence we confirmed the $\CF$-function of a massive scalar field obeys the general properties Eqs.\,\eqref{REE_F_proof} and \eqref{REEMonotonicity_proof} analytically in the large mass region and numerically in the entire region.

\subsubsection{Stationarity at UV fixed points}
Having dealt with the IR expansion of the $\CF$-function of the free massive scalar field, we look into the UV behavior around $m=0$.\footnote{See \textcite{Hertzberg:2012mn} for the case of an interacting scalar field theory.}
We see from the numerical plot in Fig.\,\ref{fig:REEmassive} that the derivative of $\CF$ is not stationary even at the UV fixed point when the free massive scalar field theory is considered as the relevant perturbation of a free massless scalar field theory by the mass term $m^2\phi^2/2$ \cite{Klebanov:2012va,Nishioka:2014kpa}.
This should be contrasted with Zamolodchikov's $c$-function that is guaranteed to be stationary at conformal fixed points.
The lack of stationarity is not peculiar to the $\CF$-function. 
Indeed, the entropic $c$-function Eq.\,\eqref{Entropic_C} of a massive free scalar field in two-dimensions is also known to be nonstationary at the UV fixed point \cite{Casini:2005zv}.

One may wonder if the nonstationarity is a generic feature for the entropic $c$- and $\CF$-functions, independent of the types of QFTs.
To inspect it a little more, let $\CO(x)$ be a relevant scalar operator of dimension $\Delta\, (\le d)$ in $d$-dimensional CFT and consider the relevant perturbation by the operator
\begin{align}
	I_E = I_\text{UV} + g_\CO\, \int \d^d x\,\sqrt{g}\,\CO(x) \ ,
\end{align}
which induces an RG flow from the UV fixed point theory described by the action $I_\text{UV}$ to an IR fixed point.
The perturbative expansion of entanglement entropy around the UV fixed point $(g_\CO =0)$ should take the form,
\begin{align}\label{CP_EE}
	S (g_\CO) = S_\text{UV} + s_1\, g_\CO + s_2\,g_\CO^2 + O(g_\CO^3) \ .
\end{align}
The leading order term is the integrated two-point function of the modular Hamiltonian $H_A$ and the relevant operator \cite{Rosenhaus:2014woa,Rosenhaus:2014zza},
\begin{align}\label{s1}
	s_1 = - 2\pi\,\int \d^d x\,\sqrt{g}\, \langle H_A\, \CO(x)\rangle \ . 
\end{align}
The modular Hamiltonian is nonlocal in general, but can be expressed as the integral of a local integrand proportional to the stress tensor for a planar or spherical entangling surface in CFT as given in Eq.\,\eqref{Modular_Hamiltonian}.
In this case, the first-order term is expected to vanish, $s_1=0$, for $\langle T_{\mu\nu}(x)\,\CO(y)\rangle = 0$ in CFT (see Eq.\,\eqref{TO_CFT}) \cite{Rosenhaus:2014woa}.
Thus this perturbative argument shows quite generally that entanglement entropy is \emph{stationary} at the UV fixed point.
We record for the reader's reference that the second-order term $s_2$ is nonvanishing, whose form is determined for a spherical entangling surface of radius $R$ on $\BR^d$ to be \cite{Faulkner:2014jva}
\begin{align}
	s_2 = - R^{2(d-\Delta)} \frac{\pi^{(d+1)/2}(d - \Delta)\Gamma\left( d/2+1 - \Delta\right)}{2\Gamma \left( d+ 3/2 - \Delta\right)} \ ,
\end{align}
for operators with $\Delta \ge d/2$.

The contradiction between the presented general argument and the numerical result for a free massive scalar field needs some explanation.
The origin of this puzzle may be traced back to our tacit assumption for the stress tensor being conformal invariant, i.e., $T^\mu_{~\mu} = 0$ in deriving the expansion Eq.\,\eqref{CP_EE} with the conformal perturbation theory.
For a free scalar field, however, we have a family of the stress tensors $T^{(\xi)}_{\mu\nu}$ parametrized by a coupling constant $\xi$ for the curvature coupling $\CR\, \phi^2$ that can be added to the action when lifting the theory from flat spacetime to a curved spacetime.
$T^{(\xi)}_{\mu\nu}$ is conformally invariant (traceless) only when $\xi = \xi_c \equiv (d-2)/\left(4 (d-1)\right)$ even on flat spacetime.
There is no a priori reason to prefer the conformal stress tensor $T^{(\xi_c)}_{\mu\nu}$ to the others in calculating entanglement entropy.
The possibility of adopting the nonconformal stress tensors was examined by \textcite{Lee:2014zaa,Casini:2014yca,Herzog:2016bhv}, resulting in the speculation that the correct value of the leading coefficient $s_1$ is given by Eq.\,\eqref{s1} with the stress tensor $T^{(\xi = 0)}_{\mu\nu}$ for the minimal coupling $\xi = 0$.

Furthermore, the stationarity of entanglement entropy is conjectured to break down when there exists an operator $\CO_{d-2}$ of dimension $\Delta = d-2$ that can couple to the scalar curvature $\CR\,\CO_{d-2}$ \cite{Herzog:2016bhv}.
It remains to be clarified under what condition the entropic $c$- and $\CF$-functions become the stronger version of a $\CC$-function.

\subsection{The entropic $\CC$-theorems in $d\ge 4$ dimensions}\label{ss:C_theorem_Higher}
We have seen that the entropic $c$- and $F$-theorems in $(1+1)$ and $(2+1)$ dimensions follow from the strong subadditivity of entanglement entropy and the Lorentz invariance of QFTs.
One may then expect the $a$-theorem in $(3+1)$ dimensions similarly follows from the same argument. 
There however are two obstacles in the generalization to the higher-dimensional cases: i) the strong subadditivity inequalities become trivial due to logarithmic divergences coming from singular surfaces of intersecting boosted spheres, and ii) there is a finite mismatch between the entanglement entropies of the wiggly spheres and the round spheres in taking the number $N$ of the boosted spheres to infinity.

To overcome these difficulties, \textcite{Casini:2017vbe,Lashkari:2017rcl} considered a difference of the entanglement entropies across a $(d-2)$-sphere of radius $R$ between a theory $\CT$ along an RG flow and the UV CFT at its UV fixed point in $d$ dimensions,
\begin{align}
	\Delta S_\CT (R) \equiv S_\CT(R) - S_\text{UV}(R) \ ,
\end{align}
and define a new entanglement measure,
\begin{align}\label{S_function}
	\CS_\CT (R) \equiv \left[ R\,\partial_R -(d-2)\right] \Delta S_\CT (R) \ .
\end{align}
The strong subadditivity Eq.\,\eqref{SSA_general} still holds for the difference $\Delta S_\CT (R)$ thanks to the Markov property of the CFT vacuum that saturates the strong subadditivity for entangling regions whose boundaries are located on a light cone \cite{Casini:2017roe}.
Furthermore, this subtraction plays a twofold role to solve the problems: i)
the UV divergences from the singular intersections are canceled out in the difference $\Delta S_\CT (R)$, and ii) the wiggly spheres can be replaced by the round spheres in the strong subadditivity as the finite mismatch of their entropies cancels out in $\Delta S_\CT (R)$ \cite{Casini:2018kzx}.\footnote{We thank Eduardo Test{\'e} for clarifying this point.}
Hence taking the large $N$ limit of Eq.\,\eqref{SSA_general} with $S$ replaced with $\Delta S_\CT$ results in the monotonicity of $\CS_\CT (R)$ in any dimension,
\begin{align}\label{S_function_Mon}
	\CS'_\CT (R) \le 0 \ .
\end{align}
This new measure reduces to the entropic $c$- and $\CF$-functions when $d=2$ and $3$, respectively, and the inequality \eqref{S_function_Mon} guarantees their monotonicities.

Now we move onto the four-dimensional case and see what we are able to draw from the inequality \eqref{S_function_Mon}.
Recalling Eqs.\,\eqref{EE_4d} and \eqref{c04d}, the entanglement entropy across a two-sphere of radius $R$ in CFT$_4$ is given by
\begin{align}\label{EE_4d_CFT}
	S(R)|_\text{CFT} = \alpha\,\frac{4\pi R^2}{\epsilon^2} - a\, \log \frac{R}{\epsilon} \ .
\end{align}
We are interested in the difference of the $a$-anomalies between the UV and IR fixed points of an RG flow and hence consider the IR CFT as a theory $\CT$ for the measure $\CS_\CT (R)$.
Evaluating the inequality \eqref{S_function_Mon} with the fixed point value Eq.\,\eqref{EE_4d_CFT} we find
\begin{align}
	\CS_\text{IR}'(R) = 2 (a_\text{IR} - a_\text{UV}) \frac{1}{R} \le 0 \ ,
\end{align}
which proves the weak version of the $a$-theorem Eq.\,\eqref{a_theorem_claim}  \cite{Casini:2017vbe}.\footnote{The dilation effective action was used by \textcite{Solodukhin:2013yha} to prove the $a$-theorem using the entanglement entropy of a spherical entangling surface.
}

In $d\ge 4$ dimensions \textcite{Giombi:2014xxa} conjectured the generalized $F$-theorem, stating the generalized $F$ coefficient defined by
\begin{align}
	\tilde F \equiv \sin \left( \frac{\pi\,d}{2}\right)\, \log\,Z[\BS^d] \ ,
\end{align}
decreases along an RG flow,
\begin{align}\label{Generalized_F}
	\tilde F_\text{UV} \ge \tilde F_\text{IR} \ .
\end{align}
We want to examine if the monotonicity Eq.\,\eqref{S_function_Mon} in $d\ge 4$ dimensions also proves the conjecture as in the lower-dimensional cases.

Noting the small $R$ limit that corresponds to the UV region of an RG flow, we found $\CS_\CT (0) = 0$ for any theory $\CT$.
Then the inequality Eq.\,\eqref{S_function_Mon} implies
\begin{align}\label{S_IR_Monotonic}
	\CS_\CT (R) \le 0 \ .
\end{align}
By choosing $\CT$ to be the IR theory, it leads to an inequality that contains the UV divergence of order $O(1/\epsilon^{d-4})$ as seen from the UV structures Eqs.\,\eqref{EE_Even} and \eqref{EE_Odd} in CFT.
Hence we are not able to derive a relation between finite values of the form, Eq.\,\eqref{Generalized_F}.

One can still hope to regularize the UV divergence in Eq.\,\eqref{S_IR_Monotonic} by the dimensional regularization and derive the finite relation \eqref{Generalized_F}.
Indeed the equality Eq.\,\eqref{EE_F} leads to
\begin{align}
	\tilde F|_\text{CFT} = \sin \left( \frac{\pi\,d}{2}\right)\, S_\text{CFT}(R) \ ,
\end{align}
and one can recast Eq.\,\eqref{S_IR_Monotonic} into the inequality for the fixed point values $\tilde F_\text{UV}$ and $\tilde F_\text{IR}$.
While the resulting inequality agrees with Eq.\,\eqref{Generalized_F} for $2< d < 4$, it becomes rather the opposite, $\tilde F_\text{IR} \ge \tilde F_\text{UV}$ for $4 < d < 6$, which is in apparent contradiction to the free field results \cite{Giombi:2014xxa}.
This conflict stems from the fact that the inequality Eq.\,\eqref{S_IR_Monotonic} holds only in the presence of the UV divergences.
It is challenging, at the time of writing this review, to derive an inequality for the dimensionally regularized entanglement entropies, which hopefully proves the generalized $F$-theorem in higher dimensions.

\subsection{Holographic RG flow}

In this section, we switch gears to consider a simple class of holographic RG flows that are dual to Lorentz invariant QFTs in $d$ dimensions and examine the entanglement entropy across a sphere under the flows \cite{Liu:2012eea,Klebanov:2012yf,Liu:2013una}.
A complementary test of the $F$-theorem in three dimensions is performed in the holographic systems.

\subsubsection{Domain wall and gapped RG flows}
The most general metric holographically describing an RG flow on a flat space $\BR^{1,d-1}$ is of Poincar{\' e} type:
\begin{align}\label{RGmetric}
\d s^2 = \frac{L^2}{z^2}\left[  \frac{\d z^2}{f(z)} - \d t^2 + \d{\vec x}^2_{d-1} \right] \ ,
\end{align}
where $\d {\vec x}^2_{d-1}$ is the metric of the $(d-1)$-dimensional flat space $\BR^{d-1}$.
We require the function $f(z)$ approaches to a constant (that we choose to $1$ here),
\begin{align}\label{fto1}
f(z) \xrightarrow[z\to 0]{}1 \ ,
\end{align}
so that the metric asymptotes to an AdS$_{d+1}$ space-time of radius $L$ near the boundary $z=0$.

We are interested in a physical situation where the metric is a solution to the Einstein equation, so hence not every function $f(z)$ satisfying Eq.\,\eqref{fto1} is of interest.
Moreover we should exclude unphysical solutions by introducing a physically sensible condition.

Suppose the metric \eqref{RGmetric} describing an RG flow is a solution in the Einstein gravity coupled to matter fields,
\begin{align}
I = \frac{1}{16\pi G_N} \int \d^{d+1} x \sqrt{-g} \left[ \CR + \CL_\text{matter} \right] \ ,
\end{align}
where $\CL_\text{matter}$ is the Lagrangian of the matter fields.
The metric Eq.\,\eqref{RGmetric} with an arbitrary function $f(z)$ can be a solution to the Einstein equation,\footnote{Our definition of the stress tensor in Lorentzian signature is $T_{\mu\nu} = (2/\sqrt{-g})\,\delta I/ \delta g^{\mu\nu}$.} $T_{\mu\nu}^\text{matter} = \CR_{\mu\nu} - \CR g_{\mu\nu}/2$, by tuning the bulk matter stress tensor $T_{\mu\nu}^\text{matter}$ or equivalently the matter Lagrangian $\CL_\text{matter}$ properly.
This construction, however, often leads to unreasonable solutions that we may want to rule out on a physical ground.
It is common practice to impose the \emph{null energy condition} for making a theory of gravity physically sensible, \cite{Hawking:1973uf}:
\begin{align}\label{NEC}
T_{\mu\nu}^\text{matter} \xi^\mu \xi^\nu \ge 0 \ .
\end{align}
This condition must be met for any future directed null vector $\xi^\mu$ ($\xi^2 = 0$).

Using the Einstein equation $T_{\mu\nu}^\text{matter} = \CR_{\mu\nu} - \CR\, g_{\mu\nu}/2$ and the null vector of the form
\begin{align}\label{NVector}
\xi^z = \sqrt{f(z)} \ , \qquad \xi^t = 1 \ , \qquad \xi^i = 0 \qquad (i\neq t, z) \ ,
\end{align}
one can derive from the null energy condition a constraint,
\begin{align}\label{fmonotonic}
f'(z) \ge 0 \ ,
\end{align}
which implies the monotonicity of $f(z)$.
This constraint Eq.\,\eqref{fmonotonic} restricts a class of solutions available to us, but
is not strong enough to fix the IR geometry in the large $z$ region.
Here we consider the following two possibilities in the $z\to \infty$ limit \cite{Liu:2012eea}:
\begin{enumerate}
\item
The IR geometry is a different fixed point from the UV. 
Namely, the metric \eqref{RGmetric} describes a domain wall between two AdS$_{d+1}$ solutions with radii $L$ and $L_\text{IR}$,
\begin{align}\label{Domain}
f(z) \xrightarrow[z\to \infty]{} \frac{L^2}{L^2_\text{IR}} > 1\ .
\end{align}
\item
The IR geometry is a confining or gapped phase when the metric described by Eq.\,\eqref{RGmetric} has the function with the IR boundary condition,
\begin{align}\label{Gapped}
f(z) \xrightarrow[z\to \infty]{}  b\, z^n \ , \qquad n>0 \ ,
\end{align}
where $b$ is a positive constant.
\end{enumerate}

Now we are concerned with a spherical entangling surface.\footnote{See \textcite{Myers:2012ed} for a similar calculation with an entangling surface of a strip type.}
It is convenient to choose the spatial metric to be in the polar coordinates:
\begin{align}
\d{\vec x}_{d-1}^2 = \d\rho^2 + \rho^2\, \d\Omega_{d-2}^2 \ .
\end{align}
Since the entangling surface is spherical, $\rho$ has to be a function of $z$ to respect the spherical symmetry.
Then the area functional is given by
\begin{align}\label{AreaFunctionalSphere}
\begin{aligned}
\CA &= L^{d-1}\,\text{Vol}(\BS^{d-2})\int_0^{z_\ast} \d z\, \CL\ ,\\
\CL &= \frac{\rho(z)^{d-2}}{z^{d-1}}\sqrt{\rho'(z)^2 + \frac{1}{f(z)}} \ ,
\end{aligned}
\end{align}
where $z_\ast$ is the maximum value of $z$ for the minimal surface.
To find a solution minimizing the area, we impose the boundary condition at the UV fixed point ($z=0$),
\begin{align}
\rho(z) = R + O(z^2) \ ,
\end{align}
while at the IR fixed point there are two types of boundary conditions depending on the topology of the minimal surface (see Fig.\,\ref{fig:Topology}).
\begin{figure}[htbp]
\centering
\begin{tikzpicture}[thick]
	\draw[->] (0,0) --++(4,0) node[right] {\large $z$};
	\draw[->] (0,0) --++(0,3) node[above] {\large $\rho$};
	\draw[very thick, blue!70] (1.5,0) node[below,black] {\large $z_\ast$} arc (0:90:1.5) ;
	\draw[very thick, orange] (0,2) to [out=0, in=180] (4,1);
	\draw[dotted] (0,0.95) node[left] {$\rho_\infty$} --++ (4,0);
\end{tikzpicture}

\caption{Two types of a minimal surface. The dashed blue and solid orange curves are disk and cylinder type surfaces, respectively.}
\label{fig:Topology}
\end{figure}
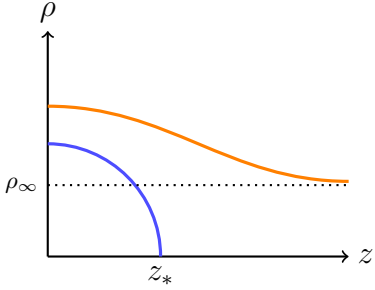
\begin{itemize}
\item
Disk type ($z_\ast < \infty$): 
the minimal surface terminates at $z=z_\ast$ where $\rho$ has the expansion of the form
\begin{align}
\rho(z) =  (z_\ast - z)^{1/2}\left[\rho_\ast + O (z_\ast - z) \right]\ .
\end{align}
\item
Cylinder type ($z_\ast = \infty$):
the minimal surface extends to the infinity as 
\begin{align}
\rho(z) = \rho_\infty + O(1/z) \ .
\end{align}
\end{itemize}
When $f(z)=1$, i.e., the AdS$_{d+1}$ spacetime, the minimal surface is of a disk type as we saw in Eq.\,\eqref{Sp_sol}:
\begin{align}\label{AdSDiskSol}
\rho_0(z) = \sqrt{R^2 - z^2} \ .
\end{align}
Hence we expect to have a disk type solution when the radius $R$ of the entangling surface is small in an asymptotically AdS spacetime.
On the other hand, a cylinder type solution can exist in a gapped geometry with the IR behavior Eq.\,\eqref{Gapped} only when $n>2$ [see Appendix C in \cite{Liu:2012eea} for details].

\subsubsection{Topology change in a gapped phase} 
The minimal surface in the asymptotically AdS spacetime Eq.\,\eqref{RGmetric} is approximated by the solution Eq.\,\eqref{AdSDiskSol} for small $R$.
If the bulk geometry allows the cylinder type minimal surface for large $R$, there must be a topology change at some critical value $R=R_c$.
This is actually a typical phenomenon for a gapped phase Eq.\,\eqref{Gapped} as one can verify as follows.

We can probe the IR region even for small $z$ by letting the parameter $b$ be very large, $b\gg 1$.
The large $b$ expansion of the equation of motion for the area functional Eq.\,\eqref{AreaFunctionalSphere} amounts to, in the leading order, either 
\begin{align}
	\rho'(z)=0 \ , 
\end{align}
or 
\begin{align}
	\rho'(z) = c\, z^{(d-1)/(d-2)} \ ,
\end{align}
for a constant $c$.
The solution to the second equation diverges for large $z$ and is not acceptable as a minimal surface
while the solution to the first implies a cylinder type minimal surface.
The topology change of the minimal surface is observed numerically in many examples \cite{Liu:2012eea,Klebanov:2012yf,Liu:2013una}.

The area of the cylinder type minimal surface may be evaluated in the system with a large gap ($b\gg 1$) by dividing the calculation into the UV part $z\leq z_b \equiv b^{-1/n}$ and the IR part $z> z_b$.
In the UV region, the solution is close to the disk solution Eq.\,\eqref{AdSDiskSol} in the AdS space while it approaches to a constant in the IR region.
Taking into account the continuity at $z=z_b$, it is reasonable to approximate the solution by the piecewise function,
\begin{align}
	\rho(z) = \begin{cases}
			\rho_0(z) & \quad (z\leq z_b) \ ,\\
			\rho_0(z_b) & \quad (z> z_b) \ .
			\end{cases}
\end{align}
Now the area consists of two parts, $\CA = \CA_{z\leq z_b} + \CA_{z\geq z_b}$, where the UV part is approximated in the large $b$ limit by
\begin{align}
\begin{aligned}
	\CA_{z\leq z_b} = \int_{\epsilon}^{z_b}\d z\,&\frac{\rho_0(z)^{d-2}(\rho_0'(z)^2 + 1)^{1/2}}{z^{d-1}} \\
		&\quad \cdot \left[ 1 - \frac{b\,z^n}{2(\rho_0'(z)^2 +1)}\right] \ ,
\end{aligned}
\end{align}
while	 the IR part becomes
\begin{align}
	\CA_{z\geq z_b} = \int_{z_b}^\infty \d z\, \frac{\rho_0(z_b)^{d-2}}{b^{1/2}\,z^{d-1 + n/2}} \ .
\end{align}
Using the dimensional regularization assuming $n$ is large enough, we have the expansion for $z_b\ll 1$,
\begin{align}\label{EE_Sphere_Gap}
	\CA &= \frac{1}{2d + n -4} \left[2 \left(\frac{R}{z_b}\right)^{d-2}  - (d-2) \left(\frac{R}{z_b}\right)^{d-4} + \cdots \right] \ .
\end{align}
The entanglement entropy across a sphere of radius $R$ is $L^{d-1}\text{Vol}(\BS^{d-2})/(4G_N)$ times larger that this area.
Compared with Eq.\,\eqref{EE_Scalar_HK} for a free massive scalar field, the parameter $z_b$ allows for interpretation as a gap scale $m$ as expected.

Similar topology change can be observed for a strip entangling region in a gapped system where the entangling surface is two parallel planes.
There are two types of extremal surfaces, one with a connected surface, and the other with two disconnected surfaces terminate on the IR boundary of the gapped geometry.
The topology change between these two types of surfaces is interpreted as a confinement/deconfinement phase transition in a confining gauge theory \cite{Nishioka:2006gr,Klebanov:2007ws,Pakman:2008ui}.

\subsubsection{Testing the $F$-theorem in holography}
As seen in Sec. \ref{ss:Ftheorem}, the proof of the $F$-theorem was based on the Lorentz invariance and the strong subadditivity of entanglement entropy that replaces the role of unitarity in QFT. 
The same argument for the proof remains to work in a holographic system as well just by resorting to the strong subadditivity of the Ryu-Takayanagi formula (for the static case in Sec. \ref{ss:HEE_Ineq}).
Nevertheless, it is worthwhile examining the validity of the $F$-theorem from  the holographic point of view to draw implications for the interplay between unitarity in QFT and its counterpart in the dual gravity.

First let us consider a holographic gapped system with a large gap
that we dealt with in the previous section.
The renormalized entanglement entropy, Eq.\,\eqref{REE}, across a sphere of large radius is calculated from Eq.\,\eqref{EE_Sphere_Gap} with $d=3$ as,
\begin{align}\label{HREE_Gap}
	\CF(R) = \frac{\pi\,L^2}{(n+2)\,G_N}\frac{z_b}{R} + O\left( (z_b/R)^{3}\right) \ .
\end{align}
This result should be compared with the typical behavior of the $\CF$-function in the large gap limit given in Eq.\,\eqref{LargeMassExpansionScalar} for a free massive scalar.
Note that the value of $\CF$ (the sphere free energy) at the UV fixed point of the holographic RG flow is $F_\text{UV} = \pi L^2/(2G_N)$ [see Eq.\,\eqref{CFT_F_value}], which is always greater than Eq.\,\eqref{HREE_Gap} in the large gap system with $z_b \ll R$.
We thus find this type of holographic RG flow is consistent with the $F$-theorem.

The next example to consider is a domain wall RG flow interpolating two closely separated fixed points without a gap \cite{Liu:2012eea}.
Such a solution can be realized by choosing the function $f(z)$ in the general metric Eq.\,\eqref{RGmetric} to be of the form
\begin{align}
\begin{aligned}
f(z) &= 1 + \eta\, g(z) \ , \qquad \eta = \frac{L^2}{L_\text{IR}^2} - 1 \ll 1 \ , 
\end{aligned}
\end{align}
and imposing the boundary conditions at the UV and IR regions,
\begin{align}
	 g(z) &\xrightarrow[z\to 0]{}0 \ ,\qquad g(z) \xrightarrow[z\to \infty]{}1 \ .
\end{align}
The null energy condition Eq.\,\eqref{fmonotonic} indicates $g(z)$ to be a monotonically increasing function,
\begin{align}\label{Mon_g}
	g'(z)\ge 0 \ .
\end{align}
In this setup, we are allowed to expand the solution of the equation of motion of the area functional Eq.\,\eqref{AreaFunctionalSphere} around the UV solution Eq.\,\eqref{AdSDiskSol},
\begin{align}
\rho(z) = \rho_0(z) + \eta\, \rho_1(z) + O(\eta^2) \ ,
\end{align}
and it follows the expansion of the area functional,
\begin{align}
\CA = A_0 + \eta \,A_1 + O(\eta^2) \ .
\end{align}

We now focus on the $d=3$ case to inspect the behavior of the $\CF$-function.
Varying Eq.\,\eqref{AreaFunctionalSphere} around the UV solution, we obtain the subleading term of the form\footnote{There are contributions from the boundaries at $z=z_\ast$ and $0$, which turns out to be negligible for $d\le 4$ \cite{Liu:2012eea}.}
\begin{align}
\begin{aligned}
A_1 &= 2\pi L^{2}\int_0^{R} \d z\, g(z)\, \frac{\delta \CL}{\delta f}\bigg|_{f=1,\,\rho=\rho_0} \ , \\
	&= - \frac{\pi L^2}{R} \int_0^R \d z\, \frac{R^2 - z^2}{z^2}\,g(z) \ .
\end{aligned}
\end{align}
This integral is UV divergent due to the contribution near the UV boundary, $z=0$,
but we can regularize it with the $\CF$-function and find the leading correction,
\begin{align}\label{DWF}
\D\CF (R) = -\eta\, \frac{\pi L^2}{2G_N} \int_0^1 \d z \, g(zR) \ .
\end{align}
which is always nonpositive due to Eq.\,\eqref{Mon_g}.
We thus find the $\CF$-function of the domain wall solution decreases as $R$ increases.
Furthermore, the leading correction matches the difference between the UV and IR free energies,
\begin{align}
\begin{aligned}
\D \CF (R\to\infty) &= -\eta\, \frac{\pi L^2}{2G_N}\ , \\
	& = F_\text{IR} - F_\text{UV} + O(\eta^2)\ .
\end{aligned}
\end{align}
Hence the domain wall RG flow realizes the strong version of the $F$-theorem.\footnote{The entanglement entropy of a holographic RG flow describing a relevantly perturbed CFT is shown to have the expansion of the form \eqref{CP_EE} with $s_1=0$ and supports the strongest version of the $F$-theorem at least around the UV fixed point when the dimension of the relevant operator is close to $3$ \cite{Taylor:2016aoi,Taylor:2016kic}.}

We note that the monotonicity of the $\CF$-function is a direct consequence of the null energy condition Eq.\,\eqref{fmonotonic} in this holographic RG flow while it follows  in the field theoretic derivation in section \ref{ss:Ftheorem} from the strong subadditivity that is substituted by the minimality condition for the Ryu-Takayanagi surface.
The derivation for Eq.\,\eqref{DWF} indicates that the minimality condition is not enough to prove the monotonicity of the $\CF$-function in holography.
To our best knowledge, no holographic proofs of the strong version of the $F$-theorem are known, hence 
it is desirable to elucidate it with the null energy condition along the line of \textcite{Freedman:1999gp,Girardello:1998pd} beyond the perturbative level.

\subsection{Supersymmetric R{\'e}nyi entropies}\label{ss:SRE}
We conclude this section with a comment on some exact results for entanglement and R{\'e}nyi entropies in interacting field theories with supersymmetries.

Calculating the exact value of entanglement entropy in QFTs is generally an intractable task unless theories are free and the entangling surface preserves a rotational symmetry around it.
The difficulty is most manifest in the replica method where the conical singularity on the entangling surface prevents us from carrying out the perturbative calculation in a conventional manner.
On the other hand, there are powerful computational techniques, called \emph{supersymmetric localizations}, in supersymmetric field theories that allow the exact calculation of the partition function and a certain class of correlation functions by reducing the infinite-dimensional path integral to a finite-dimensional one \cite{Willett:2016adv,Dumitrescu:2016ltq,Pufu:2016zxm}.
To apply supersymmetric localization, a supersymmetric field theory must be placed on manifolds preserving enough number of supersymmetries.

A simple example is a round sphere and the resulting supersymmetric partition function $Z^\text{susy}[\BS^d]$ turns out to be the partition function $Z[\BS^d]$ of the supersymmetric field theory at the IR fixed point.
Given the relation \eqref{EE_F} between the entanglement entropy across a round sphere and the sphere partition function $Z[\BS^d]$ for CFTs, it is possible to calculate the exact value of entanglement entropy for a class of supersymmetric field theories \cite{Jafferis:2011zi,Pufu:2016zxm}.

Given the success of the entanglement entropy, one may wonder if the same story applies to the R{\'e}nyi entropy, Eq.\,\eqref{Renyi_QFT}, for supersymmetric field theories as well.
Such an attempt would fail as supersymmetries are completely broken on the $n$-fold cover of the $d$-sphere $\BS^d_n$ [defined by Eq.\,\eqref{CFT_Sphere}], thus the supersymmetric localization cannot be applied.

In order to leverage the supersymmetric localization technique, we would rather tailor the definition of the R{\'e}nyi entropy so as to be compatible with supersymmetries.
Namely, we define the \emph{supersymmetric R{\'e}nyi entropy} $S_n^\text{susy}$ for supersymmetric conformal field theories by \cite{Nishioka:2013haa},
\begin{align}\label{SRE_def}
	S_n^\text{susy} \equiv \frac{1}{1-n} \log \bigg| \frac{Z^\text{susy}[\BS^d_n]}{(Z^\text{susy}[\BS^d])^n} \bigg| \ ,
\end{align}
where $Z^\text{susy}[\BS^d]$ is the supersymmetric partition function on $\BS^d_n$ with background fields turned on to preserve a part of supersymmetries.\footnote{The effect of the background fields vanishes for CFT on a round sphere, hence $Z^\text{susy}[\BS^d] = Z[\BS^d]$.}
This definition takes the same form as the R{\'e}nyi entropy Eq.\,\eqref{Renyi_QFT} across a spherical entangling surface $\BS^{d-2}$ in CFT on $\BR^d$ [see Eq.\,\eqref{F_nZ}], except for the partition function being replaced with the supersymmetric one.\footnote{
The absolute value inside the logarithm ensures that the supersymmetric R{\'e}nyi entropy is universal at least in three dimensions \cite{Closset:2012vg}.}

The explicit formula for $Z^\text{susy}[\BS^d]$ is given by an integration over finite-dimensional matrix variables that depends on the spacetime dimensions, the rank of the gauge group, and the matter contents of supersymmetric field theories, but we are not concerned with the explicit form here.
Instead of giving examples, we recap the universal aspects of the supersymmetric R{\'e}nyi entropy for $\CN= 2$ supersymmetric field theories with the global $\U(1)_R$ symmetry in three dimensions \cite{Nishioka:2013haa}:
\begin{itemize}
\item The expansion around $n=1$ takes the form
\begin{align}
	S_n^\text{susy} = S_1 + \frac{\pi^2}{16}\,\tau_{RR}\,(n-1) + O\left((n-1)^2 \right) \ ,
\end{align}
where $\tau_{RR}$ appears as the coefficients $C_J = \tau_{RR}/(4\pi^2)$ and $C_T = 3 \tau_{RR}/(2\pi^2)$ in the correlators of the $\U(1)_R$-symmetry current $j_\mu^{(R)}$ and the stress tensor given by Eqs.\,\eqref{jjCor} and \eqref{TTCor}.
Compared with the expansion Eq.\,\eqref{REexpand} of the R{\'e}nyi entropy with the first derivative Eq.\,\eqref{LeadingRE} in $d=3$ dimensions,  
\begin{align}
	S_n = S_1 + \frac{\pi^2}{8}\, \tau_{RR}\,(n-1) + O\left((n-1)^2 \right) \ ,
\end{align}
we find a factor $2$ discrepancy at the leading order coefficient.
This is due to the effect of the background gauge field for the $\U(1)_R$ symmetry that is turned on to maintain supersymmetries on $\BS^d_n$. 
Hence the supersymmetric R{\'e}nyi entropy is \emph{different} from the R{\'e}nyi entropy when $n\neq 1$ in general.
\item For theories with gauge groups of rank $N$, the $n$ dependence simplifies in the large-$N$ limit,
\begin{align} 
	S_n^\text{susy} = \frac{3n+1}{4n}\, S_1 \ ,
\end{align}
A curious fact is that the ratio $H_n\equiv S_n^\text{susy}/S_1$ satisfies the inequalities Eqs.\,\eqref{RenyiIneq1}-\eqref{RenyiIneq4} of the R{\'e}nyi entropies.
\item For a certain class of superconformal field theories allowing for the holographic dual description, the gravity dual of the supersymmetric R{\'e}nyi entropy is described by the charged topological AdS$_4$ black hole that is a half-BPS solution in the $\CN= 2$ $\U(1)$ gauged four-dimensional supergravity \cite{Nishioka:2014mwa,Huang:2014gca}.
\item The description by codimension-two defects living on the entangling surface is manifest \cite{Nishioka:2016guu}.
\end{itemize}

The supersymmetric R{\'e}nyi entropies and their gravity duals have been explored with applications in the other dimensions \cite{Mori:2015bro,Giveon:2015cgs,Huang:2014pda,Zhou:2015cpa,Hama:2014iea,Alday:2014fsa,Nian:2015xky,Zhou:2015kaj,Yankielowicz:2017xkf}.
We believe they shed new light on understanding the nature of quantum entanglement in QFT and the fundamental connection of quantum states to the holographic spacetime.

  \section{Outlook}\label{ss:Outlook}
In this review, we put emphasis on the roles of quantum inequalities of entanglement entropy in constraining the dynamics of RG flows in QFTs.
In particular, the monotonicity of $\CC$-functions built from entanglement entropy follows from the strong subadditivity, which should be contrasted with Zamolodchikov's $c$-theorem and the $a$-theorem in two- and four-dimensions whose proofs are based on unitarity in QFT.
Clearly the strong subadditivity assumes almost an equal role with unitarity when applied to QFTs, and indeed it is a consequence of the Hermicity of density matrices in finite-dimensional quantum mechanical systems \cite{Lieb:1973cp,araki1976relative}.
It is thus important to see whether the strong subadditivity can be proved by a conventional method within the canonical framework of unitary and Lorentz invariant QFTs.

Some attempts for $\CC$-theorems in higher dimensions were discussed in Sec. \ref{ss:C_theorem_Higher} where a certain class of monotonic functions were constructed out of entanglement entropy.
They result in the known $\CC$-theorems in less than four-dimensions, but do not lead to the conjectured $F$- and $a$-theorems in higher dimensions.
It deserves further investigation to see if the conjectures can be proved by a similar method to the lower-dimensional case.

It would also be interesting to consider $\CC$-theorems for RG flows across dimensions.
This possibility is anticipated on general grounds, but has attracted less attention so far \cite{Gukov:2015qea,Bobev:2017uzs}.
It is not clear at first sight what can be a measure of degrees of freedom, i.e.\, a $\CC$-function for RG flows across dimensions.
A naive speculation is to use as a measure the generalized $F$ coefficient defined by Eq.\,\eqref{Generalized_F} possibly with a dimension-dependent normalization factor.
It would be intriguing to explore this direction and find an appropriate measure that may or may not have a relation to entanglement entropy.
  
\begin{acknowledgments}
I am grateful to Shamik Banerjee, Chris Herzog, Igor Klebanov, Yuki Nakaguchi, Silviu Pufu, Ben Safdi, Tadashi Takayanagi, Amos Yarom and Kazuya Yonekura for stimulating collaborations on the various works partly reviewed in this article, and to Horacio Casini, Marina Huerta, and Noriaki Ogawa for valuable comments and discussions.
I also thank the organizers of the $18^{th}$ APCTP Winter School on Fundamental Physics, the $9^{th}$ Asian Winter School on String, Particles and Cosmology and the $6^{th}$ Advanced String School at Puri for giving me the opportunities to deliver lectures and also the participants for many comments and questions.
This work was supported in part by the JSPS Grant-in-Aid for Young Scientists (B) No.15K17628 and the JSPS Grant-in-Aid for Scientific Research (A) No.16H02182.
\end{acknowledgments}

\appendix
  \section{Real time formalism for fermions}\label{ss:RTFermion}
In this Appendix, we extend the real time formalism, developed for free bosonic systems in Sec. \ref{ss:RTapproach}, to free fermionic systems.

\subsection{Fermionic system on lattice}
We consider a quadratic Hamiltonian of fermions on lattice labeled by an integer $i=1,\cdots, N$,
\begin{align}
	H = \sum_{i,j = 1}^N \psi^\dagger_i\, M_{ij}\, \psi_j \ ,
\end{align}
with fermionic operators satisfying the anticommutation relation $\{\psi_i, \psi_j^\dagger \} = \delta_{ij}$.
The Hermitian matrix $M$ can be diagonalized to $M=U^\dagger \Delta \,U$ by a unitary matrix $U$, and then the Hamiltonian becomes
\begin{align}
	H = \sum_{l=1}^N\, \lambda_l \,\tilde\psi_l^\dagger\, \tilde\psi_l \ , \qquad \tilde\psi = U\, \psi \ ,
\end{align}
where $\lambda_l$ are real diagonal entries of $\Delta$ and the new operators $\tilde\psi_l$ satisfy the anti-commutation relation $\{\tilde\psi_l, \tilde\psi_m^\dagger \} = \delta_{lm}$.
We let $\CS^{(+)}$ and $\CS^{(-)}$ be the sets of indices for positive and negative eigenvalues, respectively.
We define the ground state as the Dirac sea 
\begin{align}
	|0\rangle \equiv \prod_{l\in \CS^{(-)}} \tilde\psi_l |\tilde 0 \rangle \ ,
\end{align}
where $|\tilde 0 \rangle$ is the state annihilated by all $\tilde\psi_l$.
Then the two-point function of the original fermionic operators $\langle \psi_i \psi_j^\dagger \rangle = \langle 0 | \psi_i \psi_j^\dagger |0 \rangle$ is
\begin{align}\label{FermC}
	C_{ij} \equiv \langle \psi_i \psi_j^\dagger \rangle = \left[ U\, \Theta (\Delta)\, U^\dagger\right]_{ij} \ ,
\end{align}
where $\Theta$ is the step function defined by
\begin{align}
	\Theta (\Delta)_{lm} = \left\{\begin{array}{cl}
						\delta_{lm} & \quad \text{for}\quad l, m \in \CS^{(+)} \ , \\
						0 & \quad \text{otherwise} \ .
					\end{array}\right.
\end{align}

Now we consider the entanglement entropy between the subregion $A$ and its compliment.
As in the bosonic case, we assume the reduced density matrix $\rho_A$ is generated by the modular Hamiltonian $H_A= \sum_{i,j \in A} \psi_i^\dagger\, G_{ij}\, \psi_j$ \cite{2003JPhA...36L.205P} as 
\begin{align}
	\rho_A = \CN\, e^{-H_A} \ ,
\end{align}
where $\CN$ is a normalization constant such that $\tr_A (\rho_A) = 1$.
We diagonalize $G$ by a unitary matrix $V$ and rewrite the modular Hamiltonian as $H_A= \sum_{I} \epsilon_I \, d_I^\dagger d_I$ where new operators $d_I = V_{Ii} \,\psi_i$ for $i\in A$ satisfy $\{ d_I, d_J^\dagger \} = \delta_{IJ}$.
In this basis, the normalization constant $\CN$ is fixed to be $\CN = \prod_{I} (1+ e^{-\epsilon_I})^{-1}$.
The two-point function $C_{ij}$ for $i,j\in A$ is evaluated with the modular Hamiltonian, which is diagonalized to be
\begin{align}
	C_{ij} = N\,\tr_A \left( \psi_i \psi_j^\dagger \, e^{-\CH}\right) = V^\dagger_{iI} \,\frac{\delta_{IJ}}{1 + e^{-\epsilon_I}}\, V_{Jj} \ .
\end{align}
Thus the eigenvalues $\epsilon_I$ can be determined by those $\nu_I$ of $C$,
\begin{align}
	\nu_I =\frac{1}{1 + e^{-\epsilon_I}} \ .
\end{align}
The trace of the $n$th power of $\rho_A$ is
\begin{align}
\begin{aligned}
	\tr_A(\rho_A^n) &= \prod_I \frac{1+ e^{-n \epsilon_I}}{(1+ e^{-\epsilon_I})^n}\ , \\
		&= \prod_I \left[ \nu_I^n + (1 - \nu_I)^n \right] \ ,
\end{aligned}
\end{align}
and we obtain the entanglement entropy Eq.\,\eqref{RenyiEE} given by the matrix $C$,
\begin{align}\label{FermCformula}
	S_A = - \tr_A \left[ C \log C + (1-C) \log (1-C) \right] \ .
\end{align}

\subsection{Free massive fermionic fields}
We proceed to deal with free massive fermions whose action is
\begin{align}
	I = \int \d^dx\sqrt{-g} \, \Psi^\dagger \Gamma^0(\i\,  \Gamma^A e_A^M \nabla_M - \i\, m ) \Psi  \ ,
\end{align}
where $A$ is the local Lorentz index in the range $A = 0,\cdots, d-1$ and $M$ stands for the coordinate indices.
The gamma matrices obey the anti-commutation relations $\{ \Gamma_A, \Gamma_B\} = 2\eta_{AB}$ for the metric $\eta = \text{diag} (-1, 1, 1, \cdots, 1)$.
$e^A$ are the vielbein one-form and $\nabla_\mu$ is the spinor covariant derivative defined by $\nabla_M \equiv \partial_M + \omega_M^{AB}[\Gamma_A, \Gamma_B]/8$ with the spin connection two-form $\omega^{AB}$ satisfying $\d e^A + \omega^{A}_{~B}\wedge e^B = 0$.

We concentrate on the case in $d\ge 3$ dimensions. 
See, for example, \textcite{Herzog:2013py} for the implementation in two dimensions.
We set the theory on the radial coordinates Eq.\,\eqref{RadialLattice} and choose the vielbein as
\begin{align}
\begin{aligned}
	e^0 &= \d t \ , \qquad e^{d-1} = \d r \ , \\
	e^a &= r\, \hat e^a\quad (a=1,\cdots,d-2) \ ,
\end{aligned}
\end{align}
where $\hat e^a$ are the vielbein on $\BS^{d-2}$.
The Hamiltonian becomes
\begin{align}
\begin{aligned}
	H = &\int \d\Omega_{d-2}\,\d r\,r^{d-2} \\
		&\cdot \Psi^\dagger \Gamma^0\left[ - \i\,  \Gamma^{d-1} \left(\partial_r + \frac{d-2}{2r}\right) -  \frac{\i}{r} \Gamma^a \hat e_a^\mu \hat \nabla_\mu  - \i\, m\right] \Psi  \ ,
\end{aligned}
\end{align}
where $\hat \nabla_\mu$ and $m$ are the spinor covariant derivative and the coordinate index on $\BS^{d-2}$, respectively.
We want to reduce this to a number of (1+1)-dimensional fermions on $r$.
We choose the basis of the gamma matrices as follows:
\begin{align}
	\Gamma^0 = \i\,{\bf 1} \otimes \sigma_3\ , \qquad 
	\Gamma^{d-1} = {\bf 1} \otimes  \sigma_2 \ , \qquad 		
	\Gamma^a = \gamma^a \otimes \sigma_1\ ,
\end{align}
where $\gamma^a$ and ${\bf 1}$ are the gamma matrices and the identity matrix in $(d-2)$-dimensions satisfying $\{\gamma_a, \gamma_b\} = 2\delta_{ab}$, and $\sigma_i~(i=1,2,3)$ the Pauli matrices.
Then there appears the Dirac operator $\nslash_{\BS^{d-2}}\equiv \gamma^a \hat e_a^\mu \hat \nabla_\mu$ on $\BS^{d-2}$ in the second term in the angle bracket. 
It is known that the eigenfunction $\psi_l$ of the Dirac operator $\nslash_{\BS^{d}}$ on a unit sphere $\BS^{d}$ is labeled by a non-negative integer $l\ge 0$ with eigenvalues \cite{Camporesi:1995fb},
\begin{align}
	\i\,  \nslash_{\BS^{d}} \chi_{l,d}^{(s)} = s \left( l + \frac{d}{2}\right) \chi_{l,d}^{(s)} \ ,
\end{align}
where $s=\pm$ and their degeneracy $g_f(d,l)$ is 
\begin{align}\label{FermiDeg}
	g_f(d,l) = \frac{2^{[d/2]}\, \Gamma(l + d )}{l! \,\Gamma (d)}  \ .
\end{align}
We expand the Dirac fermion by the eigenfunctions
\begin{align}
	\Psi_l^{(s)} =  \chi^{(s)}_{l,d-2} \otimes \left[ r^{1-d/2}\,\psi_l^{(s)}(r)\right] \ ,
\end{align}
so that $\psi_l^{(s)}(r)$ satisfies the anticommutation relation
\begin{align}
	\left\{ \psi_l^{(s)}(r), \left(\psi_{l'}^{(s')} (r')\right)^\dagger\right\} = \i\, \delta_{ss'}\delta_{ll'}\delta (r-r')\ ,
\end{align}
and rewrite the Hamiltonian as the sum over the angular modes,
\begin{align}
	H &= \sum_{l=0}^\infty  \sum_{s=\pm} g_f(d-2,l)\, H_{l}^{(s)} \ , 
\end{align}
with
\begin{align}
\begin{aligned}
	H_{l}^{(s)} =   \int \d r\, \Bigg[ &-\frac{\i}{2} \left(\psi_l^{(s)}(r)\right)^\dagger \sigma_1\, \partial_r \psi_l^{(s)}(r) \\
		&\quad + \frac{\i}{2} \partial_r \left(\psi_l^{(s)}(r)\right)^\dagger \sigma_1\,  \psi_l^{(s)}(r) \\
	&\qquad +  \frac{s}{r} \left( l + \frac{d-2}{2}\right) \left(\psi_l^{(s)}(r)\right)^\dagger  \sigma_2\, \psi_l^{(s)}(r) \\
	&\qquad\quad + m\, \left(\psi_l^{(s)}(r)\right)^\dagger \sigma_3\,  \psi_l^{(s)}(r) \Bigg]   \ .
\end{aligned}
\end{align}
To discretize the radial coordinate $r$ to lattice labeled by $j=1,2,\cdots, N$ with lattice spacing $a$, we replace 
\begin{align}
	\begin{aligned}
		r &\to j a \ , \qquad &  \delta (r - r') &\to \frac{\delta_{jk}}{a} \ , \\
		\psi_l^{(s)}(r) &\to \frac{ \psi^{(s)}_{l,j} }{\sqrt{a}}\ , \qquad & \partial_r\, \psi_l^{(s)} (r) &\to \frac{\psi^{(s)}_{l,j+1} - \psi^{(s)}_{l,j}}{a}  \ , 
	\end{aligned}
\end{align}
and then the discretized Hamiltonian reads
\begin{align}
	H_l^{(s)} = \frac{1}{a}\sum_{j,k=1}^N \left(\psi^{(s)}_{l,j}\right)^\dagger \, \left(M_l^{(s)}\right)^{j,k} \,\psi^{(s)}_{l,k} 
\end{align}
with the Hermitian matrix $M_l^{(s)}$ given by
\begin{align}\label{HermFermiM}
	\begin{aligned}
		\left(M_l^{(s)}\right)^{k,k} &= \frac{s}{k} \left( l + \frac{d-2}{2}\right) \sigma_2  + (ma)\sigma_3 \ , \\
		\left(M_l^{(s)}\right)^{k,k+1} &= - \left(M_l^{(s)}\right)^{k+1, k} = - \frac{\i}{2}\sigma_1 \ .
	\end{aligned}
\end{align}

Finally the entanglement entropy of the free massive Dirac fermion is given by the sum over the angular modes $l$,
\begin{align}
	S_A = \sum_{l=0}^\infty  \sum_{s=\pm} g_f(d-2,l)\, S_l^{(s)} \ ,
\end{align}
where $S_l^{(s)}$ is the entropy of the $l^{th}$ mode given by the formula \eqref{FermCformula} with the two-point function $C_l^{(s)}$ of the form \eqref{FermC} for the Hermitian matrix $M_l^{(s)}$, Eq.\,\eqref{HermFermiM}.

\bibliography{RMP_EE}

\end{document}